\documentclass[12pt]{article}
\usepackage{epsf}
\usepackage{graphicx}
\usepackage{lscape}
\usepackage{amssymb}
\usepackage{pazh_eng}
\usepackage{sidecap}
\usepackage{float}
\usepackage{pstricks}

\tightenlines

\def\a{$^{\mbox{\small a}}$}
\def\b{$^{\mbox{\small b}}$}

\begin{document}

\title{\bf Timing Characteristics of the Hard X-ray Emission from Bright X-ray Pulsars Based on INTEGRAL Data}

\author{\bf \hspace{-1.3cm}\copyright\, 2009  \ \
A.A.Lutovinov\affilmark{1}$^{\,*}$, S.S.Tsygankov\affilmark{2,1}}
 \affil{$^1$ {\it Space Research Institute, Russian Academy of Sciences, Profsoyuznaya ul. 84/32, Moscow 117997, Russia}\\
$^2$ {\it Max-Planck-Institut fur Astrophysik, Karl-Schwarzschild-Str. 1, Postfach 1317, D-85741 Garching, Germany}}

\vspace{2mm}

Received June 17, 2008

\sloppypar
\vspace{2mm}
\noindent

%\abstract

We review the results of a timing analysis of the observations for ten bright X-ray pulsars
(with fluxes $>100$ mCrab in the 20-100 keV energy band) that fell within the INTEGRAL field of view
from 2003 to 2007. The dependence of the pulse profile on the energy and intrinsic source luminosity has
been investigated; particular attention has been paid to searching for changes in the pulse profile near the
cyclotron frequency. The dependence of the pulsed fraction for X-ray pulsars on their luminosity and energy band has been studied in detail for the first time.


\vfill

{$^{*}$ E-mail: aal@iki.rssi.ru}
\newpage
\thispagestyle{empty}
\setcounter{page}{1}

\section*{INTRODUCTION}

According to the universally accepted theory of
accretion onto rapidly rotating neutron stars with a
strong magnetic field, the infalling matter from the
companion star is braked on the Alfven surface by the
magnetic pressure, becomes frozen into the magnetic
field, and falls along the field lines into narrow ringlike
regions at the magnetic poles of the neutron star,
producing accretion columns and releasing its gravitational
energy in the X-ray wavelength range. In that
case, as the pulsar rotates about its axis, an observer
will register pulses on the light curve differing in shape
and features, depending on the specific physical and
geometrical conditions both near the neutron star
surface and in the path of signal propagation. The
pulsed fraction will depend on the configuration of the
emitting regions, the dipole position relative to the
observer, the energy, etc.

As was pointed out in early papers on this subject
(see, e.g., Wang and Welter 1981; White et al.
1983), the observed pulse profiles differ significantly
for different sources, have a wide variety of shapes,
depending on the energy and source luminosity, and,
in some cases, can shift by as much as 180\deg~ in phase
as the energy band changes. In addition, they are also
variable on the scale of a single pulse (see, e.g., Frontera
et al. 1985; Tsygankov et al. 2007). Nevertheless,
Bulik et al. (2003) made an attempt at a ``broad''
classification of pulse profiles, according to which
each pulsar can be assigned to the subclass of either
single- or double-peaked profiles. This classification
is based on the fact that an observer sees either one
or both magnetic poles, respectively, as the neutron
star rotates. This difference may not be so distinct at
low energies, but single- and double-peaked profiles
can be clearly separated above $\sim10$ keV, where the
effect of absorption may be neglected, or, in the case of
pulsars with a cyclotron feature, above the cyclotron
energy.

A strong magnetic field ($10^{11}-10^{13}$ G) near the
emitting regions on the neutron star surface is the
source of peculiar features in the observed properties
of X-ray pulsars. In particular, the position of the
cyclotron resonance scattering feature in the pulsar
spectrum is a direct source of information about the
magnetic field strength. In this way, the magnetic
field on the X-ray pulsar Her X-1 was measured for
the first time (Truemper et al. 1978). In addition,
the properties of the accreting plasma can change
sharply at the cyclotron frequency, which can be manifested
as a change in the beam pattern of emission
(Meszaros and Nage 1985). The corresponding
changes in the pulse profile near the cyclotron energy
were observed in some of the sources (see, e.g., Tsygankov
et al. (2006) and references therein).

Basko and Sunyaev (1975, 1976) showed that the
beam pattern of emission from an accretion column
depends on the presence of a shock in it. Thus, for example,
a shock in which the infalling matter is braked
emerges above the neutron star surface at a high
luminosity ($\gtrsim10^{37}$ erg/s). The emitting plasma is
accumulated in the zone under the shock and the
emission emerges mainly through the side surfaces
of a accretion column with a fan beam pattern. At
lower luminosities, the matter can be braked near the
neutron star surface and a pencil beam pattern of
emission will be more likely due to the effects of a
strong magnetic field. At intermediate luminosities,
the beam pattern will be a combination of fan and
pencil beams. The beam pattern and, accordingly, the
pulse profile can change with energy band, since the
hotter layers of matter are closer to the neutron star
surface, where the emission formation conditions can
differ from those in higher layers (Basko and Sunyaev
1976).

Obviously, changes in the beam pattern and in
the emission formation conditions and geometry will
lead to a dependence of the pulsed fraction on both
source luminosity and energy. This was recognized
and measured by several authors back in the 1980-1990s.
 However, only with the advent of the modern
RXTE and INTEGRAL observatories, with high
time and energy resolutions (especially in the hard energy
band, where the observed emission is unaffected
by photoabsorption and depends only on the system
geometry and physical conditions in its formation
region), did systematic studies become possible. In
particular, using RXTE data, Tsygankov et al. (2007)
showed that the pulsed fraction for 4U0115+63 decreases
with increasing source luminosity and increases
with energy. The increase is not monotonic
but has features near the harmonics of the cyclotron
absorption line: its excess above the general trend
is clearly observed near the fundamental harmonic;
similar features are also observed near the higher
harmonics, but they can be slightly shifted relative to
the center of the cyclotron line harmonic. An increase
in the pulsed fraction with energy was also reported
for several other pulsars: GX 1+4 (Ferrigno et al.
2007), OAO 1657-415 (Barnstedt et al. 2008), and
EXO 2030+375 (Klochkov et al. 2008a).

This is the next paper in our series of papers
(Tsygankov et al. 2007; Lutovinov and Tsygankov
2008) devoted to the investigation of the pulse profiles
and pulsed fractions for X-ray pulsars. Using
the methods proposed and described by Tsygankov
et al. (2007), we investigated the ten (4U 0115+63,
V 0332+53, A 0535+262, Vela X-1, Cen X-3,
GX 301-2, OAO 1657-415, Her X-1, GX 1+4,
EXO 2030+375) brightest (with fluxes $>100$ mCrab)
X-ray pulsars based on INTEGRAL data in the
hard ($>20$ keV) energy band. Here, we analyze in
detail the dependence of the pulsed fraction for X-ray
pulsars on their luminosity and energy band and
the effect of resonance absorption near the cyclotron
line on it for the first time. In addition, we obtained
pulse profiles for the above pulsars in several energy
bands, relative intensity maps of pulse profiles, and
dependences of the pulsed fraction (PF) on energy for
all of the possible observed states and luminosities
and compiled the corresponding catalog. Since this
is a fairly large data set and in order not to overload
the paper, we provide here only the averaged (if the
profile did not change significantly with luminosity)
or most typical pulse profiles, intensity maps, and PF
dependences on the energy. The complete catalog is freely accessible
at http://hea.iki.rssi.ru/integral/pulsars and can
be used to construct and test radiation models for X-ray
pulsars.

\section*{OBSERVATIONS AND DATA ANALYSIS}

Data from the IBIS telescope (the ISGRI detector;
Lebrun et al. 2003) of the INTEGRAL observatory
(Winkler et al. 2003) were used to investigate the
pulse profiles in the hard X-ray energy band.

\pagebreak
%****************************************************************

%\begin{landscape}

\begin{table}[h]

 \centering
{\bf Table 1.  }{Bright X-ray pulsars from INTEGRAL data}

\vspace{2mm}
\begin{tabular}{c|c|c|c|c|c}
\hline
\hline
{\scriptsize  Pulsar}&{\scriptsize Observation,}&{\scriptsize Middle}&{\scriptsize Exposure,}&{\scriptsize Flux\a,}&{\scriptsize Luminosity\b,}\\
{\scriptsize name}&{\scriptsize orbits}&{\scriptsize of observations, MJD}&{\scriptsize ks}&{\scriptsize $\times10^{-9}$  erg s$^{-1}$ cm$^{-2}$}&{\scriptsize $10^{37}$ erg s$^{-1}$}\\
\hline
1&2&3&4&5&6\\
\hline
4U 0115+63 & 238 & 53273.6 & 97.0 &$12.9^{+0.3}_{-0.2}$ & $7.6^{+0.2}_{-0.1}$ \\
\hline
 & 272 & 53376.5 &23.7 &$78.4^{+0.8}_{-0.8}$ & $46.1^{+0.4}_{-0.5}$ \\
 & 273 & 53379.3 &42.5 &$74.0^{+0.3}_{-0.7}$ & $43.5^{+0.2}_{-0.4}$ \\
 & 274 & 53380.3 &15.9 &$70.2^{+0.7}_{-0.9}$ & $41.2^{+0.4}_{-0.5}$ \\
V 0332+53 & 278 & 53394.2  &41.9 &$45.3^{+0.7}_{-0.3}$ & $26.6^{+0.4}_{-0.2}$ \\
 & 284 & 53411.0 &107.7 &$15.0^{+0.2}_{-0.1}$ & $8.8^{+0.1}_{-0.1}$ \\
 & 285 & 53413.1 &11.6 &$14.8^{+0.6}_{-0.3}$ & $8.7^{+0.4}_{-0.2}$  \\
 & 286 & 53416.3 &16.2 &$9.9^{+2.0}_{-2.1}$ & $5.8^{+1.2}_{-1.2}$ \\
 & 287-288 & 53420.6 & 27.6 &$5.9^{+4.6}_{-0.8}$ & $3.4^{+2.7}_{-0.4}$ \\
\hline
 & 352 (low) & 53613.7  &34.1 &$15.6^{+0.1}_{-0.1}$ & $1.3^{+0.1}_{-0.1}$ \\
A 0535+262 & 352 (medium) & 53614.5 &51.0 &$18.8^{+0.1}_{-0.1}$ & $1.5^{+0.1}_{-0.1}$ \\
 & 352 (high)& 53615.4  &42.0 &$20.9^{+0.7}_{-0.6}$ & $1.7^{+0.1}_{-0.1}$ \\
\hline
 & 58 & 52734.3 & 7.2  &$10.7^{+0.7}_{-0.6}$ & $0.51^{+0.04}_{-0.03}$ \\
 & 81 & 52804.1 & 60.0  &$5.63^{+0.05}_{-0.22}$ & $0.27^{+0.01}_{-0.01}$\\
 & 82 & 52807.5 & 106.5  &$9.33^{+0.16}_{-0.29}$ & $0.45^{+0.01}_{-0.01}$ \\
 & 83 & 52809.5 & 150.8  &$9.23^{+0.12}_{-0.18}$ & $0.44^{+0.01}_{-0.01}$ \\
 & 84 & 52812.7 & 61.5  &$4.53^{+0.15}_{-0.18}$ & $0.22^{+0.01}_{-0.01}$ \\
 & 85 & 52816.7 & 79.9  &$8.07^{+0.15}_{-0.24}$ & $0.39^{+0.01}_{-0.01}$ \\
 & 86 & 52819.1 & 148.1  &$8.93^{+0.16}_{-0.25}$ & $0.43^{+0.01}_{-0.01}$ \\
 & 87 & 52822.3 & 57.8  &$4.82^{+0.06}_{-0.23}$ & $0.23^{+0.01}_{-0.01}$ \\
 & 88 & 52825.0 & 88.6  &$12.60^{+0.15}_{-0.28}$ & $0.61^{+0.01}_{-0.01}$ \\
 & 137 & 52971.4 & 111.1  &$11.68^{+0.12}_{-0.33}$ & $0.56^{+0.01}_{-0.02}$ \\
 & 138-140 & 52978.5 & 248.1  &$8.64^{+0.14}_{-0.28}$ & $0.42^{+0.01}_{-0.01}$ \\
 & 146 & 52997.5 & 5.2  &$18.6^{+0.7}_{-1.6}$ & $0.9^{+0.1}_{-0.1}$ \\
 & 149 & 53006.4 & 5.5  &$20.0^{+2.3}_{-6.6}$ & $1.0^{+0.1}_{-0.3}$ \\
 & 154 & 53021.3 & 5.9  &$14.6^{+0.6}_{-0.8}$ & $0.7^{+0.1}_{-0.1}$ \\
 & 157 & 53030.3 & 8.7  &$10.1^{+0.5}_{-1.8}$ & $0.5^{+0.1}_{-0.1}$ \\
Vela X-1 & 161 & 53042.2  & 6.8  &$7.64^{+1.11}_{-1.78}$ & $0.37^{+0.05}_{-0.09}$ \\
 & 171 & 53071.6 & 5.0  &$11.6^{+0.4}_{-0.6}$ & $0.55^{+0.02}_{-0.03}$ \\
 & 186 & 53119.2 & 5.5  &$12.2^{+1.0}_{-0.8}$ & $0.58^{+0.05}_{-0.04}$ \\
 & 203 & 53168.5 & 9.8  &$11.5^{+1.5}_{-2.8}$ & $0.55^{+0.07}_{-0.13}$ \\
 & 217 & 53209.7 & 7.1  &$12.5^{+0.9}_{-1.1}$ & $0.60^{+0.04}_{-0.05}$ \\
 & 250 & 53308.5 & 8.0  &$8.98^{+0.53}_{-0.34}$ & $0.43^{+0.03}_{-0.02}$ \\
 & 256 & 53326.6 & 8.4  &$12.2^{+0.2}_{-1.9}$ & $0.58^{+0.01}_{-0.09}$ \\
 & 262 & 53344.5 & 7.6  &$19.4^{+0.3}_{-2.1}$ & $0.93^{+0.02}_{-0.10}$ \\
 & 267 & 53361.4 & 28.4  &$10.7^{+0.6}_{-0.8}$ & $0.51^{+0.03}_{-0.04}$ \\
 & 275 & 53383.6 & 10.4  &$7.56^{+1.4}_{-2.2}$ & $0.36^{+0.07}_{-0.11}$ \\
 & 301 & 53460.9 & 4.9  &$13.3^{+0.5}_{-0.8}$ & $0.64^{+0.02}_{-0.04}$ \\
 & 373-383 & 53693.1  & 783.0  &$9.66^{+0.10}_{-0.15}$ & $0.46^{+0.01}_{-0.01}$ \\
 \end{tabular}
\end{table}
%\end{landscape}

\pagebreak

%\begin{landscape}

\begin{table}[h]

\begin{tabular}{c|c|c|c|c|c}
\hline
\,\,\,\,\,\,\,\,\,\,\,\,\,\,1\,\,\,\,\,\,\,\,\,\,\,\,\,\,&\,\,\,\,\,\,\,\,\,\,\,\,\,\,2\,\,\,\,\,\,\,\,\,\,\,\,\,\,&\,\,\,\,\,\,\,\,\,\,\,\,\,\,3\,\,\,\,\,\,\,\,\,\,\,\,\,\,&\,\,\,\,\,\,\,\,\,\,\,\,\,\,4\,\,\,\,\,\,\,\,\,\,\,\,\,\,&\,\,\,\,\,\,\,\,\,\,\,\,\,\,5\,\,\,\,\,\,\,\,\,\,\,\,\,\,&\,\,\,\,\,\,\,\,\,\,\,\,\,\,6\,\,\,\,\,\,\,\,\,\,\,\,\,\,\\
\hline
 & 398 & 53751.1 & 4.7  &$23.7^{+0.3}_{-0.4}$ & $1.13^{+0.01}_{-0.02}$ \\
 & 433-440 & 53865.9 & 749.5  &$8.58^{+0.10}_{-0.13}$ & $0.41^{+0.01}_{-0.01}$ \\
 & 445 & 53891.6 & 24.7  &$6.09^{+0.43}_{-0.34}$ & $0.29^{+0.02}_{-0.02}$ \\
 & 446 & 53894.8 & 7.0  &$6.85^{+0.52}_{-0.71}$ & $0.33^{+0.03}_{-0.03}$ \\
\hline
 & 192 & 53136.1 & 94.5  &$10.6^{+0.2}_{-0.3}$ & $8.1^{+0.1}_{-0.2}$ \\
 & 194-197 & 53146.8 & 349.3  &$5.05^{+0.14}_{-0.13}$ & $3.88^{+0.11}_{-0.10}$ \\
 & 198-200 & 53157.0 & 264.0  &$5.43^{+0.07}_{-0.04}$ & $4.17^{+0.05}_{-0.03}$ \\
Cen X-3 & 201-205 & 53165.5 & 175.8  &$4.16^{+0.11}_{-0.07}$ & $3.20^{+0.08}_{-0.05}$ \\
 & 325 & 53533.9 & 44.7  &$3.54^{+0.12}_{-0.04}$ & $2.72^{+0.09}_{-0.03}$ \\
 & 327 & 53539.9 & 80.1  &$2.04^{+0.09}_{-0.09}$ & $1.57^{+0.07}_{-0.07}$ \\
 & 330 & 53549.0  & 62.6  &$5.37^{+0.12}_{-0.11}$ & $4.12^{+0.09}_{-0.08}$ \\
\hline
 & 46 & 52700.5 & 4.0  &$11.7^{+0.2}_{-0.2}$ & $3.94^{+0.06}_{-0.06}$ \\
 & 76 & 52789.5 & 44.8  &$3.77^{+0.07}_{-0.11}$ & $1.27^{+0.02}_{-0.04}$ \\
 & 77 & 52791.1 & 29.5  &$2.03^{+0.02}_{-0.05}$ & $0.68^{+0.01}_{-0.02}$ \\
 & 78 & 52796.0 & 24.6  &$4.19^{+0.04}_{-0.09}$ & $1.41^{+0.01}_{-0.03}$ \\
 & 91 & 528343.0 & 34.2  &$2.65^{+0.11}_{-0.14}$ & $0.89^{+0.04}_{-0.05}$ \\
 & 150 & 53009.8  & 33.4  &$4.72^{+0.02}_{-0.07}$ & $1.59^{+0.01}_{-0.02}$ \\
 & 176 & 53087.4 & 103.9  &$2.00^{+0.01}_{-0.04}$ & $0.67^{+0.01}_{-0.01}$ \\
 & 213 & 53198.0 & 3.3  &$2.91^{+0.02}_{-0.03}$ & $0.98^{+0.01}_{-0.01}$ \\
 & 267 & 53361.1 & 3.6  &$4.79^{+0.08}_{-0.09}$ & $1.62^{+0.03}_{-0.03}$ \\
GX 301-2 & 268 & 53362.5 & 4.9  &$8.09^{+0.08}_{-0.14}$ & $2.73^{+0.03}_{-0.05}$ \\
 & 283 & 53407.4 & 3.2  &$12.5^{+0.1}_{-0.2}$ & $4.22^{+0.04}_{-0.06}$ \\
 & 322 & 53524.9 & 112.4  &$2.88^{+0.01}_{-0.03}$ & $0.97^{+0.01}_{-0.01}$ \\
 & 323 & 53528.2 & 104.4  &$7.07^{+0.01}_{-0.05}$ & $2.38^{+0.01}_{-0.02}$ \\
 & 325 & 53533.4 & 31.4  &$5.65^{+0.03}_{-0.07}$ & $1.91^{+0.01}_{-0.02}$ \\
 & 326 & 53536.8 & 73.3  &$2.70^{+0.01}_{-0.04}$ & $0.91^{+0.01}_{-0.01}$ \\
 & 330 & 53549.4 & 63.6  &$3.71^{+0.02}_{-0.05}$ & $1.25^{+0.01}_{-0.02}$ \\
 & 407 & 53778.4 & 3.1  &$11.9^{+0.1}_{-0.3}$ & $4.02^{+0.05}_{-0.11}$ \\
 & 444 & 53888.8 & 3.1  &$3.60^{+0.09}_{-0.13}$ & $1.21^{+0.03}_{-0.05}$ \\
 & 463 & 53946.1 & 3.1  &$10.6^{+0.1}_{-0.2}$ & $3.59^{+0.05}_{-0.07}$ \\
 & 517 & 54108.5  & 8.6  &$6.01^{+0.09}_{-0.11}$ & $2.02^{+0.03}_{-0.04}$ \\
\hline
 & 36-38 & 52671.8 & 142.1  &$3.26^{+0.05}_{-0.17}$ & $1.60^{+0.03}_{-0.08}$ \\
 & 46-47 & 52701.1  & 145.4  &$3.73^{+0.07}_{-0.13}$ & $1.83^{+0.03}_{-0.06}$\\
 & 50-52 & 52714.6 & 169.4  &$3.69^{+0.07}_{-0.17}$ & $1.81^{+0.03}_{-0.08}$ \\
 & 60-61 & 52742.8 & 66.3  &$4.13^{+0.08}_{-0.17}$ & $2.03^{+0.04}_{-0.08}$ \\
 & 103 & 52870.0 & 110.6  &$5.03^{+0.10}_{-0.25}$ & $2.47^{+0.05}_{-0.12}$ \\
 & 105-110 & 52883.3 & 151.9  &$3.58^{+0.09}_{-0.20}$ & $1.76^{+0.04}_{-0.10}$ \\
 & 116 & 52908.4 & 48.6  &$4.92^{+0.08}_{-0.24}$ & $2.42^{+0.04}_{-0.12}$ \\
 & 119 & 52916.7 & 53.1  &$3.83^{+0.08}_{-0.27}$ & $1.88^{+0.04}_{-0.13}$ \\
 & 175 & 53085.2 & 36.4  &$1.35^{+0.06}_{-0.12}$ & $0.66^{+0.03}_{-0.06}$ \\
 \end{tabular}
\end{table}
%\end{landscape}

\pagebreak

%\begin{landscape}

\begin{table}[h]

\begin{tabular}{c|c|c|c|c|c}
\hline
\,\,\,\,\,\,\,\,\,\,\,\,\,\,1\,\,\,\,\,\,\,\,\,\,\,\,\,\,&\,\,\,\,\,\,\,\,\,\,\,\,\,\,2\,\,\,\,\,\,\,\,\,\,\,\,\,\,&\,\,\,\,\,\,\,\,\,\,\,\,\,\,3\,\,\,\,\,\,\,\,\,\,\,\,\,\,&\,\,\,\,\,\,\,\,\,\,\,\,\,\,4\,\,\,\,\,\,\,\,\,\,\,\,\,\,&\,\,\,\,\,\,\,\,\,\,\,\,\,\,5\,\,\,\,\,\,\,\,\,\,\,\,\,\,&\,\,\,\,\,\,\,\,\,\,\,\,\,\,6\,\,\,\,\,\,\,\,\,\,\,\,\,\,\\
\hline
 & 224 & 53232.5 & 75.6  &$4.32^{+0.07}_{-0.13}$ & $2.12^{+0.03}_{-0.06}$ \\
 & 233 & 53256.9 & 17.9  &$1.01^{+0.06}_{-0.12}$ & $0.50^{+0.03}_{-0.06}$ \\
 & 283-284 & 53409.9 & 72.2  &$2.22^{+0.06}_{-0.15}$ & $1.09^{+0.03}_{-0.07}$ \\
 & 285-289 &53420.4  & 231.7  &$3.51^{+0.08}_{-0.16}$ & $1.73^{+0.04}_{-0.08}$ \\
OAO 1657-415  & 290-294 & 53434.1 & 230.0  &$2.62^{+0.05}_{-0.11}$ & $1.29^{+0.03}_{-0.05}$ \\
 & 295-299 & 53450.5 & 182.9  &$3.83^{+0.08}_{-0.21}$ & $1.88^{+0.04}_{-0.10}$ \\
 & 303-307 & 53474.3 & 168.3  &$3.39^{+0.06}_{-0.20}$ & $1.67^{+0.03}_{-0.10}$ \\
 & 345 & 53594.9 & 24.5  &$1.17^{+0.03}_{-0.65}$ & $0.58^{+0.02}_{-0.32}$ \\
 & 347 & 53599.7 & 81.2  &$2.43^{+0.06}_{-0.09}$ & $1.19^{+0.03}_{-0.04}$ \\
 & 350 & 53608.7 & 91.4  &$1.21^{+0.03}_{-0.06}$ & $0.60^{+0.02}_{-0.03}$ \\
 & 364 & 53650.1 & 83.6  &$6.19^{+0.13}_{-0.36}$ & $3.04^{+0.06}_{-0.18}$ \\
 & 409 & 53785.1 & 69.1  &$4.27^{+0.08}_{-0.26}$ & $2.01^{+0.04}_{-0.13}$ \\
 & 412 & 53794.8 & 44.7  &$2.73^{+0.08}_{-0.14}$ & $1.34^{+0.04}_{-0.07}$ \\
 & 472-473 & 53974.8 & 56.4  &$6.05^{+0.14}_{-0.30}$ & $2.97^{+0.07}_{-0.15}$ \\
 & 525 & 54131.4 & 91.6  &$5.96^{+0.12}_{-0.24}$ & $2.93^{+0.06}_{-0.12}$ \\
 & 531 & 54149.5 & 79.7  &$3.38^{+0.08}_{-0.13}$ & $1.66^{+0.04}_{-0.06}$ \\
 & 536 & 54164.6 & 81.7  &$2.91^{+0.07}_{-0.12}$ & $1.43^{+0.03}_{-0.06}$ \\
 & 540 & 54176.1 & 49.4  &$3.47^{+0.06}_{-0.17}$ & $1.71^{+0.03}_{-0.08}$ \\
\hline
 & 339 & 53575.8  & 132.3  &$3.65^{+0.02}_{-0.09}$ & $1.58^{+0.01}_{-0.04}$ \\
Her X-1 & 340 & 53577.8 & 19.5  &$5.09^{+0.03}_{-0.23}$ & $2.20^{+0.01}_{-0.10}$ \\
 & 341 & 53582.4 & 40.4  &$4.10^{+0.16}_{-0.21}$ & $1.77^{+0.07}_{-0.09}$ \\
 & 342 & 53584.2  & 67.9  &$1.75^{+0.06}_{-0.10}$ & $0.76^{+0.03}_{-0.04}$ \\
\hline
 & 46 & 52699.4  &100.0 &$3.55^{+0.01}_{-0.12}$ & $1.53^{+0.01}_{-0.05}$ \\
 & 60-64 & 52746.9  & 149.3 &$0.63^{+0.01}_{-0.03}$ & $0.27^{+0.01}_{-0.01}$ \\
 & 119-122 & 52923.1  & 283.3  &$1.38^{+0.01}_{-0.04}$ & $0.59^{+0.01}_{-0.02}$ \\
 & 164-165 & 53053.8  & 207.2  &$1.83^{+0.02}_{-0.06}$ & $0.79^{+0.01}_{-0.03}$ \\
 & 168 & 53064.1 & 74.0  &$2.78^{+0.01}_{-0.13}$ & $1.12^{+0.01}_{-0.05}$ \\
 & 172-173 & 53078.5  & 152.4  &$1.28^{+0.02}_{-0.05}$ & $0.55^{+0.01}_{-0.02}$ \\
 & 175  & 53084.9  & 38.9  &$1.21^{+0.01}_{-0.07}$ & $0.52^{+0.01}_{-0.03}$ \\
 & 178-179 & 53095.3 & 231.4  &$1.84^{+0.01}_{-0.05}$ & $0.80^{+0.01}_{-0.02}$ \\
GX 1+4 & 181 & 53103.7 & 95.4  &$2.28^{+0.02}_{-0.06}$ & $0.99^{+0.01}_{-0.03}$ \\
 & 183 & 53108.5  & 48.3  &$1.31^{+0.01}_{-0.07}$ & $0.57^{+0.01}_{-0.03}$ \\
 & 185 & 53114.8 & 29.9  &$4.02^{+4.02}_{-0.12}$ & $1.74^{+1.74}_{-0.05}$ \\
 & 225-234 & 53248.5 & 691.2  &$2.84^{+0.01}_{-0.05}$ & $1.23^{+0.01}_{-0.02}$\\
 & 235-237 & 53267.3 & 291.7  &$3.45^{+0.01}_{-0.05}$ & $1.49^{+0.01}_{-0.02}$\\
 & 240-241 & 53281.1 & 250.8  &$3.49^{+0.01}_{-0.06}$ & $1.51^{+0.01}_{-0.02}$ \\
 & 242-246 & 53291.2 & 238.6  &$1.62^{+0.01}_{-0.06}$ & $0.70^{+0.01}_{-0.03}$ \\
 & 297-299 & 53453.7 & 158.1  &$0.61^{+0.01}_{-0.03}$ & $0.26^{+0.01}_{-0.01}$ \\
 & 303-305 & 53471.3 & 181.4  &$1.64^{+0.01}_{-0.04}$ & $0.71^{+0.01}_{-0.02}$ \\
 & 306-308 & 53479.2 & 277.9  &$1.91^{+0.01}_{-0.06}$ & $0.83^{+0.01}_{-0.03}$ \\
 & 475-477 & 53985.2 & 211.5  &$2.63^{+0.01}_{-0.03}$ & $1.14^{+0.01}_{-0.01}$ \\
\hline
 \end{tabular}
\end{table}
%\end{landscape}

\pagebreak

%\begin{landscape}

\begin{table}[h]

\begin{tabular}{c|c|c|c|c|c}
\hline
\,\,\,\,\,\,\,\,\,\,\,\,\,\,1\,\,\,\,\,\,\,\,\,\,\,\,\,\,&\,\,\,\,\,\,\,\,\,\,\,\,\,\,2\,\,\,\,\,\,\,\,\,\,\,\,\,\,&\,\,\,\,\,\,\,\,\,\,\,\,\,\,3\,\,\,\,\,\,\,\,\,\,\,\,\,\,&\,\,\,\,\,\,\,\,\,\,\,\,\,\,4\,\,\,\,\,\,\,\,\,\,\,\,\,\,&\,\,\,\,\,\,\,\,\,\,\,\,\,\,5\,\,\,\,\,\,\,\,\,\,\,\,\,\,&\,\,\,\,\,\,\,\,\,\,\,\,\,\,6\,\,\,\,\,\,\,\,\,\,\,\,\,\,\\
\hline
 & 67 & 52761.8 &64.4 & $2.52^{+0.03}_{-0.04}$ & $1.52^{+0.02}_{-0.03}$ \\
 & 80 & 52801.2 &129.2 &$1.73^{+0.03}_{-0.04}$ & $1.03^{+0.02}_{-0.02}$ \\
 & 159-160 & 53040.9 &249.4  &$2.40^{+0.01}_{-0.04}$ & $1.44^{+0.01}_{-0.02}$ \\
 &  190& 53130.2 &122.8  &$3.80^{+0.02}_{-0.05}$ & $2.28^{+0.01}_{-0.03}$ \\
EXO 2030+375 & 218 & 53214.2 &99.0 &$1.13^{+0.03}_{-0.04}$ & $0.68^{+0.01}_{-0.02}$ \\
 & 251-255 & 53319.2 &339.4 &$3.02^{+0.02}_{-0.04}$ & $1.81^{+0.01}_{-0.02}$ \\
 & 437-438 & 53870.3 &126.4 &$1.97^{+0.02}_{-0.05}$ & $1.18^{+0.01}_{-0.03}$ \\
 & 452-453 & 53914.9 &16.5 &$21.7^{+0.5}_{-0.6}$ & $13.0^{+0.3}_{-0.3}$\\
 & 462 & 53942.9 &70.1  & $29.3^{+0.6}_{-1.2}$ & $17.6^{+0.4}_{-0.7}$ \\
 & 470 & 53967.6 &140.1  &$28.3^{+0.7}_{-1.1}$ & $17.0^{+0.4}_{-0.7}$ \\
 & 486 & 54015.2 &133.6  &$4.29^{+0.31}_{-0.37}$ & $2.61^{+0.18}_{-0.24}$ \\
\hline
\hline
\end{tabular}
\vspace{3mm}

\begin{tabular}{ll}
\a & In the 3-100 keV energy band. \\
\b & In the 3-100 keV energy band assuming the distance to \\
 & the source to be known (see table 2).\\
\end{tabular}
\end{table}
%\end{landscape}

%=========================================================================

\clearpage

 The
IBIS data were processed for a timing analysis using
software developed and maintained in
the National Astrophysical Institute in Palermo, Italy
(http://www.pa.iasf.cnr.it/\~\, ferrigno/INTEGRALsoftware.html); 
a description of the data processing
technique can be found in Segreto and Ferrigno
(2007) and Mineo et al. (2006). Since one
of our main objectives was to search for a correlation
between timing and spectral characteristics of
the emission, we also performed a comprehensive
spectral analysis (in particular, we searched for and
determined the parameters of cyclotron features) for
the entire set of sources under study using the same
software as that applied for our timing analysis. In
accordance with the user's guide for the INTEGRAL
data processing software, the corresponding component
at a 2\% level was added in the XSPEC package
to take into account the systematic uncertainty in our
spectral analysis.

A log of pulsar observations is presented in Table
1. Its columns give (1) the source names, (2)
INTEGRAL observatory orbit numbers, (3) dates of observations (in MJD),
(4) exposure times, (5) source fluxes, and (6) source
luminosities (calculated by assuming the distance to
the source to be known) in the $3-100$ keV energy
band. The published sources of distances to the objects
under study are presented in Table 2.

\begin{table}[h]

 \centering
{\bf Table 2}

\vspace{2mm}
\begin{tabular}{c|c|c}
\hline
Source name&Distance, kpc&Reference \\
\hline
4U 0115+63 & 7  &  Negueruela and Okazaki (2001) \\
V 0332+53  &  7  &   Negueruela et al. (1999) \\
A0535+262  &  2.6  &  Janot-Pacheco et al. (1987) \\
Vela X-1  &  2  &   Sadakane et al. (1985) \\
Cen X-3  &  8  &   Krzeminski (1974) \\
GX 301-2  &  5.3   &  Kaper et al. (1995) \\
OAO 1657-415  &  6.4  &   Chakrabarty et al. (2002) \\
Her X-1  &  6  &  Howarth and Wilson (1983) \\
GX 1+4  &  6  &  Chakrabarty and Roche (1997) \\
EXO 2030+375  &  7.1  &  Wilson et al. (2002) \\
\hline

\end{tabular}
\end{table}

It should be noted that since different authors used
different models (Gaussian or Lorentzian profiles) in
fitting the cyclotron line, there is a spread in its energies
for the same sources and their states in the
literature. Here, we used a Lorentzian profile in the
form

\begin{equation}\label{ac2}
CYCL(E)=exp\biggl(\frac{-\tau_{cycl}(\sigma_{cycl}E/E_{cycl})^2}{(E-E_{cycl})^2+\sigma_{cycl}^2}\biggr),
\end{equation}

where $E_{cycl}$ is the line center, $\tau_{cycl}$ is the line depth,
and $\sigma_{cycl}$ is the line width (Mihara et al. 1990). In
our spectral analysis, we added this component to the
``standard'' fit for X-ray pulsars (a power law with a
high-energy cutoff; White et al. 1983).

%********************************************
\begin{figure*}[t]
\centerline{\includegraphics[width=16cm,bb=5 100 575 800,clip]{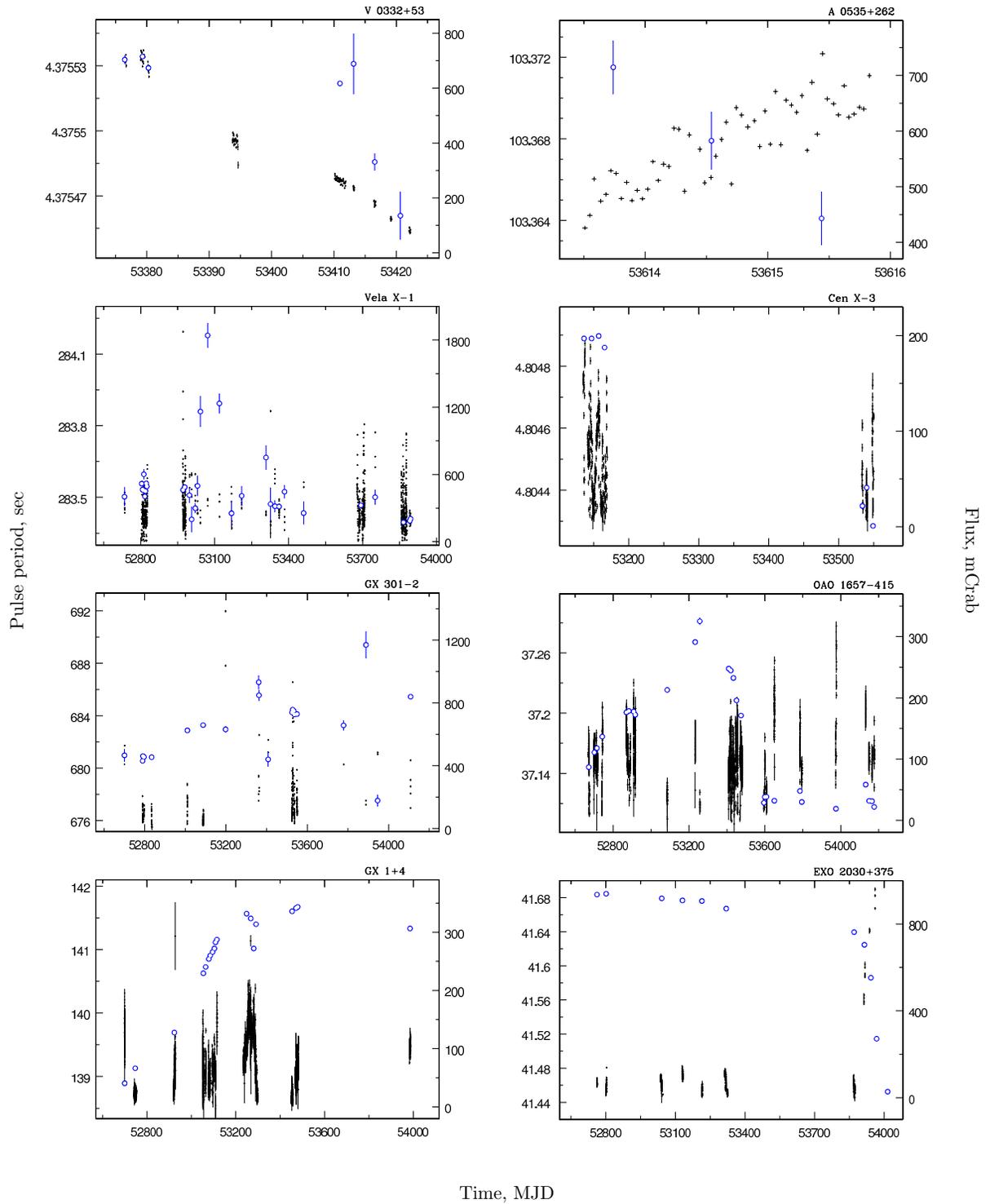}}

\vfill

\caption{Light curves for the pulsars under study (in the 20-60 keV energy band; crosses) and time dependences of their periods
(open circles). The error in the period was determined by the method described by Filippova et al. (2004) and corresponds
to $1\sigma$.}\label{pperiods}
\end{figure*}
%********************************************

The time dependences of the observed source flux
in the 20-60 keV energy band and the pulse period
are shown in Fig. 1. The pulsar 4U 0115+63 was
observed by INTEGRAL for a fairly short time interval
in September 2004, in which its intensity and
pulse period underwent no significant changes; for the
pulsar Her X-1, the corresponding periods and fluxes
were obtained by Klochkov et al. (2008b).

Our main results are the pulse profiles in wide
energy channels (20-30, 30-40, 40-50, 50-70,
70-100 keV), the relative intensity maps in 
``energy -- pulse phase'' coordinates from IBIS data, and the
dependencies of PF (calculated as $PF=\frac{I_{max}-I_{min}}{I_{max}+I_{min}}$, where
$I_{max}$ and $I_{min}$ are the source intensities at the maximum
and minimum of the pulse profile) on
energy and luminosity. The intensity maps were
obtained by convolving the pulsar light curve corrected
for the background emission in narrow energy
channels (the width was chosen to provide an optimal
significance and was about 5 keV) whose centroid
shifted from channel to channel by 2-3 keV. Each
profile was constructed in units relative to the mean
count rate in a given channel. We give the resulting
map normalized to unity (all intensities were divided
by the maximum value over the entire map). Such a
representation reflects well the evolution of the profile
structure (the change in the relative contribution of
different peaks etc.). For each source, the results
mentioned above are either averaged if the pulse
profile did not change significantly with luminosity
or most typical for different states (the source state
is given in the corresponding captions to the figures).
The dotted lines in the corresponding figures indicate
the positions of the harmonics of the cyclotron absorption
line (if it is observed in the spectrum). Note
that the background is difficult to properly subtract at
high energies ($>50-60$ keV, depending on the source
spectrum and intensity) and, therefore, the formal
values of PF can exceed 100\% in some cases due to
insufficient statistics. In current paper all relative 
intensity maps are presented in a black-and-white style; 
more brighter colors correspond to regions of higher 
intensities. The color maps are available in the 
electronic publication.

To illustrate the operation of the method used and
to estimate the accuracy and validity of our timing
analysis, we performed the corresponding calculations
for the pulsar NP0532 in the Crab Nebula.
This source was observed by INTEGRAL once in
several months to calibrate the instruments. Figure 2
shows the pulse profile in the wide 20-50 keV energy
channel and its relative intensity map. The profile has
a characteristic double-peaked shape with a slight
increase in the relative intensity of the second (relative
to the zero phase we chose) peak with energy, which
is clearly seen from the intensity map (Fig. 2b).

%********************************************
\begin{figure*}[t]
\centerline{\includegraphics[width=8cm,bb=30 350 550 705,clip]{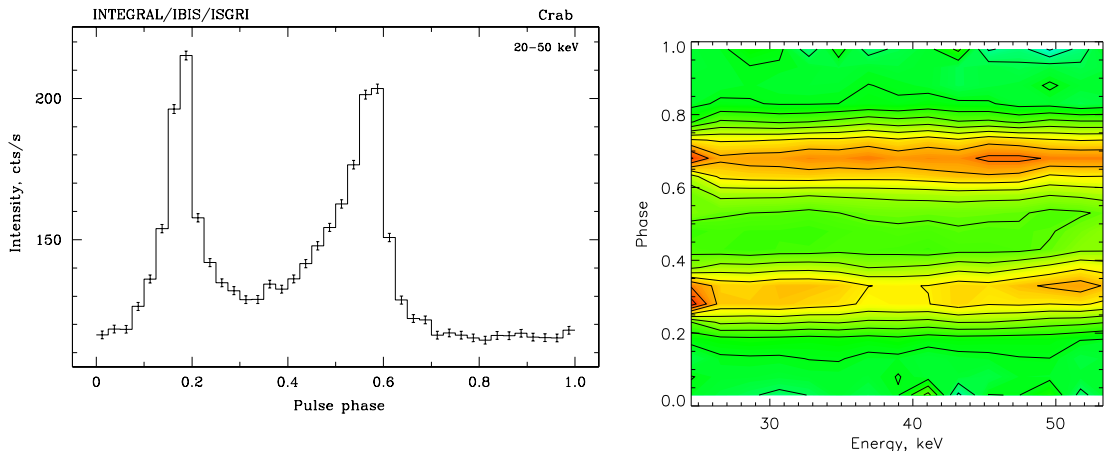}
\includegraphics[width=7cm,bb=30 350 550 710]{figures1/crab_0422_0422_rev_contour.ps2}}

\vfill

\caption{ (a) Pulse profile for the pulsar NP0532 in the Crab Nebula from INTEGRAL data in the 20-50 keV energy band.
(b) Relative intensity map in ``energy -- pulse phase'' coordinates (for more detail, see the text).}\label{crabppr}
\end{figure*}
%********************************************

In Fig. 3a, the filled circles indicate the time dependence
of the intrinsic pulsar period that was measured
and reduced to the Solar system barycenter by
the methods used here; the solid line represents an extrapolation
of the pulsar period measurements in the
radio frequency band by the Jodrell Bank Observatory
in the form $P=0.033522231+0.00000003633565\times(T-52014)$
where $T$ is the time in MJD (Lyne et al.,
1993; http://www.jb.man.ac.uk/pulsar/crab.html).
Figure 3b shows the deviations of the periods measured
by INTEGRAL from the extrapolation. The
dashed line indicates the deviations of the actual
period measurements by the Jodrell Bank Observatory
from the same extrapolation. We see excellent
agreement between the radio and gamma-ray measurements.
Note also the detection of a pulsar glitch
by both observatories at MJD $\sim53070$.

%********************************************
\begin{figure*}[t]
\centerline{\includegraphics[width=12cm,bb=75 310 515 690,clip]{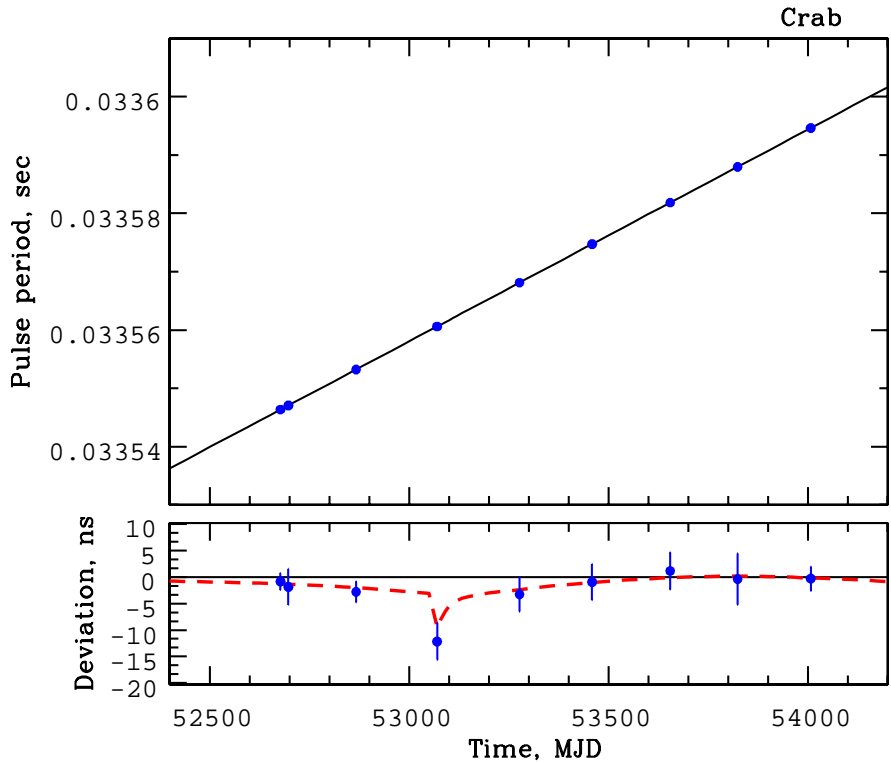}}

\vfill

\caption{ (a) Time dependence of the pulse period for NP0532 in the Crab Nebula from INTEGRAL data (filled circles); the solid
line represents an extrapolation of the Jodrell Bank radio measurements. (b) The corresponding deviations of the INTEGRAL
measurements from the extrapolation (filled circles) and the radio measurements (dashed line). In both ranges, a glitch is clearly
seen at MJD 53070.}\label{crabper}
\end{figure*}
%********************************************

Comparison of the data obtained suggests that the
pulse period determined for the pulsar NP0532 by
INTEGRAL coincides with a high accuracy (a typical
difference $\sim1\times10^{-9}$ s) with that measured in the
radio frequency band.

\pagebreak

\section*{RESULTS}

We studied X-ray pulsars in binary systems of various
classes; however, all of these sources are united
by the fact that a relatively high ($>100$ mCrab) flux
is observed from them in the hard X-ray energy band
and one or more harmonics of the cyclotron absorption
resonance line are present in the spectra of most
of them (see, e.g., Coburn et al. 2002; Filippova et al.
2005). Since our analysis required high statistics,
the chosen sources either had long exposure times
or were investigated during their outburst activity.
Below, we briefly present the results obtained for each
X-ray pulsar; our attention was focused on analyzing
the dependences of PF on the energy and source
luminosity.

{\it 4U\,0115+63}

The transient X-ray pulsar
4U 0115+63 was observed by INTEGRAL
during an intense outburst in September 2004 at
the decay phase, when the source luminosity was
$\sim7.5\times10^{37}$ erg/s. The pulse profile characteristic
of such a luminosity level in hard energy channels,
along with the relative intensity map and the energy
dependence of PF, are shown in Fig. 4. Note that the
pulse profile properties and the PF variation over a
wide range of pulsar luminosities, energies, and time
were studied in detail by Tsygankov et al. (2007)
based on the RXTE data obtained during the 1999
and 2004 outbursts. In particular, these authors found
that a ``wavelike'' behavior of the profile with changing
energy is observed over the entire luminosity range of
the pulsar 4U 0115+63. The effect lies in the fact that
each line of the level of equal intensity of the main
peak does not lie at one phase for different energies
but slightly shifts alternatively in one and the other
directions. Interestingly, such a phase variability of
the main peak is repeatable in energy and its period
roughly coincides with the energy separation between
the harmonics of the cyclotron absorption line in the
pulsar spectrum (Fig. 4). The detection of a strong
dependence of PF on energy and luminosity, according
to which PF decreases with increasing luminosity
and increases with energy, is no less important and
interesting. In the latter case, this increase is not
uniform but has local maxima near the cyclotron line
harmonics (see, e.g., Figs. 3 and 4 from Tsygankov
et al. 2007). Such a behavior of PF is confirmed by
INTEGRAL data (Fig. 4).

Note that 4U 0115+63 was the first X-ray pulsar
for which such peculiarities of PF were detected.
Subsequently, similar dependences were found for
several other X-ray pulsars with cyclotron lines in
their spectra (see below).

{\it V\,0332+53}

Like 4U 0115+63, the X-ray pulsar
V 0332+53 was observed during an intense
outburst in a wide range of fluxes from $\sim 70$ to
$\sim700$ mCrab (20-60 keV energy band), which
corresponds to a spread in luminosities from $\sim
4\times10^{37}$ to $\sim 4\times10^{38}$ erg/s (assuming the distance
to the pulsar to be $d\sim7$ kpc). Three harmonics of
the cyclotron absorption line were identified in its
spectrum (Coburn et al. 2005); the position of the
fundamental harmonic is not constant with luminosity
(Mihara et al. 1998). Tsygankov et al. (2006)
were the first to show that the cyclotron energy in the
source spectrum decreases linearly with increasing
luminosity during outbursts. Dramatic changes in the
pulse profile with luminosity and energy, especially
near the cyclotron feature, were also found. In this
study, we obtained a set of pulse profiles and relative
intensity maps for the entire INTEGRAL data set
(see http://hea.iki.rssi.ru/integral/pulsars). Here, as
an illustration, we present only the results obtained
for high (orbit 273, Fig. 5) and low (orbit 284, Fig. 6)
states, where changes in the profile with luminosity
are clearly seen.

Despite the above dramatic changes in the pulse
profile, the energy dependences of PF are very similar
in both cases: a general increase with energy and
local features (maxima) near the cyclotron absorption
line (Figs. 5 and 6). Interestingly, their positions are
almost independent of the source luminosity and are
shifted toward the higher energies with the center at
$\sim32$ keV.

{\it A0535+262}

The X-ray pulsar A0535+262 was
observed by INTEGRAL near the maximum of its
outburst in August-September 2005. Although the
total exposure time was not very long, we arbitrarily
divided it into three parts, depending on the source
luminosity (see Table 1) and searched for pulsations
and constructed the pulse profiles in each of them.
The derived pulse profile is a fairly wide ($\sim0.8$ period)
peak with two maxima and an indistinct dip in intensity
between them (Fig. 7). The width of the dip increases
with energy, while the amplitude of the second
maximum decreases, as can be clearly seen from the
relative intensity map(Fig. 7). In softer channels, the
pulse profile is considerably more complex (Caballero
et al. 2008).

The source spectrum exhibits two harmonics of
the cyclotron absorption line ($\sim45$ and $\sim100$ keV);
 the position of the fundamental harmonic
measured using data from the Suzaku observatory remains
almost constant in a wide range of luminosities
(Terada et al. 2006). However, Caballero et al. (2008)
pointed out a possible change in the position of the
cyclotron line from $\sim50$ keV during the precursor to
$\sim46$ keV during the main outburst. It is seen from 
the intensity map that the relative contribution of peaks 
in the profile is changed when crossing the cyclotron 
energy, that is typical for other pulsars from current review.

The source PF increases with energy; no distinct
features are observed near the cyclotron line harmonics,
which is most likely because the line depth is not so 
high and the line is fairly wide. 

A change in the pulse period of the source with
its intensity during the outburst is clearly seen from
Fig. 1. Such a spin-up of the neutron star can be
roughly described in terms of the theory for angular
momentum transport from the disk to the neutron
star (Ghosh and Lamb 1979):

%=========================================================================
\begin{equation}\label{gosh}
\dot \nu \varpropto \mu^{2/7} n(\omega_s) L^{6/7} = \mu^{2/7} n(\omega_s)
(4\pi d^2 F)^{6/7},
\end{equation}
%========================================================================

where $\mu$ is the magnetic moment of the neutron star
with magnetic field $B$ and radius $R$, $n(\omega_s)$ is a dimensionless
function of the fastness parameter $\omega_s$, $d$ is the
distance to the system, and $F$ is its X-ray flux. Determining
the magnetic field of the neutron star from a
spectral analysis, we can estimate the distance to the
binary system from Eq. (2) to be $d\simeq3.1\pm0.7$ kpc,
which agrees, within the error limits, with the results
of optical observations (Janot-Pacheco et al. 1987).

{\it Vela X-1}

During almost the entire period of its
INTEGRAL observations, the X-ray pulsar Vela X-1
 exhibited an intense outburst activity (see, e.g.,
Krivonos et al. 2003) accompanied by significant
changes in its pulse period (Fig. 1). At the same time,
the shape of the source spectrum was almost constant,
as were the positions of the two cyclotron line
components at energies $\sim 24$ and $\sim 50$ keV (Filippova
et al. 2005). Note that the absorption line at $\sim 50$ keV
had long believed to be the fundamental harmonic of
the cyclotron line and only the RXTE (Kreykenbohm
et al. 2002) and INTEGRAL (Filippova et al. 2005)
observations revealed a weaker lower harmonic at
energies 24-26 keV.

The source pulse profile has a fairly complex morphology
and depends strongly on the energy band:
at low energies, the profile is an asymmetric five-peaked
one with its gradual simplification to a double-peaked
sinusoidal profile above $\sim10$ keV (see, e.g.,
Kreykenbohm et al. (2002) and references therein).
The evolution of the pulse profile in hard ($>20$ keV)
energy channels is shown in Fig. 8. Emission pulsations
from the source are detected up to $\sim100$ keV
with a double-peaked profile: the first peak is triangular
in shape with a gradual rise and a sharp
decay; the second peak is almost symmetric relative
to its maximum. We see from the intensity map that
the relative intensity of the first peak increases with
energy, having a maximum near the first cyclotron
line harmonic, while its maximum approaches that of
the first peak in phase (Fig. 8). PF for Vela X-1 is
rather high compared to other pulsars, $\sim60$\% in the
20-30 keV energy band, and gradually increases with
energy, having a local maximum near 50 keV, where
the second cyclotron line harmonic is observed in the
source spectrum (Fig. 8).

{\it Cen X-3}

Although the X-ray pulsar Cen X-3 
is believed to be a persistent source, its intensity
can change dramatically (by more than a factor
of 10) both during the orbital cycle and on long time
scales. These changes can be accompanied by significant
changes in the pulse period (Fig. 9). Nagase
et al. (1992) detected a possible cyclotron absorption
line in the source spectrum at energy $\sim 30$ keV,
which was confirmed by Santangelo et al. (1998).
Subsequently, Burderi et al. (2000) showed that the
position of the cyclotron line depends strongly on the
pulse phase and can change with the range from $\sim 28$ 
to $\sim 36$ keV. Despite dramatic changes in the intensity
of Cen X-3 during its INTEGRAL observations, we
found no significant changes in the position of the
cyclotron line with a mean energy of $\sim31\pm1$ keV
associated with them.

The source pulse profile depends on the orbital
phase and energy band: in a high state, the profile is
double-peaked at low energies; the relative intensity
of the peaks changes with increasing energy and the
main peak disappears almost completely at high energies
(above $\sim 20$ keV) (Nagase et al. 1992). The X-ray
pulsar 4U 0115+63 exhibits a similar behavior (Tsygankov
et al. 2007). PF for the source is rather high
($\sim60$\% at 20-30 keV) and increases monotonically
with energy without any features (Fig. 9).

{\it GX 301-2}

The pulse profile for the X-ray pulsar
GX 301-2 has been adequately investigated over
a wide energy range (see, e.g., Borkus et al. 1998;
Tsygankov et al. 2004). It has a double-peaked sinusoidal
shape with a constant ratio of the peak intensities at 
different energies. Analysis of the behavior
of the cyclotron line revealed a strong variability of
its position in the pulsar spectrum, depending on the
pulse phase and intrinsic source luminosity (La Barbera
et al. 2005). Thus, for example, the cyclotron line
energy in a state with a low luminosity 
($L_x\sim0.8\times10^{37}$ erg/s) was $\sim45$ keV, while it was $\sim53$ keV at
maximum luminosity ($L_x\sim2\times10^{37}$ erg/s). Similar
changes in the cyclotron energy with luminosity
were also revealed by the INTEGRAL observatory
(Filippova et al. 2005).

Despite significant changes in the flux from the
source, especially near the periastron and apoastron,
its INTEGRAL pulse profile remains fairly stable (see
Fig. 10 and http://hea.iki.rssi.ru/integral/pulsars),
while PF increases with energy almost linearly. Below,
we will show that, in this case, it depends significantly
on the source luminosity.

{\it OAO 1657-415}

Despite the fairly long INTEGRAL
exposure time for this pulsar and the hard
spectrum of this object, as yet no cyclotron absorption
line has been reliably detected in its spectrum
(Filippova et al. 2005; Barnstedt et al. 2008). This
may be indicative of a fairly strong magnetic field near
the neutron star surface, just as is observed with INTEGRAL in other
pulsars (see, e.g., Tsygankov and Lutovinov 2005a,
2005b). Note also that the pulsar exhibited a dramatic
variability, both aperiodic one and one related to the
orbital cycle, during its INTEGRAL observations.
However, the pulse period changes revealed no correlation
between them and the changes in the source
flux (Fig. 11).

The pulse profile is in the shape of a wide asymmetric
peak with two maxima in the standard X-ray
energy band whose peculiarity is a dip at phase 0.4-0.5 
(see, e.g., Lutovinov et al. 1994). The amplitude
of the second maximum decreases as we pass to the
harder energies, the dip is essentially ``smeared'', and
the pulse profile turns into an asymmetric one with
a maximum near phase 0.25 (Fig. 11); PF increases
with energy monotonically, without any features (see
Fig. 11 and Barnstedt et al. (2008)).

{\it Her X-1}

The X-ray pulsar Her X-1 is the first
source with a cyclotron resonance scattering feature
detected in its spectrum (Truemper et al. 1978).
Over several decades, the properties of this source
have been studied by many authors based on data
from various X-ray observatories. However, only in
recent years has serious progress been achieved in
this question: in particular, Staubert et al. (2007)
showed that, in contrast to bright pulsars, the energy
of the fundamental cyclotron line harmonic in Her X-1 
is directly proportional to the source luminosity,
while using data from the Suzaku observatory, Enoto
et al. (2008) have recently confirmed the presence of
a second cyclotron line harmonic in the spectrum at
$\sim73$ keV, the possible existence of which was previously
claimed by Di Salvo et al. (2004) based on
BeppoSAX data. This discovery makes this object the
sixth in the list of pulsars with the detected second
cyclotron frequency harmonic.

The pulsar Her X-1 is one of the few X-ray sources
with variability simultaneously on three different time
scales: the pulse period, the orbital period, and the
super-orbital period related to accretion disk precession.
The source pulse profile depends strongly on
both energy (Fig. 12) and phase of the super-orbital
period (http://hea.iki.rssi.ru/integral/pulsars); it was
analyzed in detail (including phase-resolved spectroscopy)
on the basis of INTEGRAL data (Klochkov
et al. 2008b). Here, we will note the behavior of PF
with energy: apart from its increase with energy, a
local feature at energies $\sim 30-45$ keV (Fig. 12) located
near the cyclotron absorption line is present
on the plot. This feature can be interpreted either as
a local maximum shifted relative to the fundamental
harmonic toward the lower energies or as a local PF
minimum shifted toward the higher energies.

{\it GX 1+4}

Ferrigno et al. (2007) claimed the possible
detection of a cyclotron feature in the source
spectrum based on INTEGRAL data at $\sim 34$ keV.
Previously, the magnetic field was estimated only by
indirect methods (see, e.g., Cui 1997); its strength
turned out to be an order of magnitude larger than the
above value. Note that based on first 2.5 years of the INTEGRAL observations,
Filippova et al. (2005) detected no cyclotron feature
in the source spectrum.

As regards the changes in the pulse profiles, many
authors (see, e.g., Greenhill et al. (1998) and references
therein) showed that the profile clearly depends
not only on the source intensity but also on
the spin-up. According to IBIS data, the source pulse
profile is single-peaked with a slight tendency for
the separation into two sub-peaks with increasing
energy (Figs. 13 and 14). In some cases (Fig. 13),
the separation into two sub-peaks is observed even
in the softest available energy channel. A detailed
analysis of the double-peaked pulse profile structure
based on INTEGRAL data can be found in Ferrigno
et al. (2007). PF is practically independent of the energy 
and source luminosity; note only its local increase 
near the energy ~45 keV (Fig. 13) in March 2003.

Throughout the history of its studies, the pulsar
GX1+4 had unique spin-up and spin-down characteristics.
During its INTEGRAL observations, the
pulsar spun down almost uniformly approximately
until May 2004, following which the pulse period remained
relatively constant. At the spin-down phase,
the period increased with a mean rate of $\sim$1.8
s/yr.

{\it EXO\,2030+375}

Like the sources 4U 0115+63
and V 0332+53, the X-ray pulsar EXO 2030+375
is a member of a binary system with a Be star
and manifests itself during outbursts; in contrast
to the two previous objects, the outburst observed
from EXO 2030+375, can be both normal, with an
intensity $\sim50-150$ mCrab, and giant, when the flux
from the source increases almost to 1 Crab. During
one of such giant outbursts ($\sim MJD 53900-54100$),
the INTEGRAL observatory detected a significant
spin-up of the neutron star (Fig. 1), which, in addition,
correlated linearly with the rise in the source
luminosity (see also Klochkov et al. 2007).

EXO 2030+375 is one of the first X-ray pulsars
to be systematically studied in the area of the dependence
of the pulse profile on the intrinsic luminosity
(Parmar et al. 1989). The authors found that the profile
changes both with luminosity and with energy in
the standard X-ray energy band. Significant changes
in the pulse profile with source luminosity are also
observed in the hard energy band (Figs. 15 and 16),
with the energy dependences of PF being similar in
both cases. For a bright state during an outburst, this
dependence is given in Klochkov et al. (2008a). These
authors also detected a possible cyclotron absorption
line in the source spectrum. This feature manifests
itself only in a narrow range of pulse phases at energy
$\sim63$ keV. Unfortunately, such a high energy does not
allow the features in the dependence of PF on energy
at the cyclotron frequency observed in other pulsars
to be detected at a statistically significant level.

\section*{CONCLUSIONS}

We analyzed in detail the timing characteristics of
the hard ($>20$ keV) X-ray emission from ten bright
X-ray pulsars observed by the INTEGRAL observatory.
In particular, we have investigated the dependence
of PF on the luminosity and energy band
for the first time and showed that PF increases with
energy for all pulsars; in many cases, this increase
is not monotonic, exhibiting local features near the
cyclotron line harmonics, which is most likely attributable
to the effect of resonance absorption. In
addition, we compiled a catalog of pulse profiles in
five hard energy channels (20-30, 30-40, 40-50,
50-70, and 70-100 keV) for all the observed states
of the pulsars under study and their luminosity levels
(http://hea.iki.rssi.ru/integral/pulsars). A 
characteristic feature of most of the double-peaked pulse
profiles is a decrease in the relative intensity of one
of the peaks with increasing energy and decreasing
source luminosity. Such a behavior and the increase
in PF with energy noted above can be qualitatively
explained in terms of a simple geometric model proposed
by Tsygankov et al. (2007) and Lutovinov and
Tsygankov (2008). In this model, accretion columns
whose height depends on the pulsar luminosity are
formed at the poles of a neutron star with a dipole
magnetic field; the rotation axis of the neutron star is
inclined with respect to the magnetic dipole axis and
the observer in such a way that the latter can see the
accretion column over its entire height at one of the
poles and only its upper part at the other pole (see
Fig. 17). Since the temperature of the accretion column
increases toward its base (Basko and Sunyaev
1976), high-energy photons are emitted from regions
close to the neutron star surface, while ``soft'' photons
are formed in the upper part of the column. Thus, an
observer will see two formation regions of soft emission
and only one formation region of hard emission,
which will correspond to a change of the pulse profile
from double-peaked to single-peaked with increasing
energy. As the luminosity decreases, the column
height also decreases; the visible part of the second
column decreases and it can disappear altogether
from the observer's field of view at some time, which
also leads to the transition from a double-peaked
pulse profile to a single-peaked one. The observed
increase in PF with energy can also be naturally
explained in terms of this model: at higher energies,
the contrast between the minimum and maximum
visible surfaces of the accretion columns is highest.
Note that a similar model was proposed by David
et al. (1998) to explain the dependence of the pulse
profile for the X-ray pulsar GX1+4 on its intensity.

It should be noted that the simple model described
above gives only qualitative (in the first approximation)
explanations for some of the observed effects.
During further (quantitative) modeling, the temperature
distribution along the column, the beam pattern
of emission, its dependence on energy and luminosity,
the bending of light rays in strong gravitational fields
near the neutron star surface, etc. should be included
in the model.

The increase in PF with energy may also be related
to the characteristic shape of the spectra for X-ray
pulsars with an exponential cutoff at high energies.
The local PF features near the cyclotron line
harmonics may be associated with changes in the
opacity of matter at these energies and its strong
dependence on the viewing angle. In that case, the cyclotron
line parameters (energy, optical depth, equivalent
width, etc.) will change significantly over one
rotation (pulse) of the neutron star (Tsygankov et al.
2009, prepared for publication).

{\it The Dependence of PF on Source Intensity}

As we showed above, the pulse profiles for X-ray
pulsars often depend on their intensity, which,
in turn, can change PF. To test and illustrate this
assumption, we constructed the dependence of PF in
the 25-45 keV energy band on the source intensity
in the same band for the entire set of available data
(in contrast to Fig. 1, the results of this analysis are
also presented for the pulsar Her X-1, which was
observed in different states, and are not presented for
the pulsar A0535+63, for which the dynamic range
of observed INTEGRAL fluxes was too small for any
changes in PF to be observed).
 
X-ray pulsars for which PF decreases with source
luminosity can be separated from the set of results
presented in Fig. 18: GX 301-2 and OAO1657-415.
Such a behavior of PF can be explained in terms of the
model described above. Pulsars that exhibit a fairly
large spread in PF at close luminosities but show no
general trend in the wide dynamic range of observed
fluxes represent another group: Vela X-1, Cen X-3,
and GX 1+4. In this case, the source luminosity is
most likely insufficient for the formation of an accretion
column and the emission is formed near the
neutron star surface (the latter is indirectly confirmed
by the absence of any change in the position of the cyclotron
line with luminosity for Cen X-3 and Vela X-1); 
the observed changes in PF may be related to local
inhomogeneities of the stellar wind or accretion flows.
For the X-ray pulsar Her X-1, the sharp decrease in
PF observed during a low on state may stem from the
fact that at this moment the emission is partially scattered
in the corona at the edge of a precessing accretion
disk (Jones and Forman 1976). The PF dependence
obtained for the pulsar V 0332+53 is most difficult
to explain: PF decreases with increasing luminosity
at low luminosities, as is expected from the model
described above, but it increases with luminosity at
high luminosities.

{\bf ACKNOWLEDGMENTS}

We thank U. Poutanen, M. Revnivtsev, A. Serber,
and V. Suleimanov for useful discussions of results
and valuable remarks. This work was supported
by grant no. NSh-5579.2008.2 from the President
of Russia, the ``Origin and Evolution of Stars and
Galaxies'' Program of the Presidium of the Russian
Academy of Sciences, and the Russian Foundation
for Basic Research (projects no. 07-02-01051 and 08-08-13734). We
used data from the INTEGRAL Science Data Center
(Versoix, Switzerland) and the Russian INTEGRAL
Science Data Center (Moscow, Russia). We
are grateful to the software developers from the National
Astrophysical Institute in Palermo (Italy). This
work was performed in part during visits to the International
Space Science Institute (ISSI, Bern), to
which we express our deep gratitude.

{\bf REFERENCES}
 
1. A. La Barbera, A. Segreto, A. Santangelo, et al.,
Astron. Astrophys. 438, 617 (2005).

2. J. Barnstedt, R. Staubert, A. Santangelo, et al., Astron.
Astrophys. 486, 293 (2008).

3. M. M. Basko and R. A. Sunyaev, Astron. Astrophys.
42, 311 (1975).

4. M. M. Basko and R. A. Sunyaev, Mon. Not. R. Astron.
Soc. 175, 395 (1976).

5. V. V. Borkus, A. S. Kaniovsky, R. A. Sunyaev et al.,
Pis'ma Astron. Zh. 24, 83 (1998) [Astron. Lett. 24,
60 (1998)].

6. T. Bulik, D. Gondek-Rosinska, A. Santangelo, et al.,
Astron. Astrophys. 404, 1023 (2003).

7. L. Burderi, T. Di Salvo, N. R. Robba, et al., Astrophys.
J. 530, 429 (2000).

8. I. Caballero, A. Santangelo, P. Kretschmar, et al.,
Astron. Astrophys. 480, L17 (2008).

9. D. Chakrabarty and P. Roche, Astrophys. J. 489, 254
(1997).

10. D. Chakrabarty, Z. Wang, A.M. Juett, et al., Astrophys.
J. 573, 789 (2002).

11. W. Coburn, W. A. Heindl, R. E. Rothschild, et al.,
Astrophys. J. 580, 394 (2002).

12. W. Coburn, P. Kretschmar, I. Kreykenbohm, et al.,
Astron. Telegrams 381, 1 (2005).

13. W. Cui, Astrophys. J. 482, 163 (1997).

14. P. David, P. Laurent, M. Denis, et al., Astron. Astrophys.
332, 165 (1998).

15. T. Enoto, K. Makishima, Y. Terada, et al., Publ. Astron.
Soc. Japan 60, S57 (2008).

16. C. Ferrigno, A. Segreto, A. Santangelo, et al., Astron.
Astrophys. 462, 995 (2007).

17. E. V. Filippova, S. S. Tsygankov, A. A. Lutovinov, and
R. A. Sunyaev, Pis'ma Astron. Zh. 31, 729 (2005)
[Astron. Lett. 31, 729 (2005)].

18. F. Frontera, D. dal Fiume, E. Morelli, and G. Spada,
Astrophys. J. 298, 585 (1985).

19. P. Ghosh and F. Lamb, Astrophys. J. 234, 296 (1979).

20. J. G. Greenhill, D. Galloway, and M. C. Storey, Publ.
Astron. Soc. Australia 15, 254 (1998).

21. I. D. Howarth and B. Wilson, Mon. Not. R. Astron.
Soc. 204, 1091 (1983).

22. E. Janot-Pacheco, C. Motch, and M. Mouchet, Astron.
Astrophys. 177, 91 (1987).

23. C. Jones and W. Forman, Astrophys. J. 209, 131
(1976).

24. L. Kaper, H. J. G. Lamers, E. Ruymaekers, et al.,
Astron. Astrophys. 300, 446 (1995).

25. D. Klochkov, D. Horns, A. Santangelo, et al., Astron.
Astrophys. 464, 45 (2007).

26. D. Klochkov, A. Santangelo, R. Staubert, and C. Ferrigno,
Astron. Astrophys. 491, 833 (2008a).

27. D. Klochkov, R. Staubert, K. Postnov, et al., Astron.
Astrophys. 482, 907 (2008b).

28. I. Kreykenbohm, W. Coburn, J. Wilms, et al., Astron.
Astrophys. 395, 129 (2002).

29. R. Krivonos, N. Produit, I. Kreykenbohm, et al., Astron.
Telegrams 211, 1 (2003).

30. W. Krzeminski, Astrophys. J. 192, L135 (1974).

31. F. Lebrun, J. P. Leray, P. Lavocat, et al., Astron. Astrophys.
411, L141 (2003).

32. A. A. Lutovinov, S. A. Grebenev, M. N. Pavlinsky, and
R. A. Sunyaev, Pis'ma Astron. Zh. 20, 631 (1994)
[Astron. Lett. 20, 538 (1994)].

33. A. A. Lutovinov and S. S. Tsygankov, AIP Conf. Proc.
1054, 191 (2008).

34. A. Lyne, R. Pritchard, and F. Graham-Smith, Mon.
Not. R. Astron. Soc. 265, 1003 (1993).

35. P. Meszaros and W. Nage, Astrophys. J. 299, 138
(1985).

36. T. Mihara, K. Makishima, T. Ohashi, et al., Nature
346, 250 (1990).

37. T. Mihara, K. Makishima, and F. Nagase, Adv. Space
Res. 22, 987 (1998).

38. T. Mineo, C. Ferrigno, L. Foschini, et al., Astron.
Astrophys. 450, 617 (2006).

39. F. Nagase, R. H. D. Corbet, C. S. R. Day, et al.,
Astrophys. J. 396, 147 (1992).

40. I. Negueruela and A. T. Okazaki, Astron. Astrophys.
369, 108 (2001).

41. I. Negueruela, P. Roche, J. Fabregat, and M. J. Coe,
Mon. Not. R. Astron. Soc. 307, 695 (1999).

42. A. N. Parmar, N. E.White, and L. Stella, Astrop hys.
J. 338, 373 (1989).

43. K. Sadakane, R. Hirata, J. Jugaku, et al., Astrophys.
J. 288, 284 (1985).

44. T. Di Salvo, A. Santangelo, and A. Segreto, Nucl.
Phys. B Proc. Suppl. 132, 446 (2004).

45. A. Santangelo, S. del Sordo, A. Segreto, et al., Astron.
Astrophys. 340, 55 (1998).

46. A. Segreto and C. Ferrigno, in Proc. of the 6th
INTEGRAL Workshop The Obscured Universe, Ed.
by S. Grebenev, R. Sunyaev, and C. Winkler, ESA
SP-662 (Noordwijk, 2007), p. 633; arXiv:0709.4132
(2007).

47. R. Staubert, N.I. Shakura, K.A. Postnov, et al., Astron.
Astrophys. 465, 25 (2007).

48. Y. Terada, T. Mihara, M. Nakajima, et al., Astrophys.
J. 648, 139 (2006).

49. J. Truemper,W. Pietsch, C. Reppin, et al., Astrophys.
J. 219, 105 (1978).

50. S. S. Tsygankov and A. A. Lutovinov, Pis'ma Astron.
Zh. 31, 99 (2005a) [Astron. Lett. 31, 88 (2005)].

51. S. S. Tsygankov and A. A. Lutovinov, Pis'ma Astron.
Zh. 31, 427 (2005b) [Astron. Lett. 24, 380 (1998)].

52. S. S. Tsygankov, A. A. Lutovinov, S. A. Grebenev,
et al., Pis'ma Astron. Zh. 30, 596 (2004) [Astron. Lett.
30, 540 (2004)].

53. S. S. Tsygankov, A. A. Lutovinov, E. M. Churazov,
and R. A. Sunyaev, Mon. Not. R. Astron. Soc. 371,
19 (2006).

54. S. S. Tsygankov, A. A. Lutovinov, E. M. Chupazov,
and R. A. Sunyaev, Pis'ma Astron.Zh. 33, 417 (2007)
[Astron. Lett. 33, 368 (2007)].

55. Y.-M. Wang and G. L. Welter, Astron. Astrophys.
102, 97 (1981).

56. N. White, J. Swank, and S. Holt, Astrophys. J. 270,
771 (1983).

57. C. A. Wilson, M. H. Finger, M. J. Coe, et al., Astrophys.
J. 570, 287 (2002).

58. C. Winkler, T. J.-L. Courvoisier, G. Di Cocco, et al.,
Astron. Astrophys. 411, L1 (2003).

\clearpage

%********************************************
\begin{minipage}[t]{9cm}
\begin{figure}[H]
\begin{center}
\rput(8.0,-2.6){\scalebox{1}{\includegraphics[width=7cm,bb=45 310 515 725,clip]{figures1/4u0115.all_238_238_rev_contour_col.ps2}}}
\rput(5.5,0.8){\mbox{\bf f}}
\rput(5.5,-6.5){\mbox{\bf g}}
\end{center}
\end{figure}

\end{minipage}

\begin{minipage}[h]{9cm}
\begin{figure}[H]
\begin{center}
\rput(8.3,-7.6){\scalebox{1}{\includegraphics[width=8cm,bb=20 275 560 672,clip]{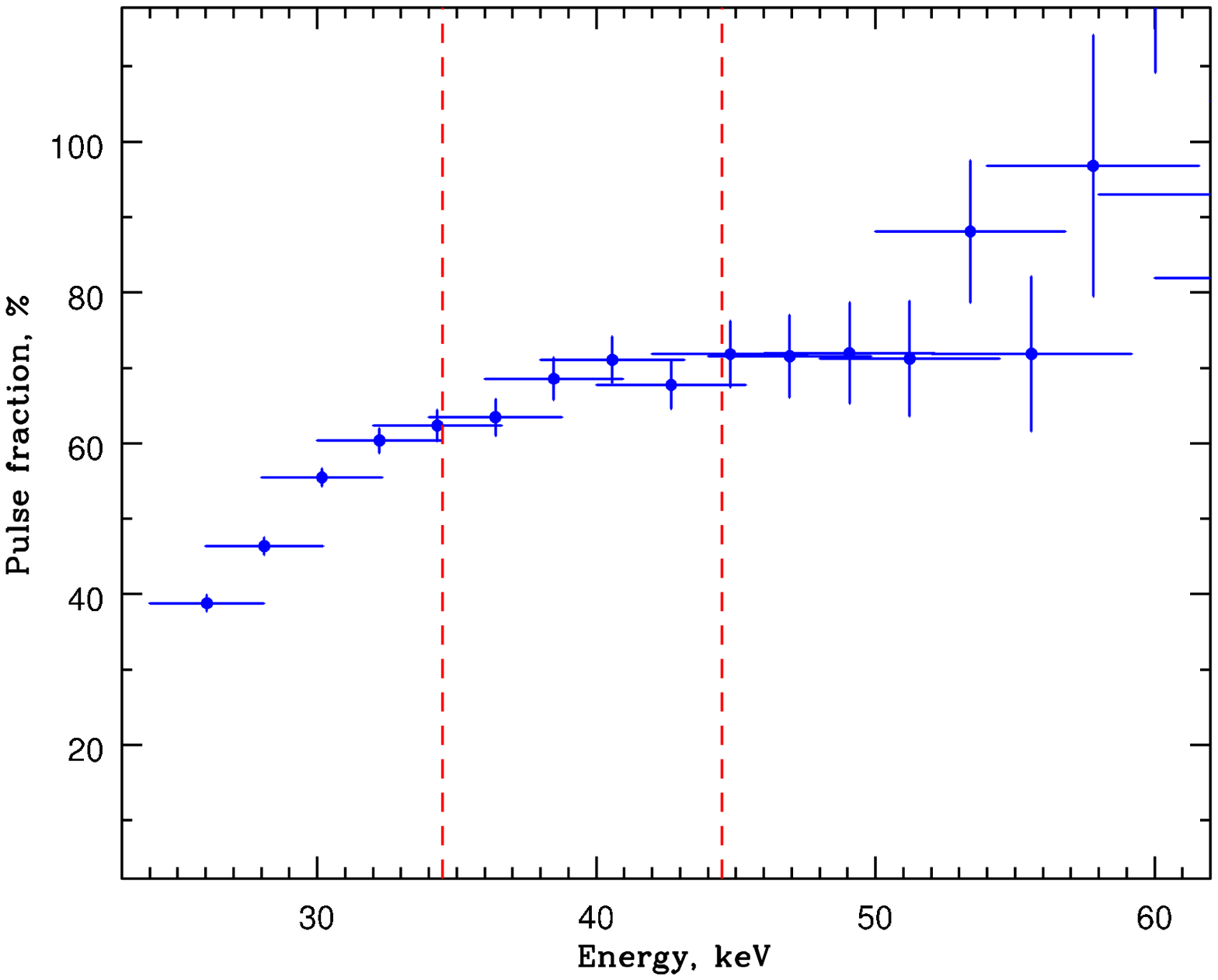}}}
\end{center}
\end{figure}

\end{minipage}

\begin{minipage}[t]{10cm}
\begin{figure}[H]
\begin{center}
\rput(-2,-3.3){\scalebox{1}{\includegraphics[width=11cm,bb=60 175 500 698,clip]{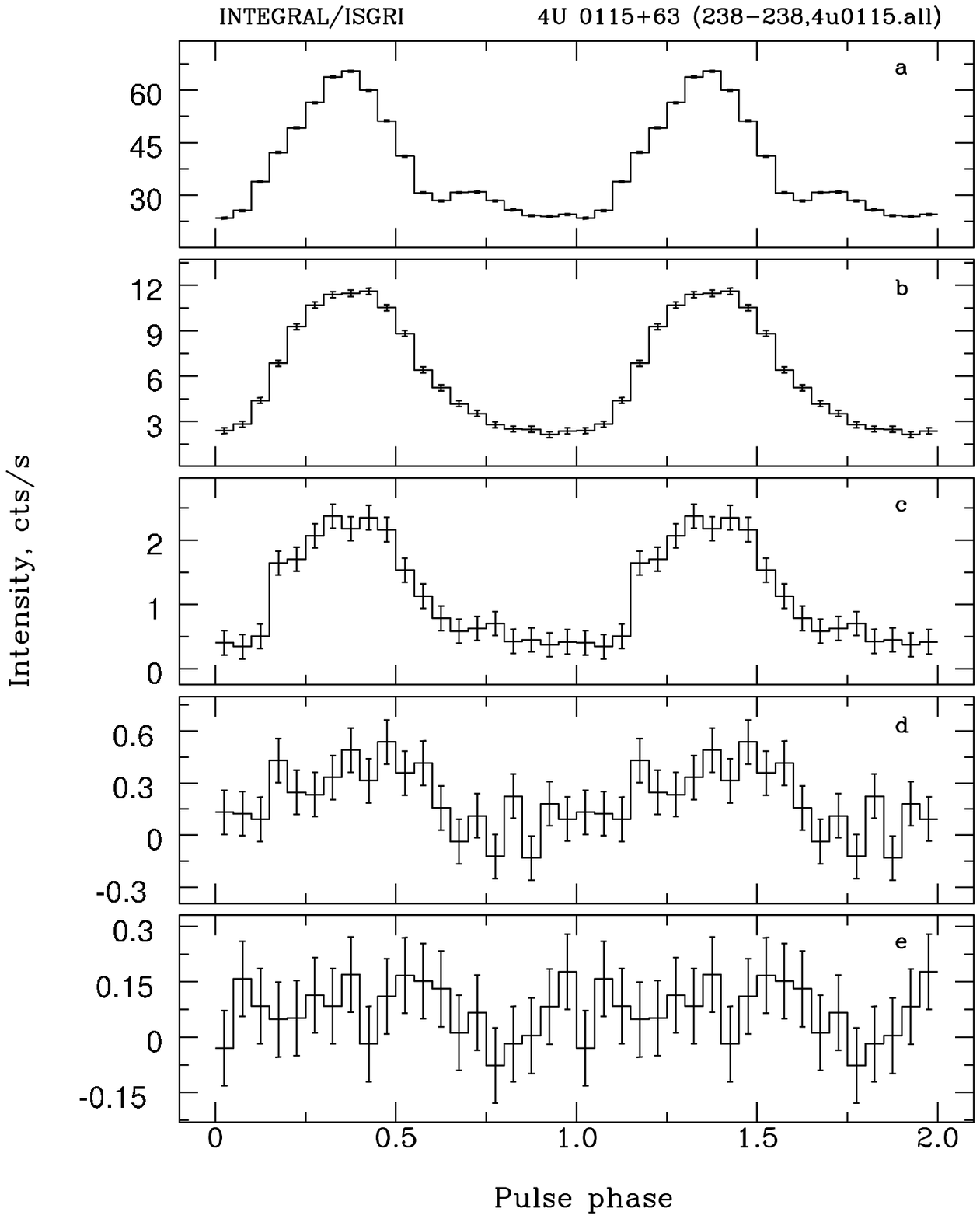}}}
\end{center}
\end{figure}

\end{minipage}

\vspace{85mm}

\begin{minipage}[b]{15cm}
\begin{figure}[H]
\begin{center}
\rput(-2.,-3.3){\scalebox{1}{\includegraphics[width=1cm,bb=60 175 500 698,clip]{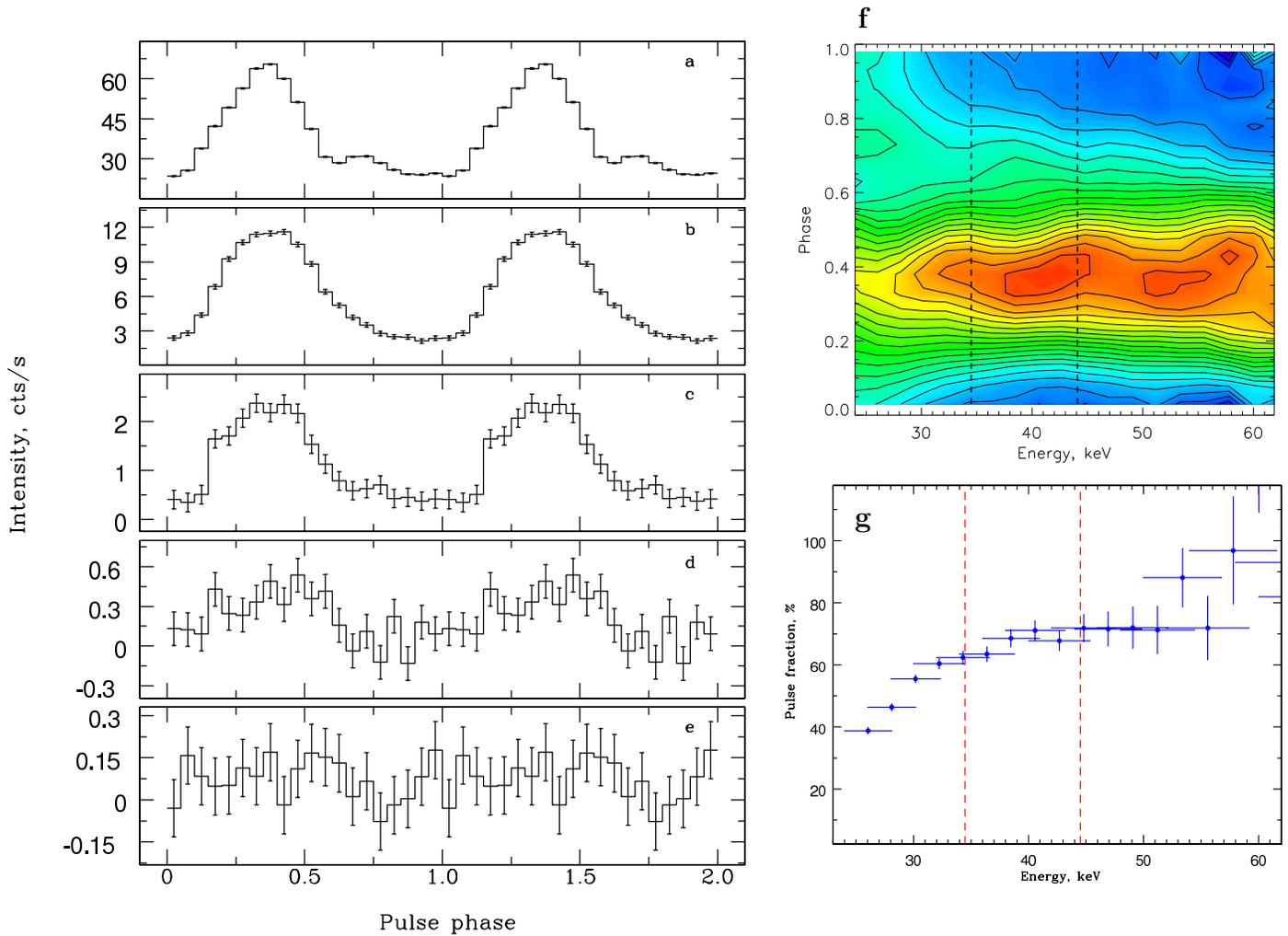}}}
\caption{ Pulse profile for the X-ray pulsar 4U 0115+63 in the (a) 20-30, (b) 30-40, (c) 40-50, (d) 50-70, and (e) 70-100 keV
energy bands from IBIS/INTEGRAL data, orbit 238 (see Table 1). (f) The
corresponding relative intensity map in ``energy -- 
pulse phase'' coordinates and (g) the energy dependence of PF.}\label{u0115}
\end{center}
\end{figure}
\end{minipage}
%********************************************

\clearpage

%********************************************
\begin{minipage}[t]{9cm}
\begin{figure}[H]
\begin{center}
\rput(8.0,-2.6){\scalebox{1}{\includegraphics[width=7cm,bb=45 310 515 725,clip]{figures1/v0332.all_273_273_rev_contour_col.ps2}}}
\rput(5.5,0.8){\mbox{\bf f}}
\rput(5.5,-6.5){\mbox{\bf g}}
\end{center}
\end{figure}

\end{minipage}

\begin{minipage}[h]{9cm}
\begin{figure}[H]
\begin{center}
\rput(8.3,-7.6){\scalebox{1}{\includegraphics[width=8cm,bb=20 275 560 672,clip]{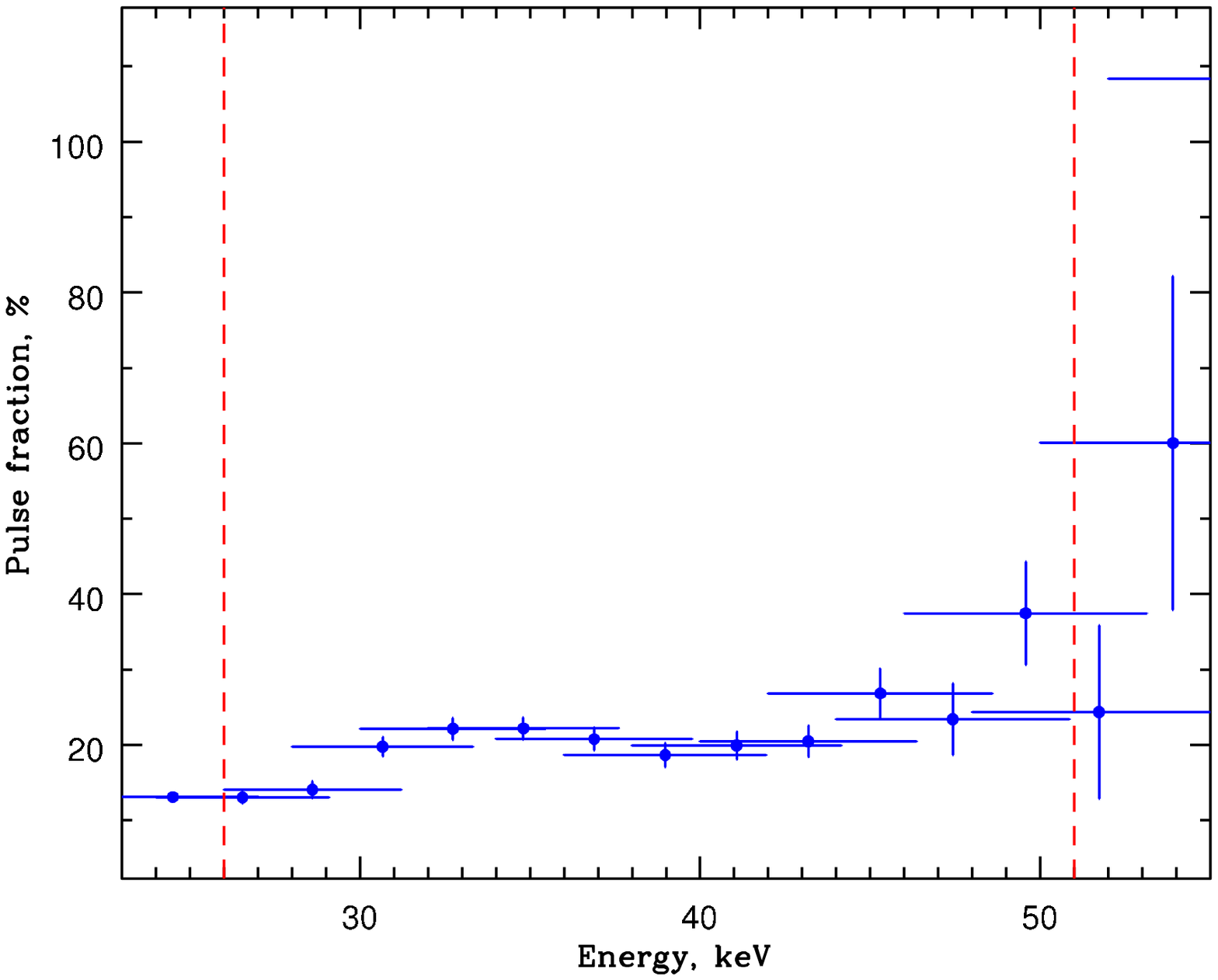}}}
\end{center}
\end{figure}

\end{minipage}

\begin{minipage}[t]{10cm}
\begin{figure}[H]
\begin{center}
\rput(-2,-3.3){\scalebox{1}{\includegraphics[width=11cm,bb=60 175 500 698,clip]{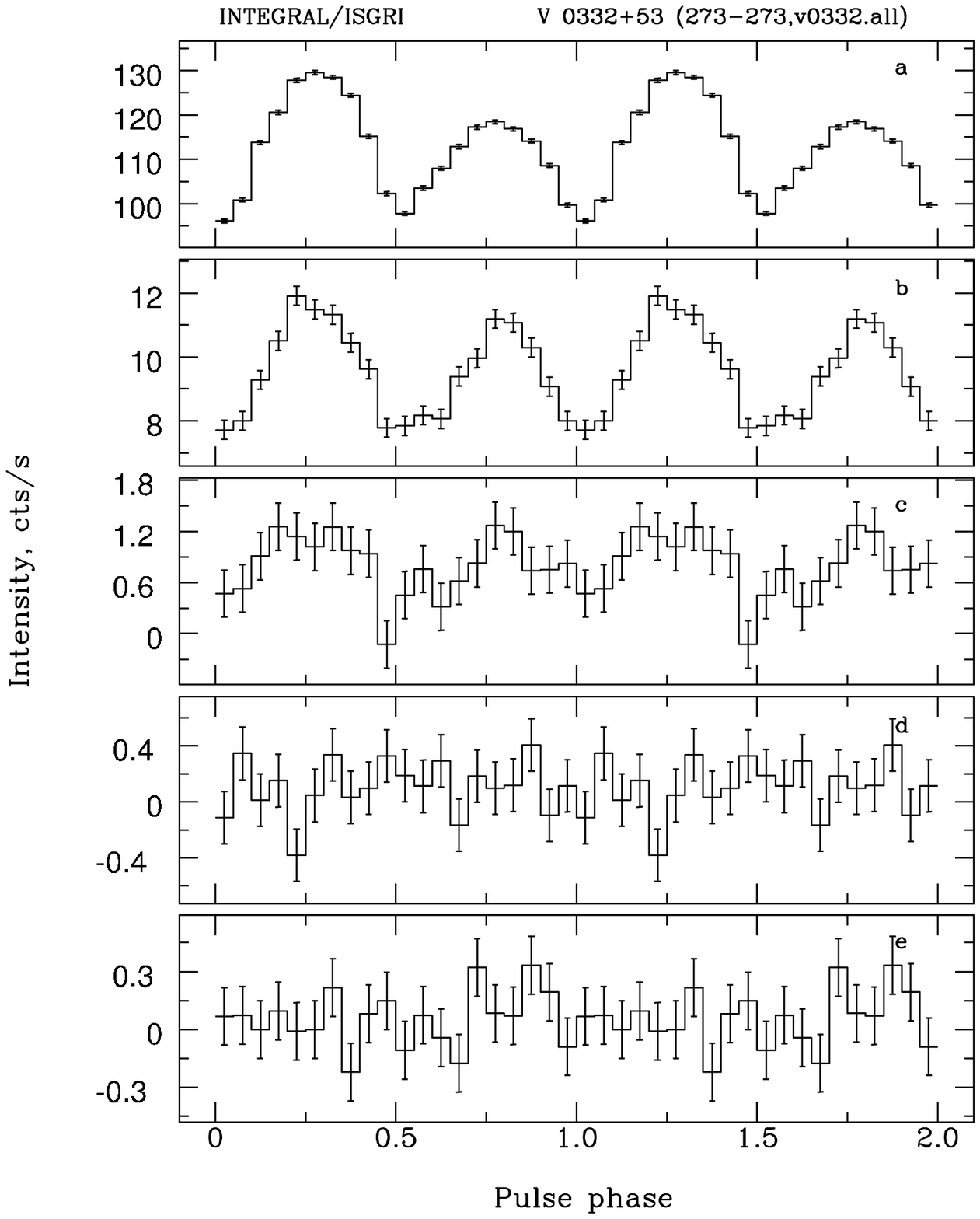}}}
\end{center}
\end{figure}

\end{minipage}

\vspace{85mm}

\begin{minipage}[b]{15cm}
\begin{figure}[H]
\begin{center}
\rput(-2,-3.3){\scalebox{1}{\includegraphics[width=1cm,bb=60 175 500 698,clip]{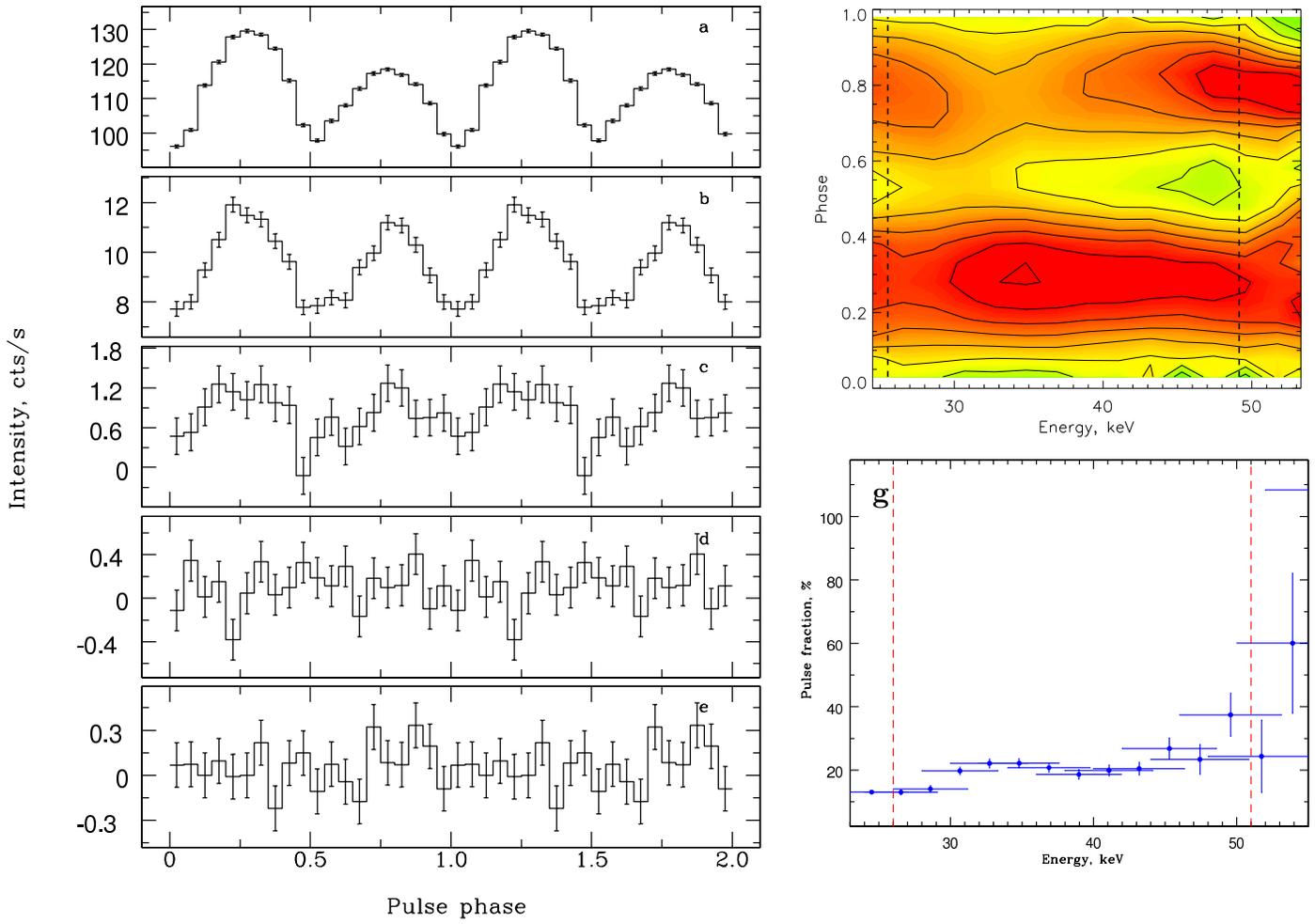}}}
\caption{ Same as Fig. 4 for the source V0332+53, orbit 273, a high state.}\label{v03hicont}
\end{center}
\end{figure}
\end{minipage}
%********************************************

\clearpage

%********************************************
\begin{minipage}[t]{9cm}
\begin{figure}[H]
\begin{center}
\rput(8.0,-2.6){\scalebox{1}{\includegraphics[width=7cm,bb=45 310 515 725,clip]{figures1/v0332.all_284_284_rev_contour_col.ps2}}}
\rput(5.5,0.8){\mbox{\bf f}}
\rput(5.5,-6.5){\mbox{\bf g}}
\end{center}
\end{figure}

\end{minipage}

\begin{minipage}[h]{9cm}
\begin{figure}[H]
\begin{center}
\rput(8.3,-7.6){\scalebox{1}{\includegraphics[width=8cm,bb=20 275 560 672,clip]{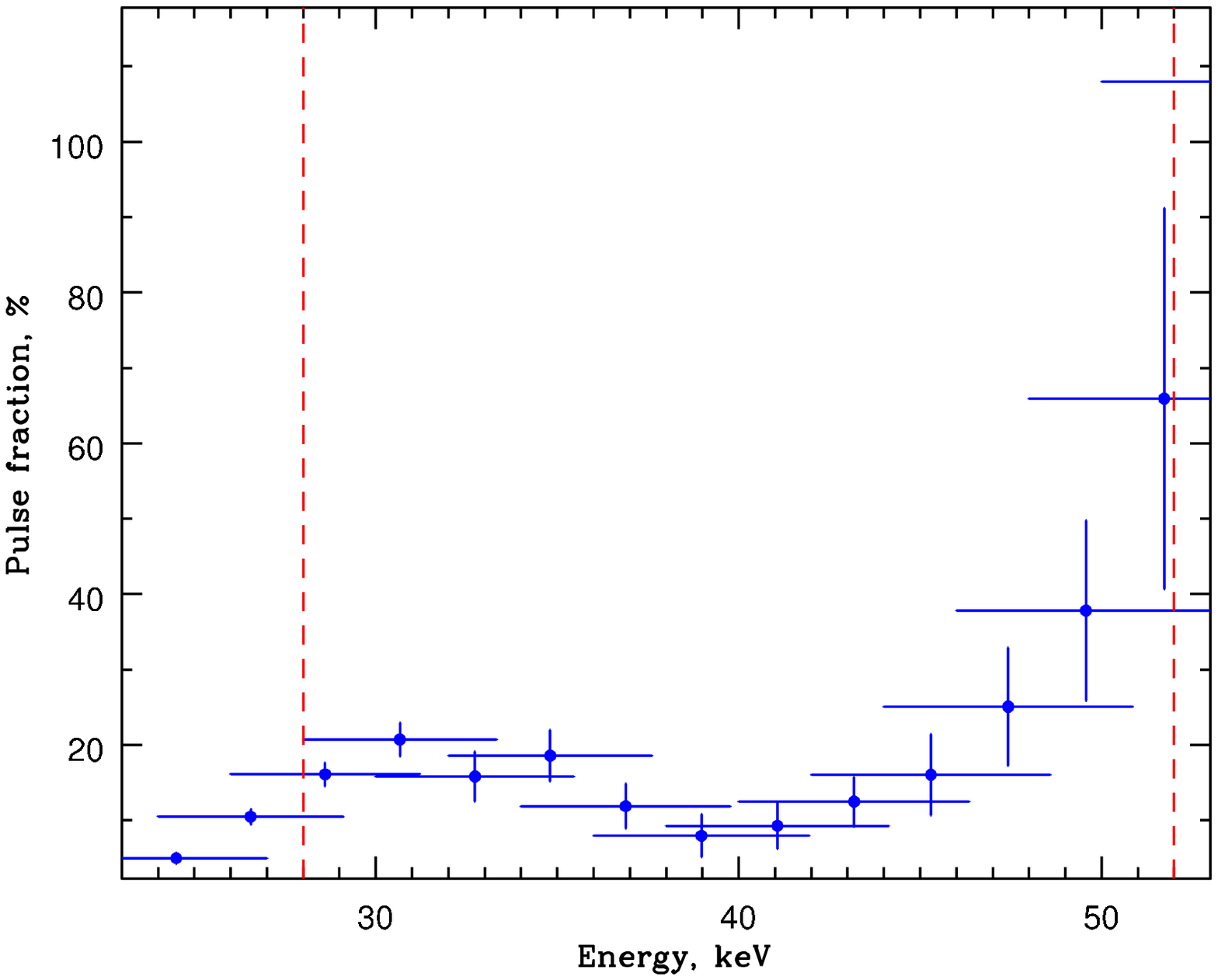}}}
\end{center}
\end{figure}

\end{minipage}

\begin{minipage}[t]{10cm}
\begin{figure}[H]
\begin{center}
\rput(-2,-3.3){\scalebox{1}{\includegraphics[width=11cm,bb=60 175 500 698,clip]{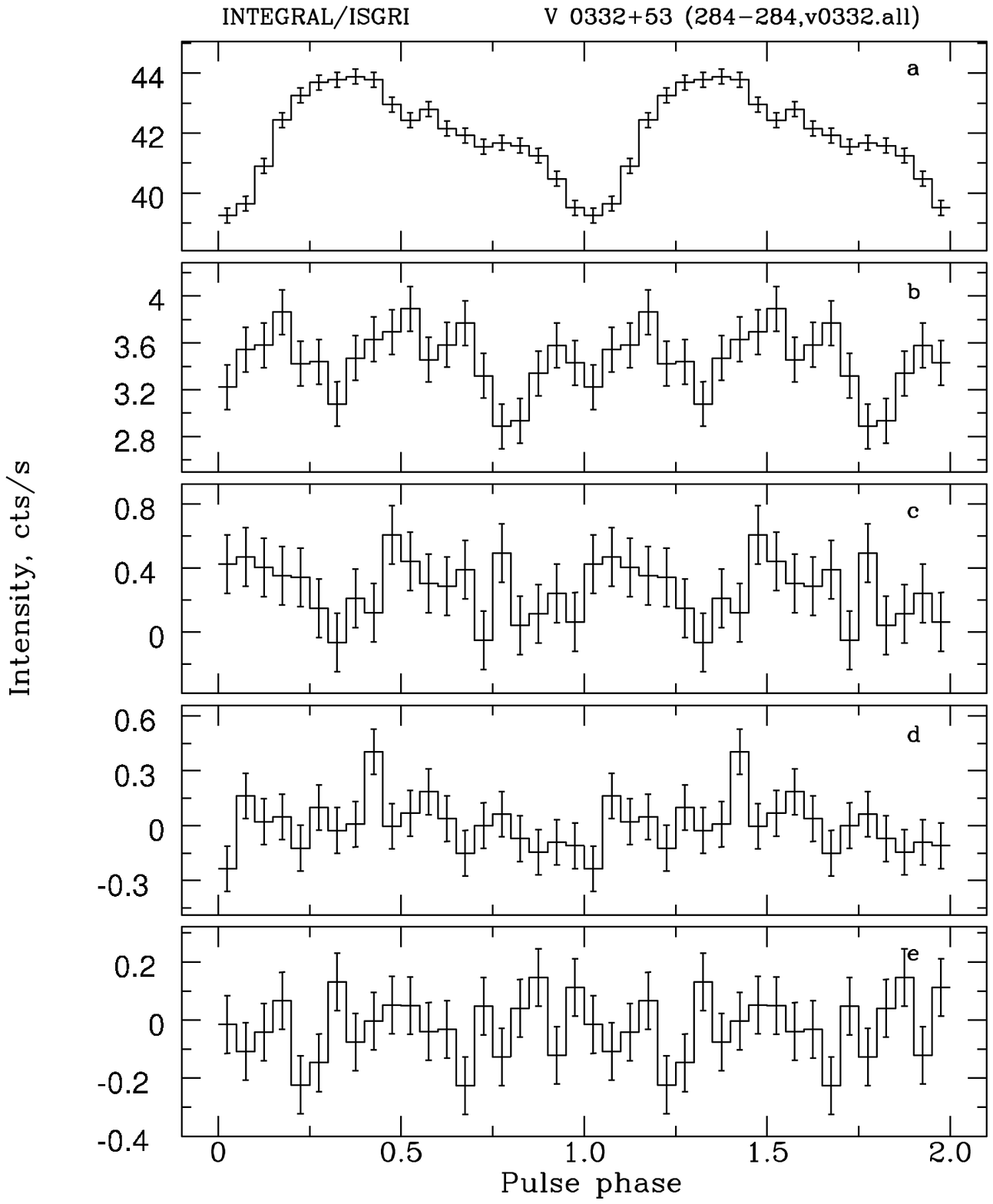}}}
\end{center}
\end{figure}

\end{minipage}

\vspace{85mm}

\begin{minipage}[b]{15cm}
\begin{figure}[H]
\begin{center}
\rput(-2,-3.3){\scalebox{1}{\includegraphics[width=1cm,bb=60 175 500 698,clip]{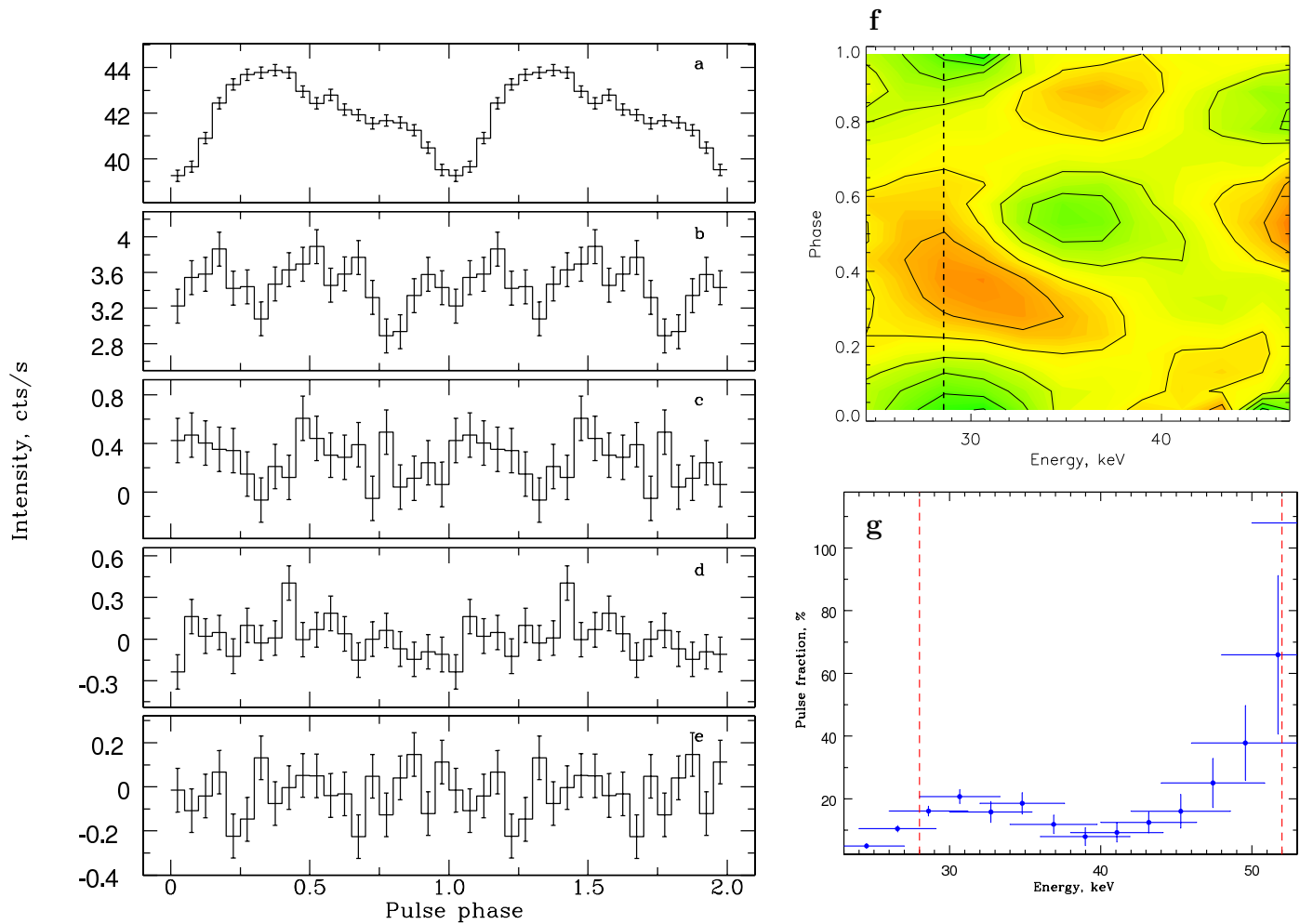}}}
\caption{ Same as Fig. 4 for the source V0332+53, orbit 284, a low state.}\label{v03locont}
\end{center}
\end{figure}
\end{minipage}
%********************************************

\clearpage

%********************************************
\begin{minipage}[t]{9cm}
\begin{figure}[H]
\begin{center}
\rput(8.0,-2.6){\scalebox{1}{\includegraphics[width=7cm,bb=45 310 515 725,clip]{figures1/a0535_all_352_352_rev_contour_col.ps2}}}
\rput(5.5,0.8){\mbox{\bf f}}
\rput(5.5,-6.5){\mbox{\bf g}}
\end{center}
\end{figure}

\end{minipage}

\begin{minipage}[h]{9cm}
\begin{figure}[H]
\begin{center}
\rput(8.3,-7.6){\scalebox{1}{\includegraphics[width=8cm,bb=20 275 560 672,clip]{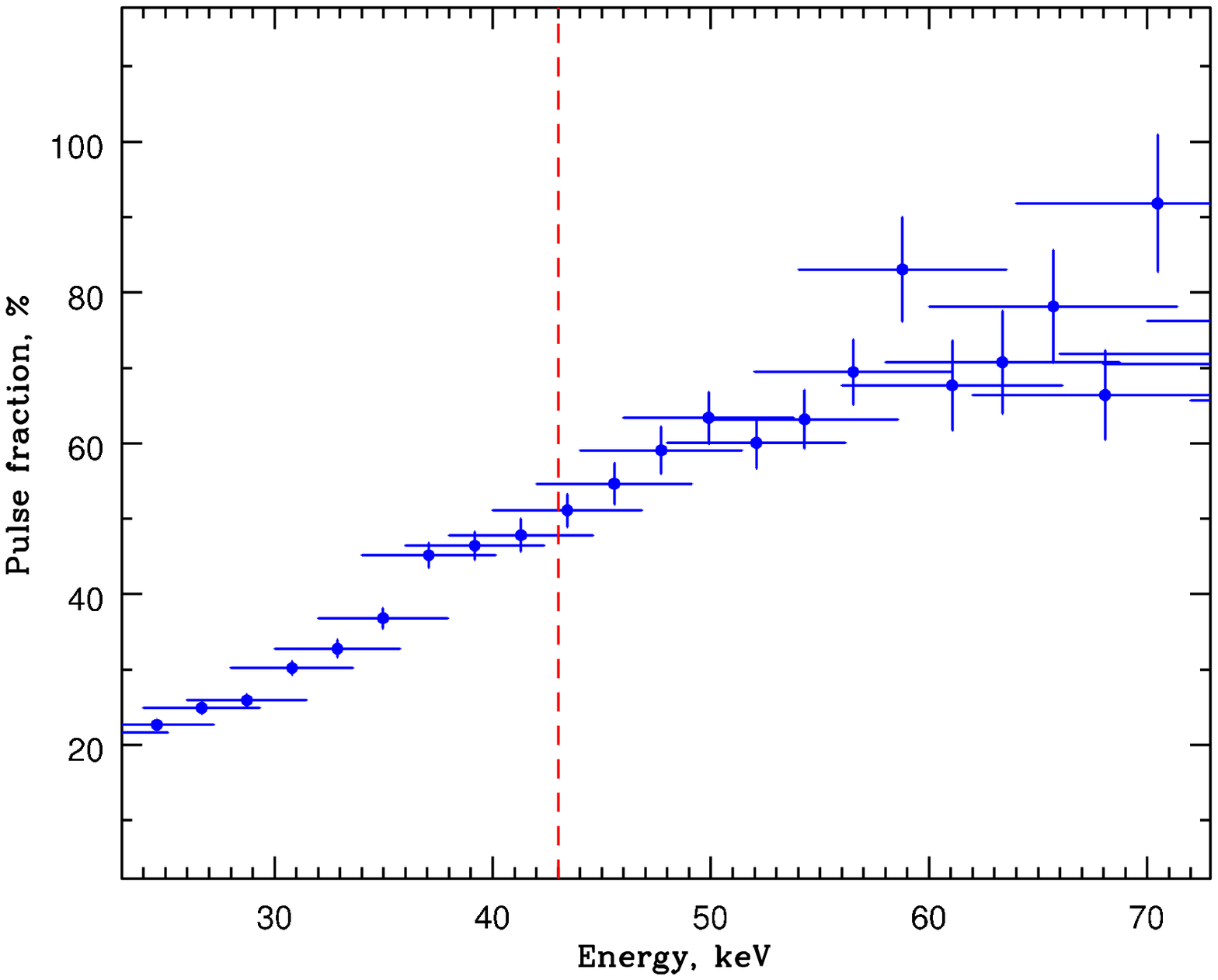}}}
\end{center}
\end{figure}

\end{minipage}

\begin{minipage}[t]{10cm}
\begin{figure}[H]
\begin{center}
\rput(-2,-3.3){\scalebox{1}{\includegraphics[width=11cm,bb=60 175 500 698,clip]{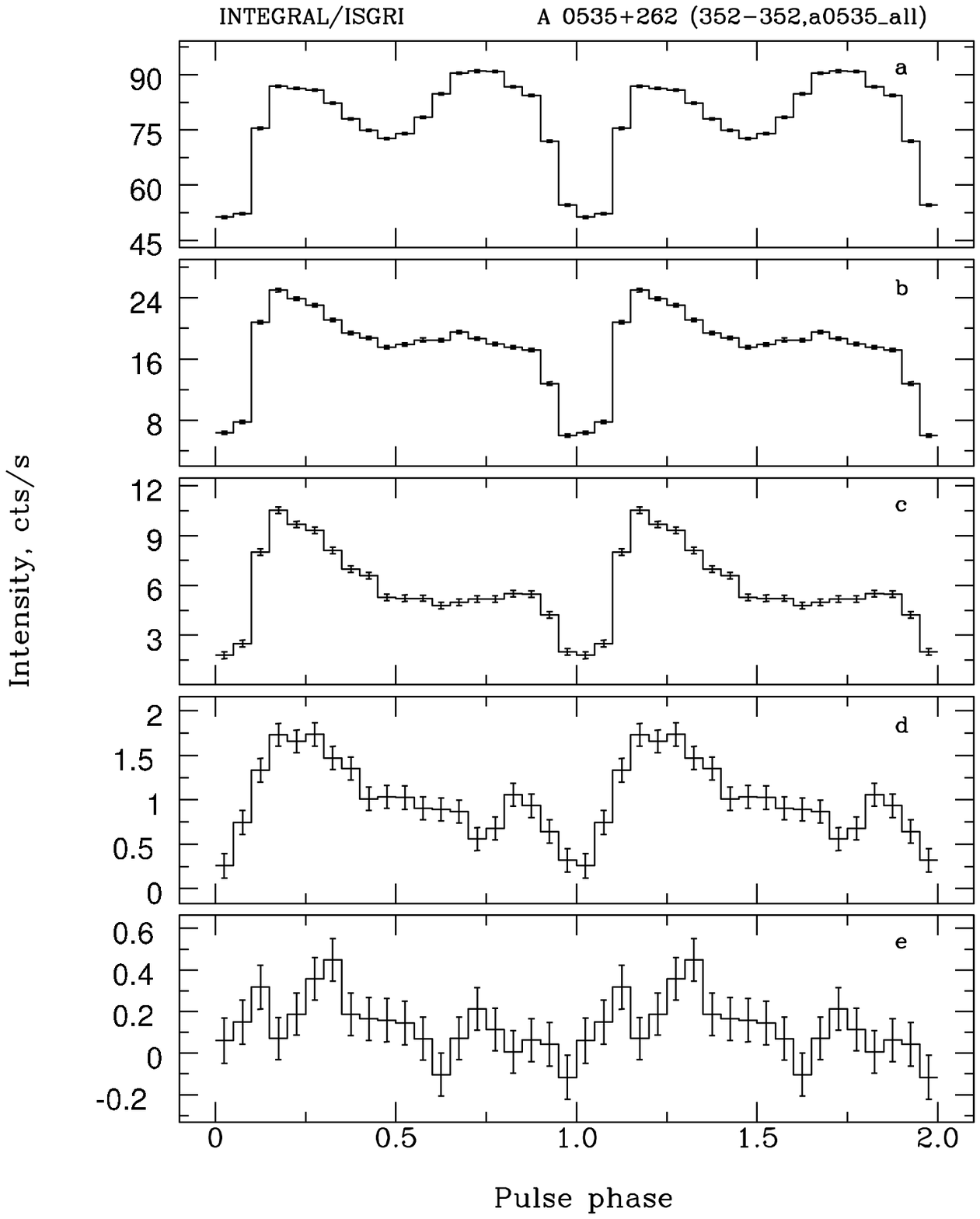}}}
\end{center}
\end{figure}

\end{minipage}

\vspace{85mm}

\begin{minipage}[b]{15cm}
\begin{figure}[H]
\begin{center}
\rput(-2,-3.3){\scalebox{1}{\includegraphics[width=1cm,bb=60 175 500 698,clip]{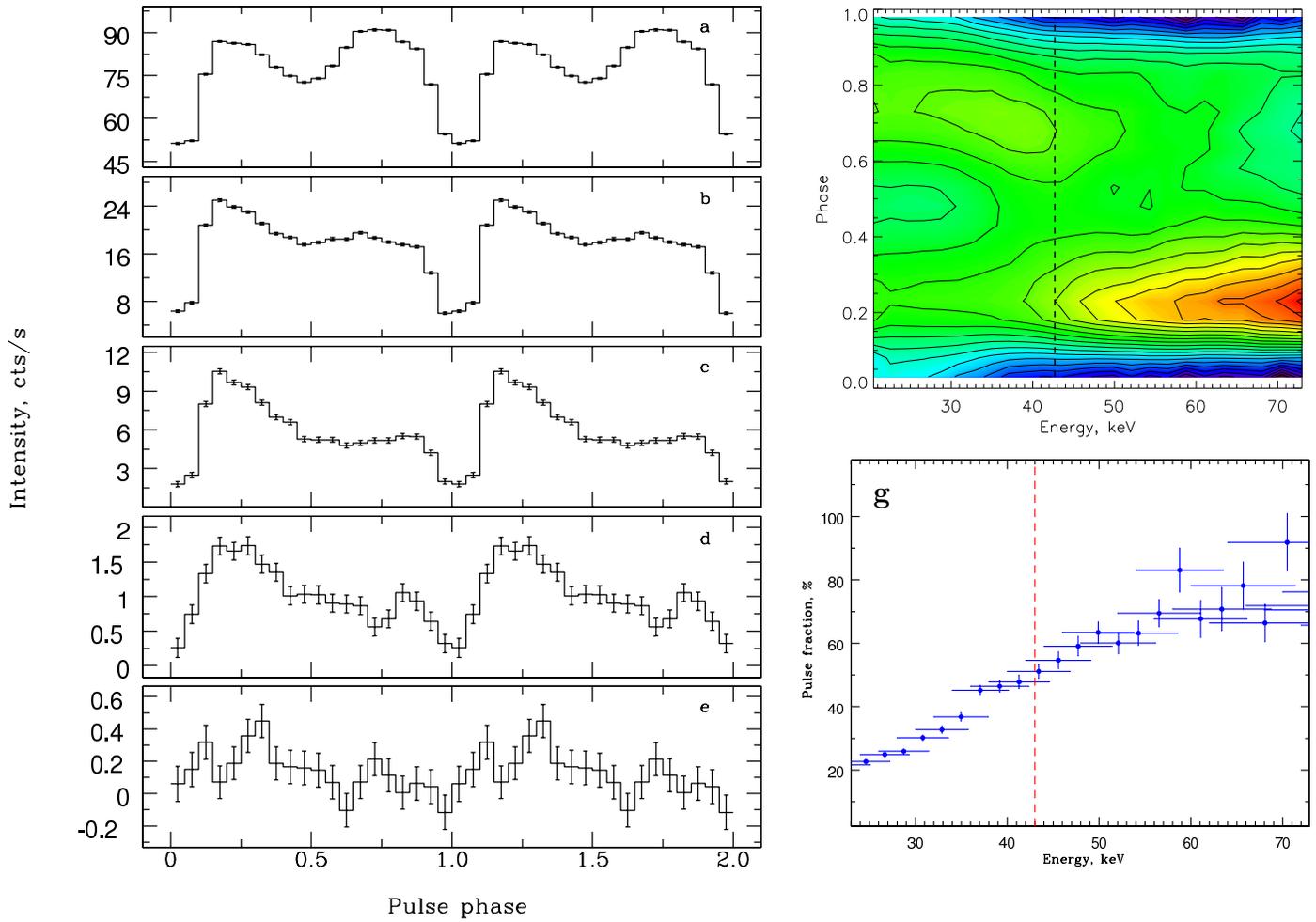}}}
\caption{ Same as Fig. 4 for the source A0535+262, orbit 352, averaged.}\label{a0535}
\end{center}
\end{figure}
\end{minipage}
%********************************************

\clearpage

%********************************************
\begin{minipage}[t]{9cm}
\begin{figure}[H]
\begin{center}
\rput(8.0,-2.6){\scalebox{1}{\includegraphics[width=7cm,bb=45 310 515 725,clip]{figures1/velax1_all_370_390_rev_contour_col.ps2}}}
\rput(5.5,0.8){\mbox{\bf f}}
\rput(5.5,-6.5){\mbox{\bf g}}
\end{center}
\end{figure}

\end{minipage}

\begin{minipage}[h]{9cm}
\begin{figure}[H]
\begin{center}
\rput(8.3,-7.6){\scalebox{1}{\includegraphics[width=8cm,bb=20 275 560 672,clip]{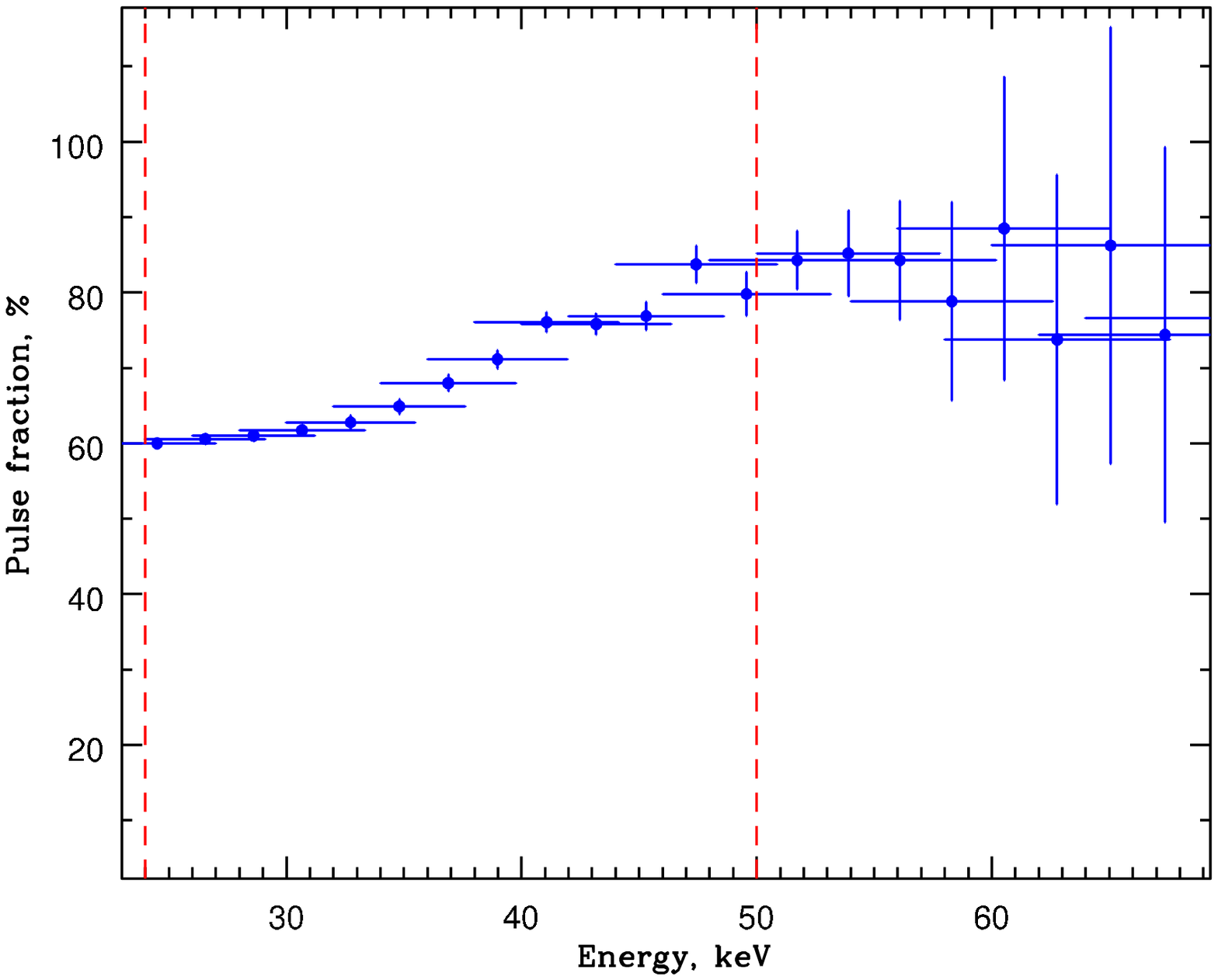}}}
\end{center}
\end{figure}

\end{minipage}

\begin{minipage}[t]{10cm}
\begin{figure}[H]
\begin{center}
\rput(-2,-3.3){\scalebox{1}{\includegraphics[width=11cm,bb=60 175 500 698,clip]{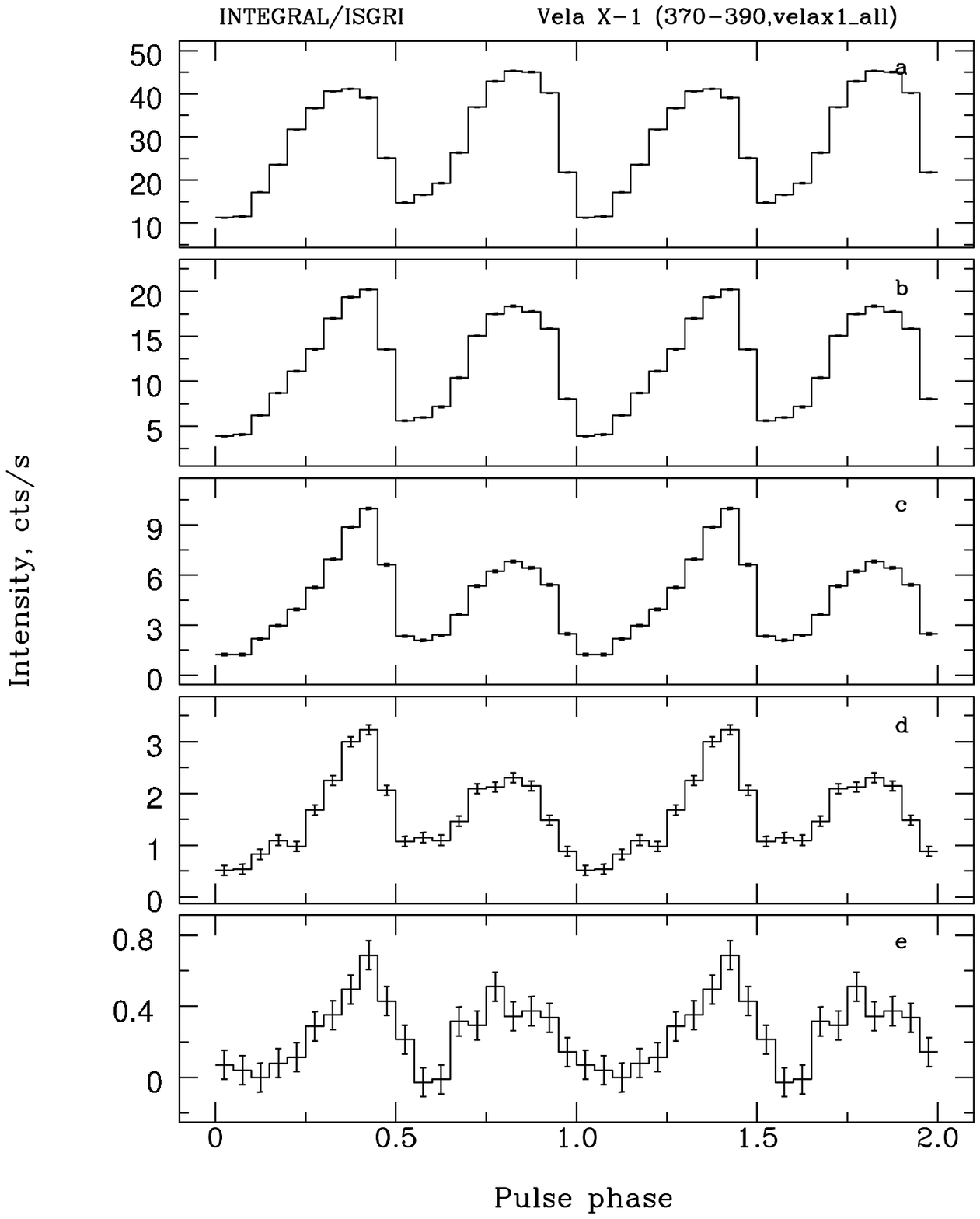}}}
\end{center}
\end{figure}

\end{minipage}

\vspace{85mm}

\begin{minipage}[b]{15cm}
\begin{figure}[H]
\begin{center}
\rput(-2,-3.3){\scalebox{1}{\includegraphics[width=1cm,bb=60 175 500 698,clip]{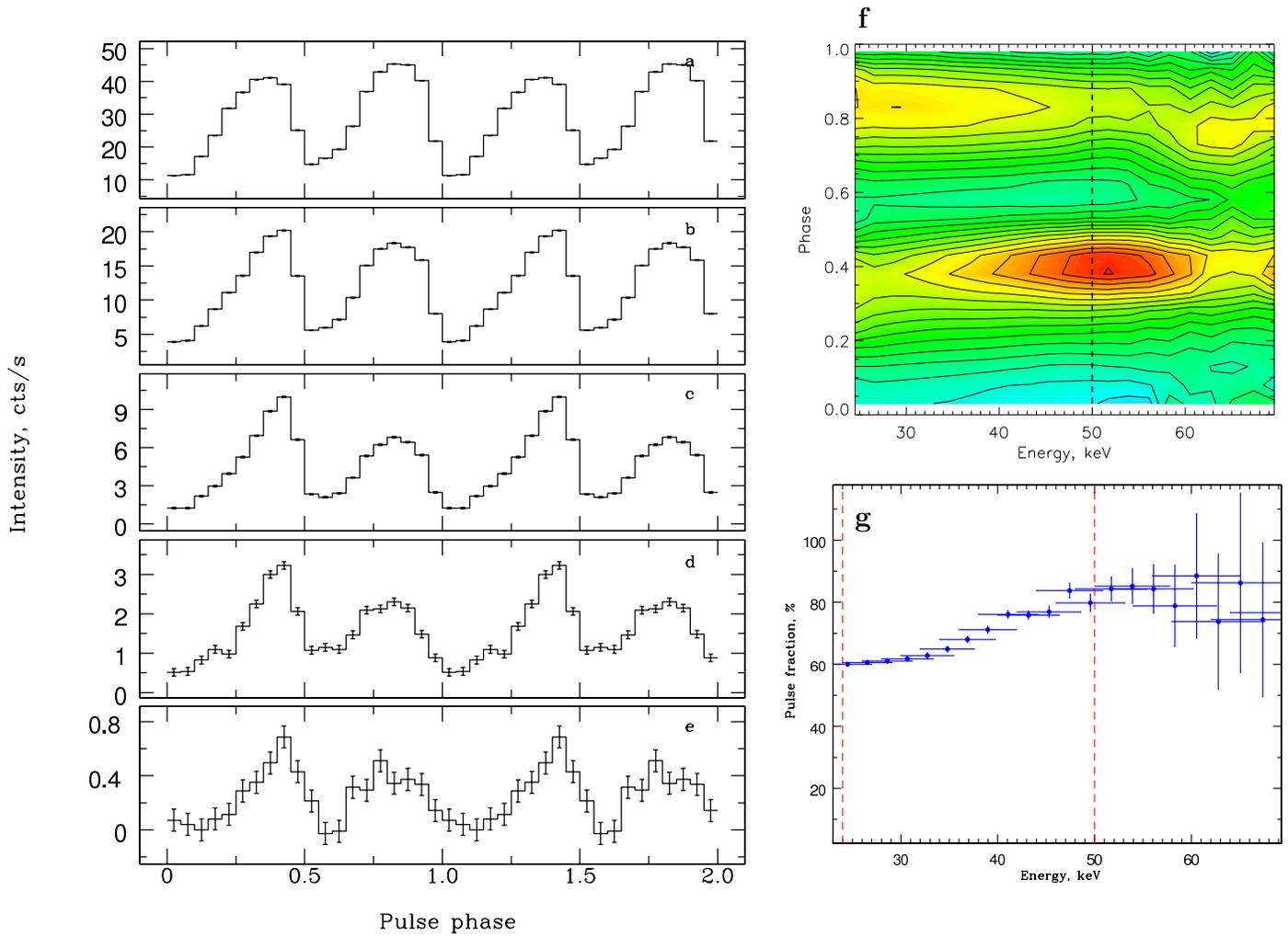}}}
\caption{ Same as Fig. 4 for the source Vela X-1, orbits 373-383, averaged.}\label{velax1}
\end{center}
\end{figure}
\end{minipage}
%********************************************

\clearpage

%********************************************
\begin{minipage}[t]{9cm}
\begin{figure}[H]
\begin{center}
\rput(8.0,-2.6){\scalebox{1}{\includegraphics[width=7cm,bb=45 310 515 725,clip]{figures1/cenx3_high_190_193_rev_contour_col.ps2}}}
\rput(5.5,0.8){\mbox{\bf f}}
\rput(5.5,-6.5){\mbox{\bf g}}
\end{center}
\end{figure}

\end{minipage}

\begin{minipage}[h]{9cm}
\begin{figure}[H]
\begin{center}
\rput(8.3,-7.6){\scalebox{1}{\includegraphics[width=8cm,bb=20 275 560 672,clip]{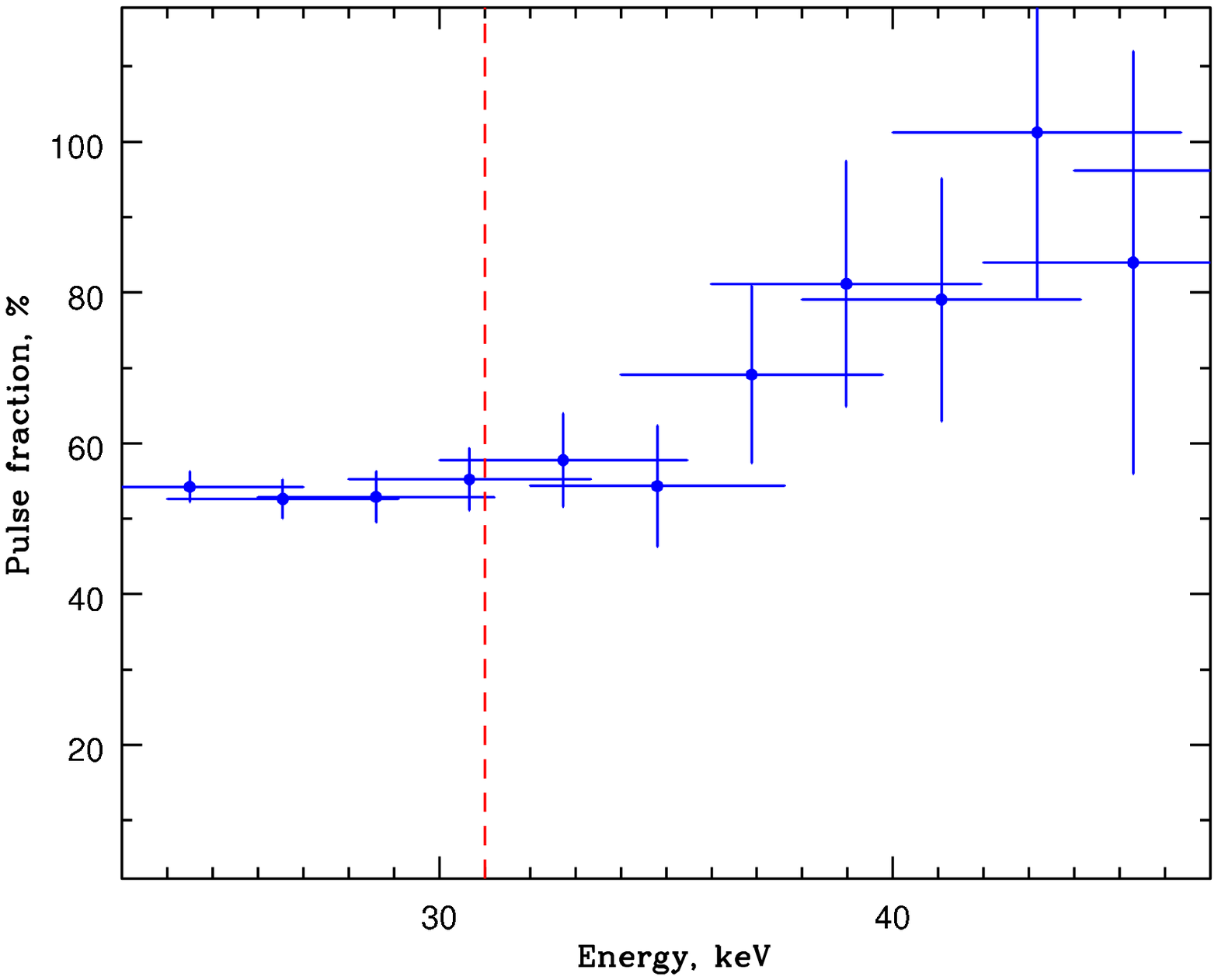}}}
\end{center}
\end{figure}

\end{minipage}

\begin{minipage}[t]{10cm}
\begin{figure}[H]
\begin{center}
\rput(-2,-3.3){\scalebox{1}{\includegraphics[width=11cm,bb=60 175 500 698,clip]{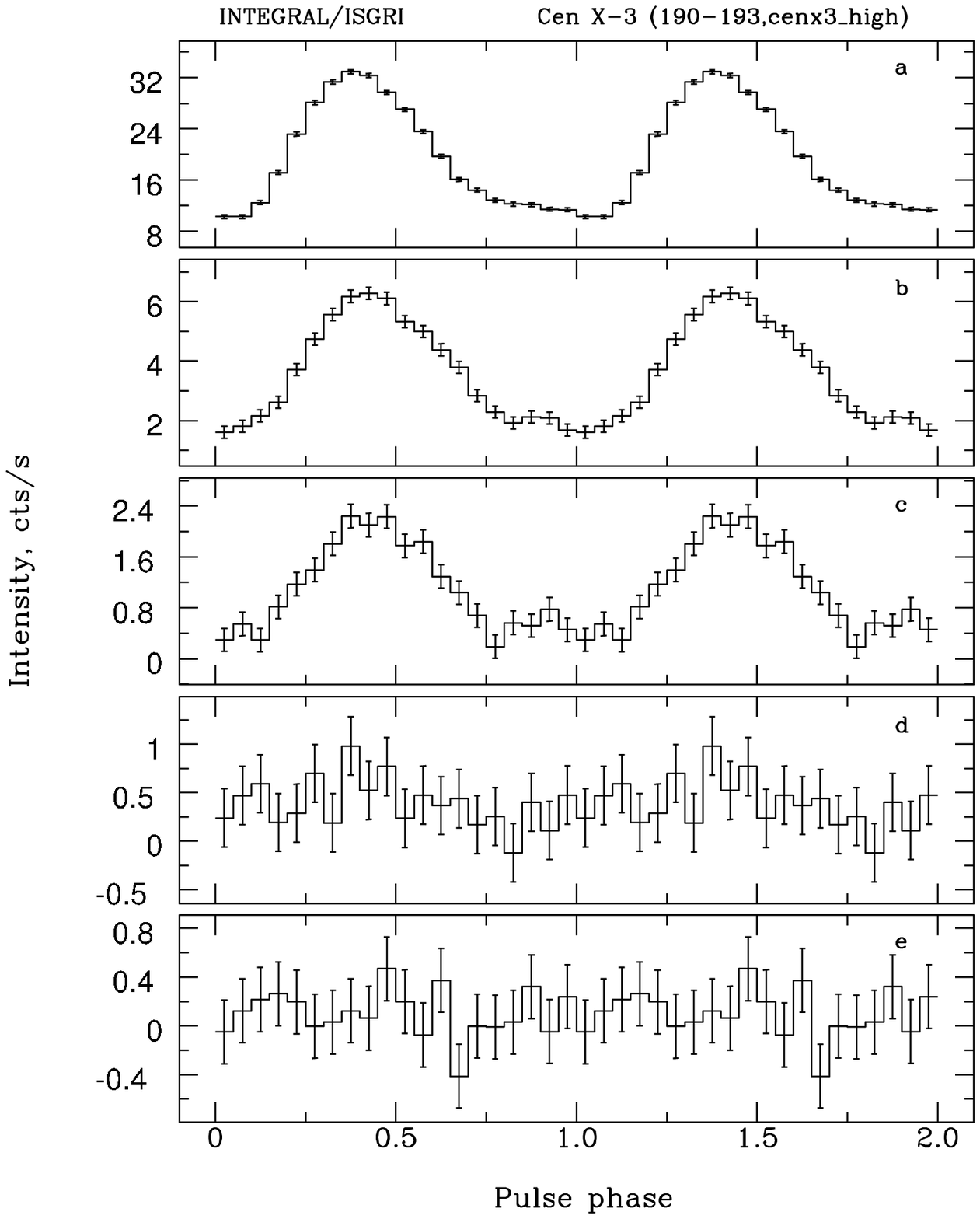}}}
\end{center}
\end{figure}

\end{minipage}

\vspace{85mm}

\begin{minipage}[b]{15cm}
\begin{figure}[H]
\begin{center}
\rput(-2,-3.3){\scalebox{1}{\includegraphics[width=1cm,bb=60 175 500 698,clip]{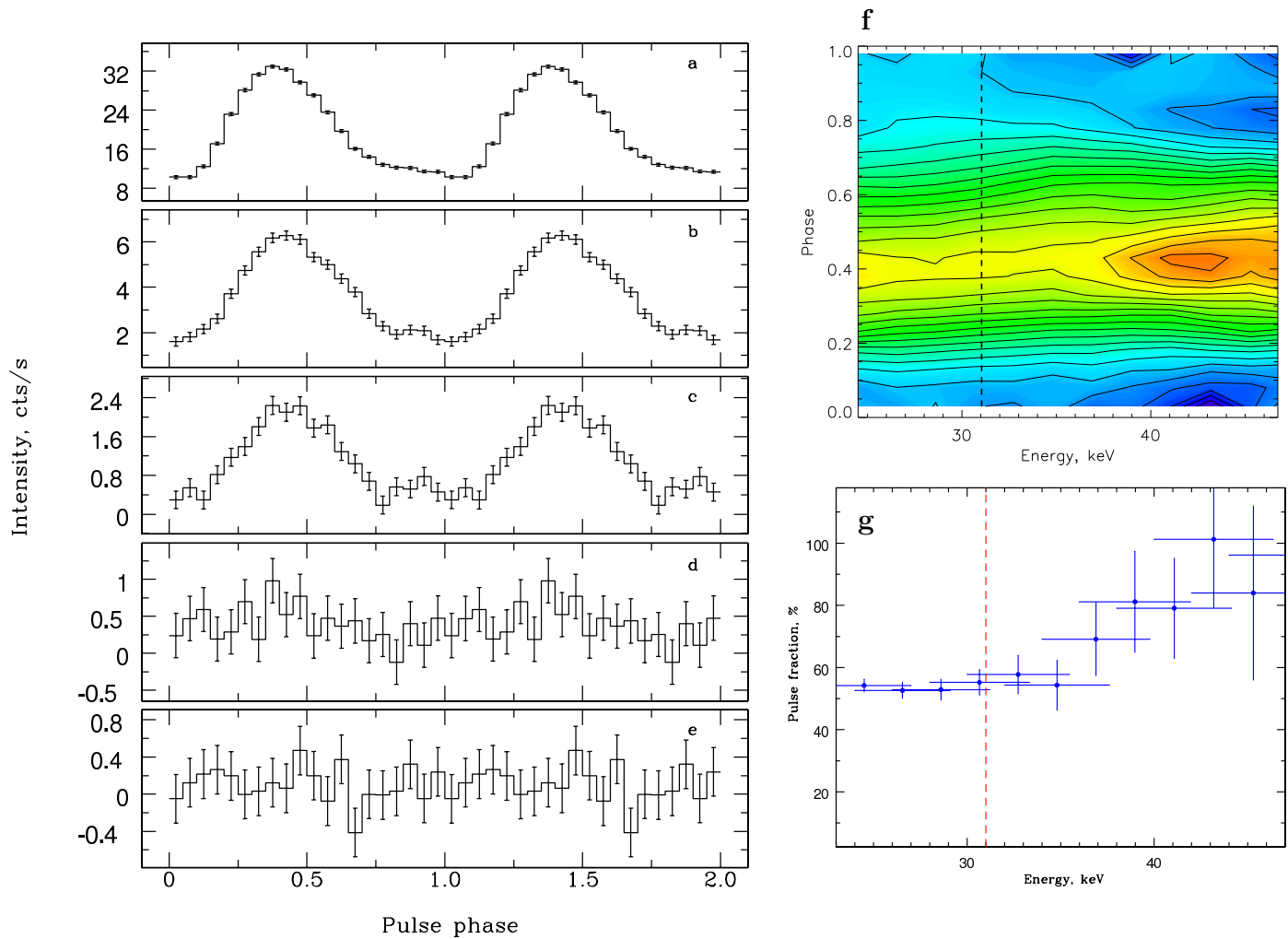}}}
\caption{ Same as Fig. 4 for the source Cen X-3, orbit 192, averaged.}\label{cenx3}
\end{center}
\end{figure}
\end{minipage}
%********************************************

\clearpage

%********************************************
\begin{minipage}[t]{9cm}
\begin{figure}[H]
\begin{center}
\rput(8.0,-2.6){\scalebox{1}{\includegraphics[width=7cm,bb=45 310 515 725,clip]{figures1/gx301_all_323_323_rev_contour_col.ps2}}}
\rput(5.5,0.8){\mbox{\bf f}}
\rput(5.5,-6.5){\mbox{\bf g}}
\end{center}
\end{figure}

\end{minipage}

\begin{minipage}[h]{9cm}
\begin{figure}[H]
\begin{center}
\rput(8.3,-7.6){\scalebox{1}{\includegraphics[width=8cm,bb=20 275 560 672,clip]{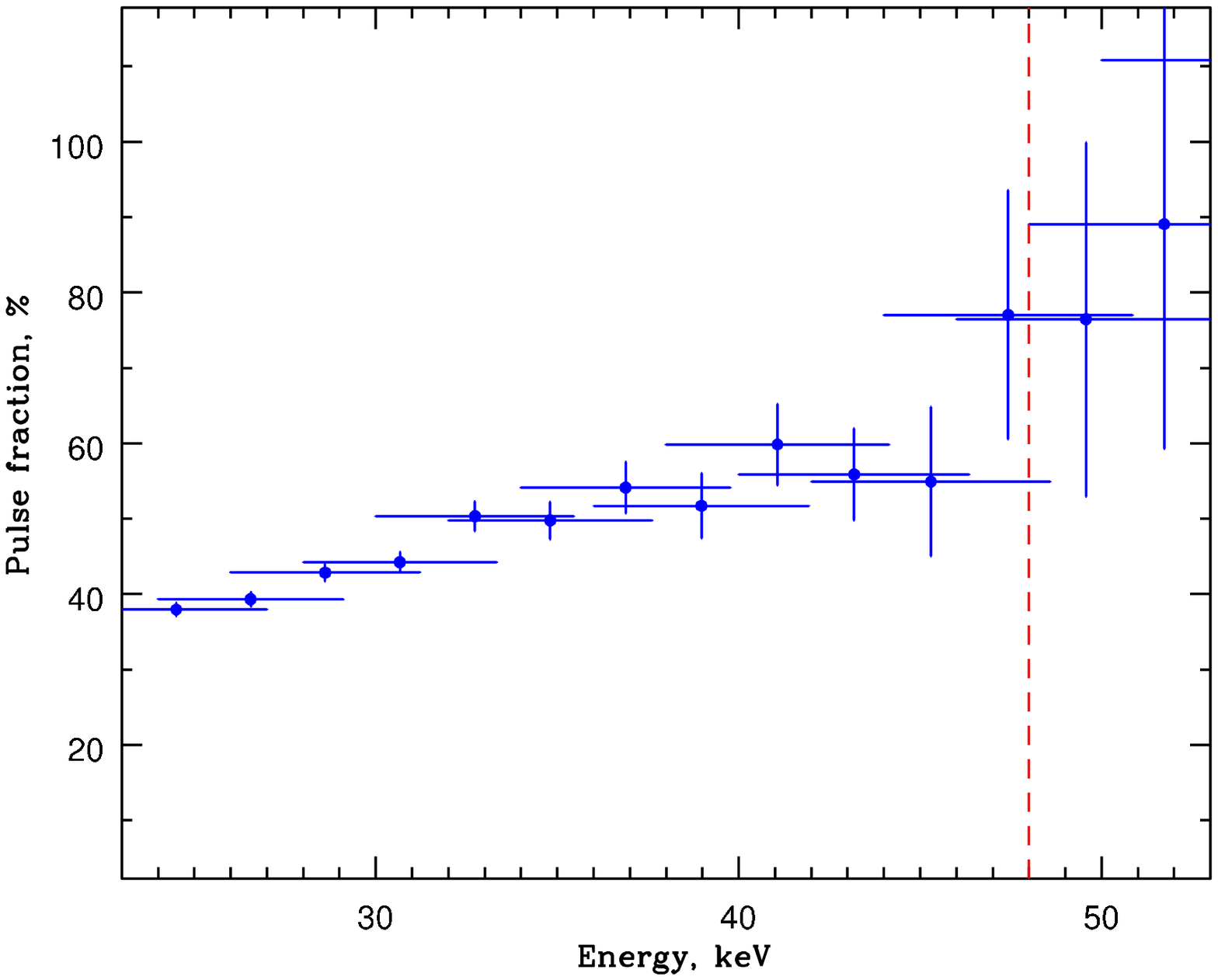}}}
\end{center}
\end{figure}

\end{minipage}

\begin{minipage}[t]{10cm}
\begin{figure}[H]
\begin{center}
\rput(-2,-3.3){\scalebox{1}{\includegraphics[width=11cm,bb=60 175 500 698,clip]{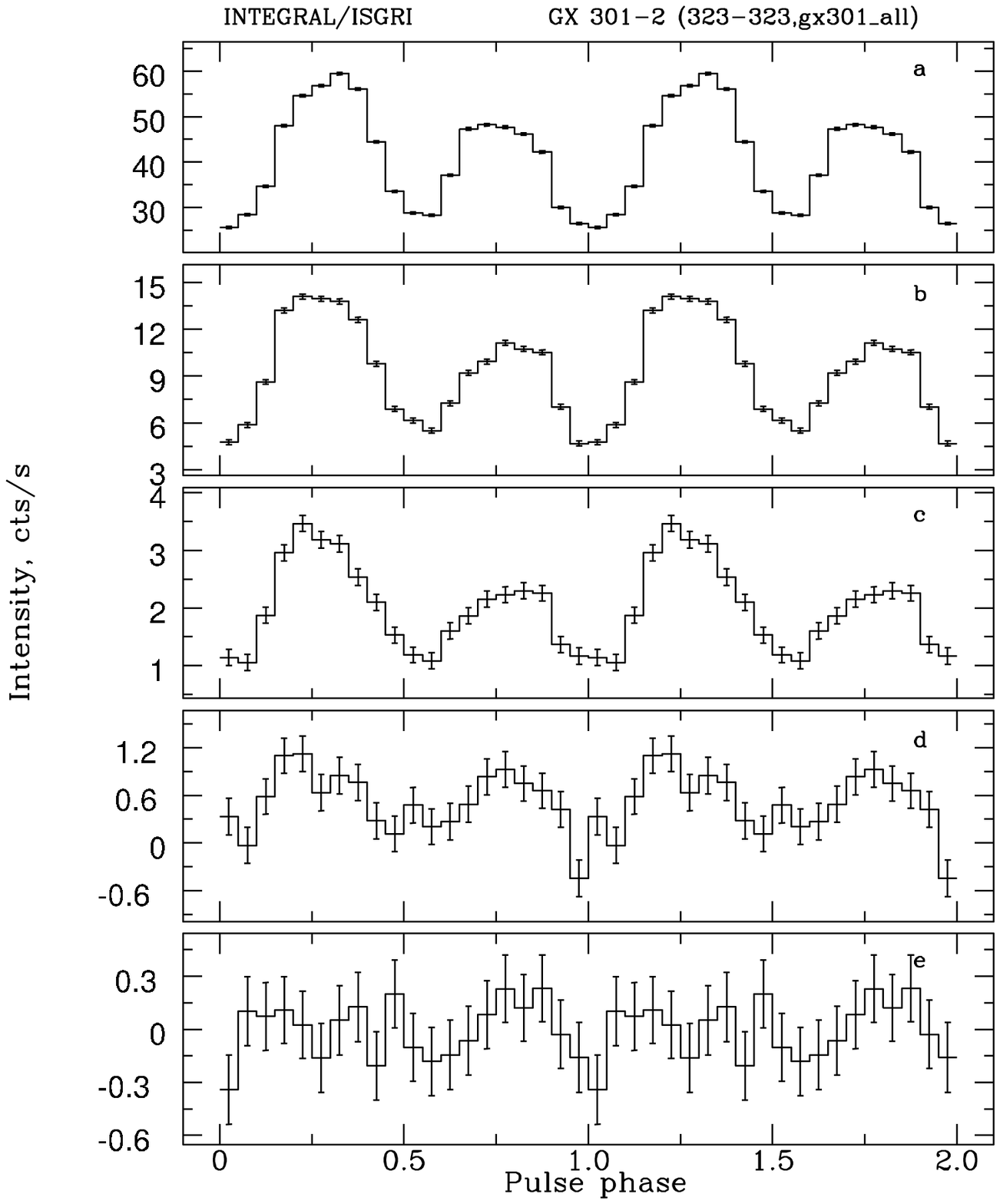}}}
\end{center}
\end{figure}

\end{minipage}

\vspace{85mm}

\begin{minipage}[b]{15cm}
\begin{figure}[H]
\begin{center}
\rput(-2,-3.3){\scalebox{1}{\includegraphics[width=1cm,bb=60 175 500 698,clip]{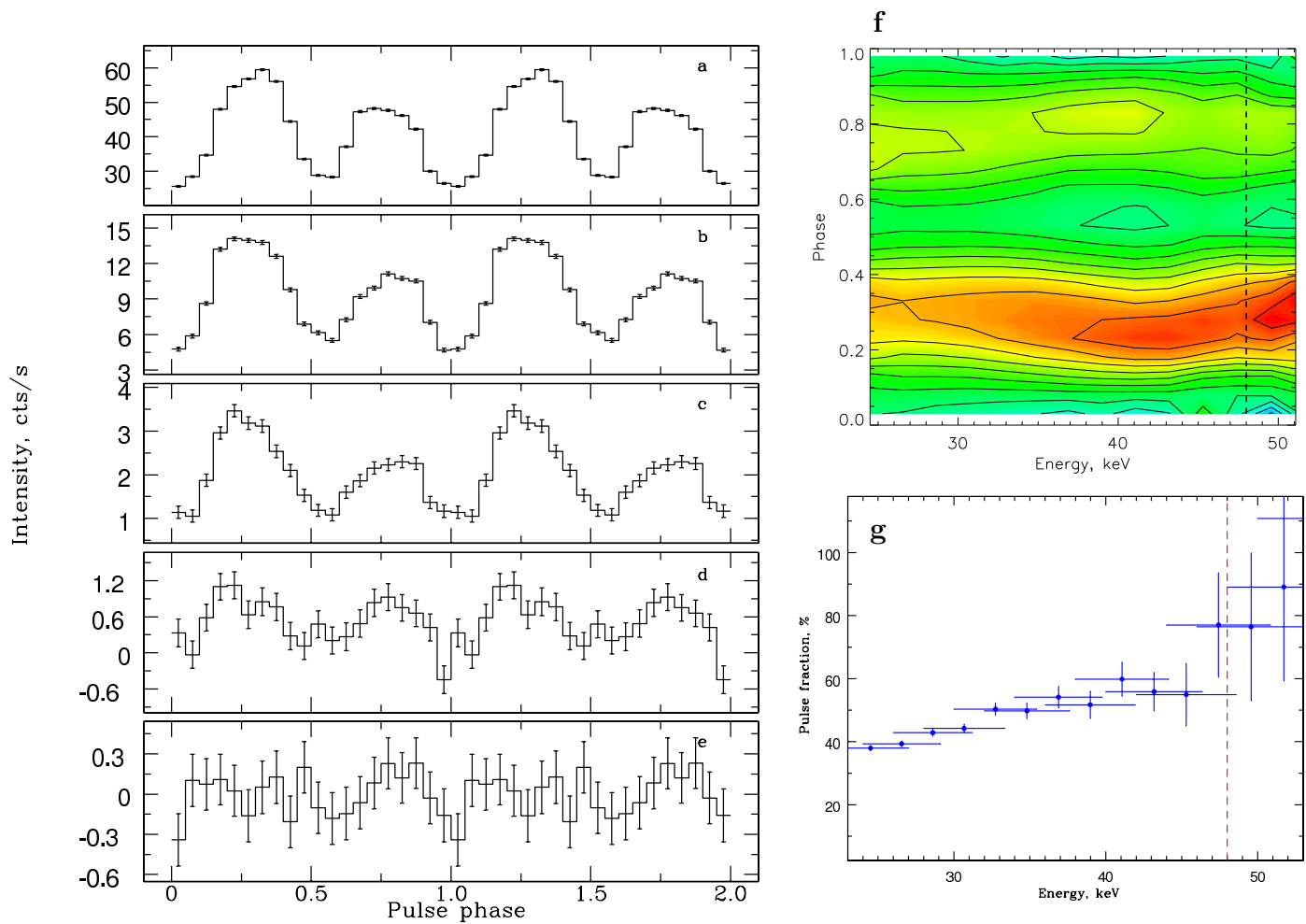}}}
\caption{ Same as Fig. 4 for the source GX 301-2, orbit 323.}\label{gx301}
\end{center}
\end{figure}
\end{minipage}
%********************************************

\clearpage

%********************************************
\begin{minipage}[t]{9cm}
\begin{figure}[H]
\begin{center}
\rput(8.0,-2.6){\scalebox{1}{\includegraphics[width=7cm,bb=45 310 515 725,clip]{figures1/oao1657_high_364_365_rev_contour_col.ps2}
}}
\rput(5.5,0.8){\mbox{\bf f}}
\rput(5.5,-6.5){\mbox{\bf g}}
\end{center}
\end{figure}

\end{minipage}

\begin{minipage}[h]{9cm}
\begin{figure}[H]
\begin{center}
\rput(8.3,-7.6){\scalebox{1}{\includegraphics[width=8cm,bb=20 275 560 672,clip]{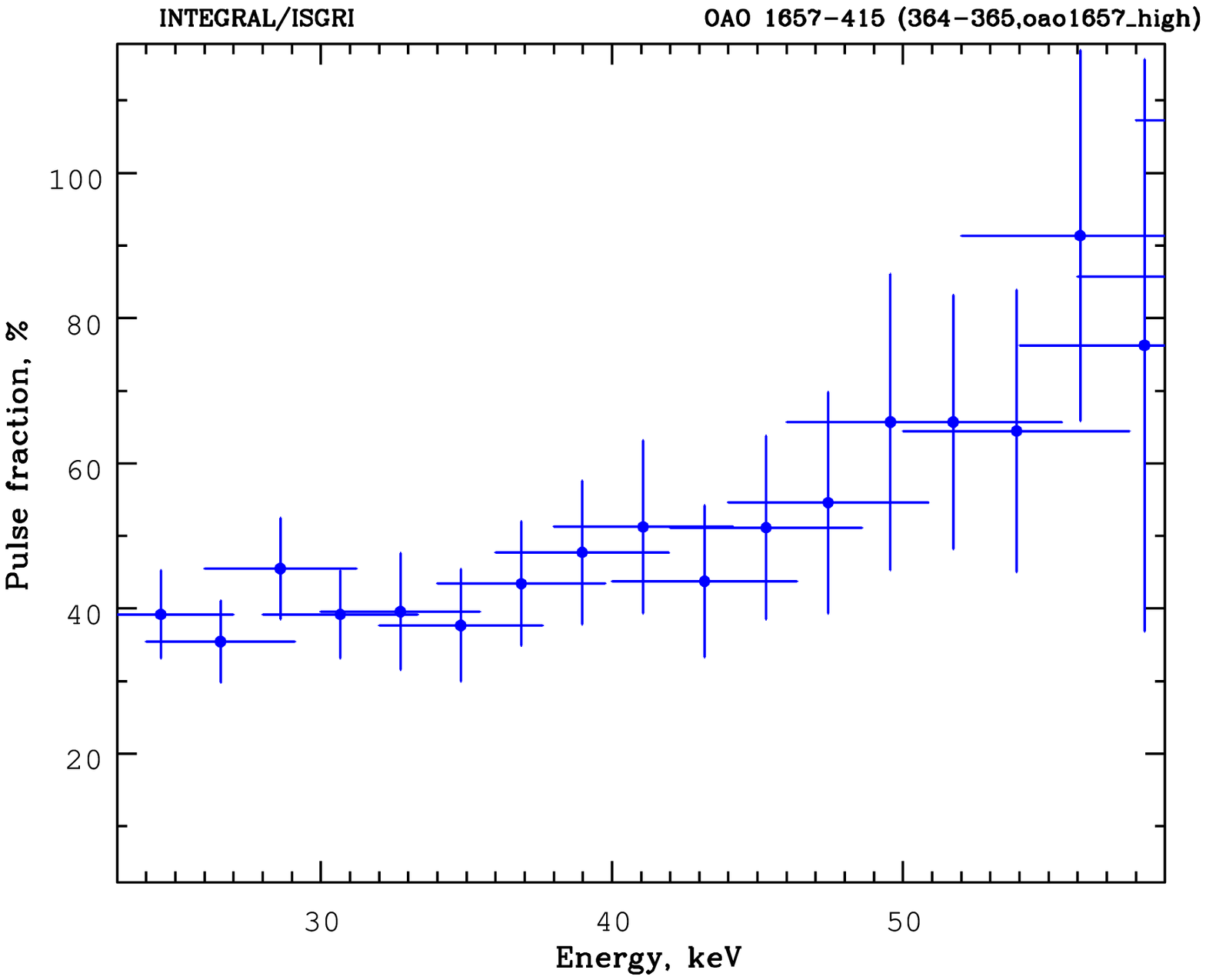}}}
\end{center}
\end{figure}

\end{minipage}

\begin{minipage}[t]{10cm}
\begin{figure}[H]
\begin{center}
\rput(-2,-3.3){\scalebox{1}{\includegraphics[width=11cm,bb=60 175 500 698,clip]{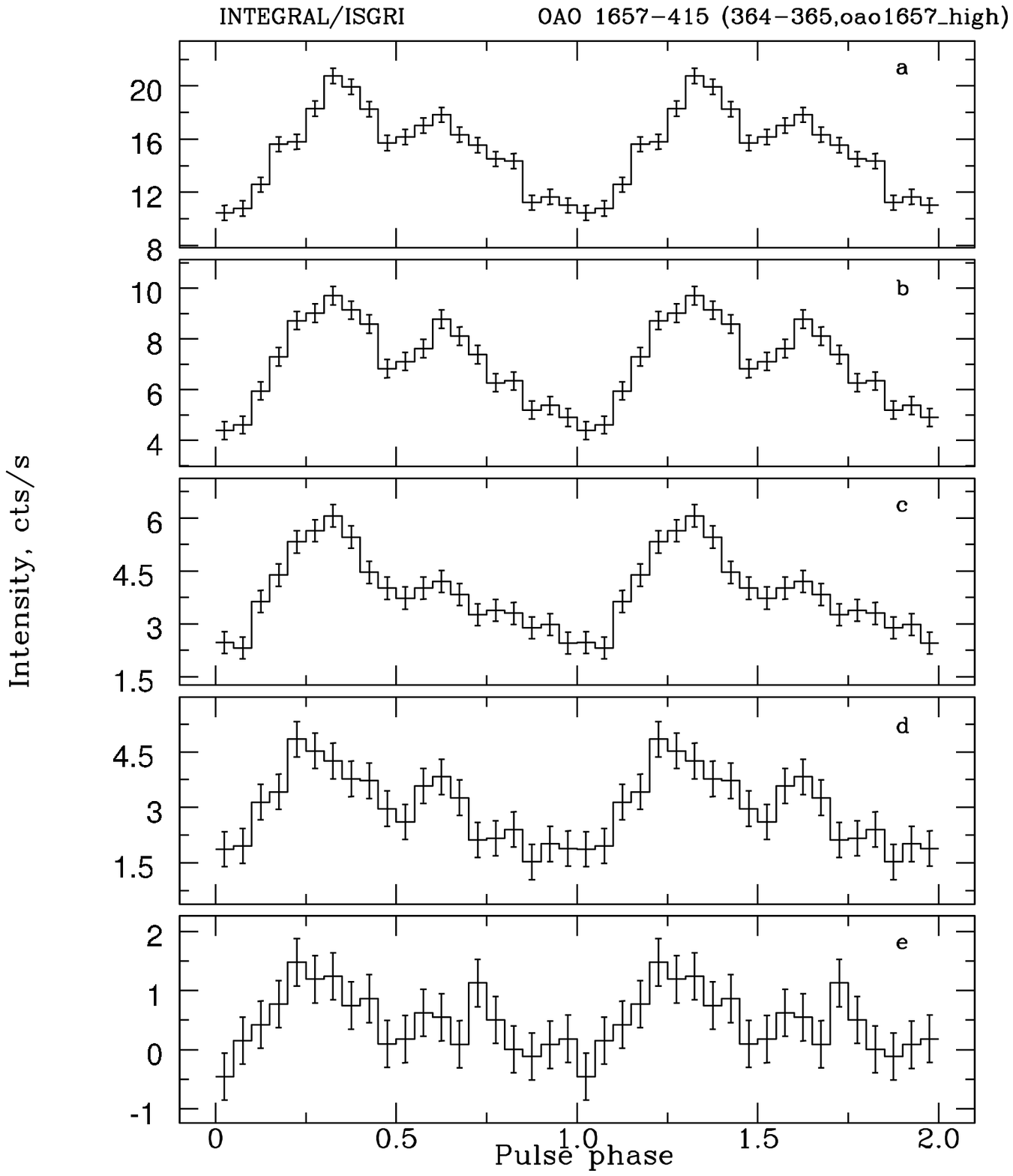}}}
\end{center}
\end{figure}

\end{minipage}

\vspace{85mm}

\begin{minipage}[b]{15cm}
\begin{figure}[H]
\begin{center}
\rput(-2,-3.3){\scalebox{1}{\includegraphics[width=1cm,bb=60 175 500 698,clip]{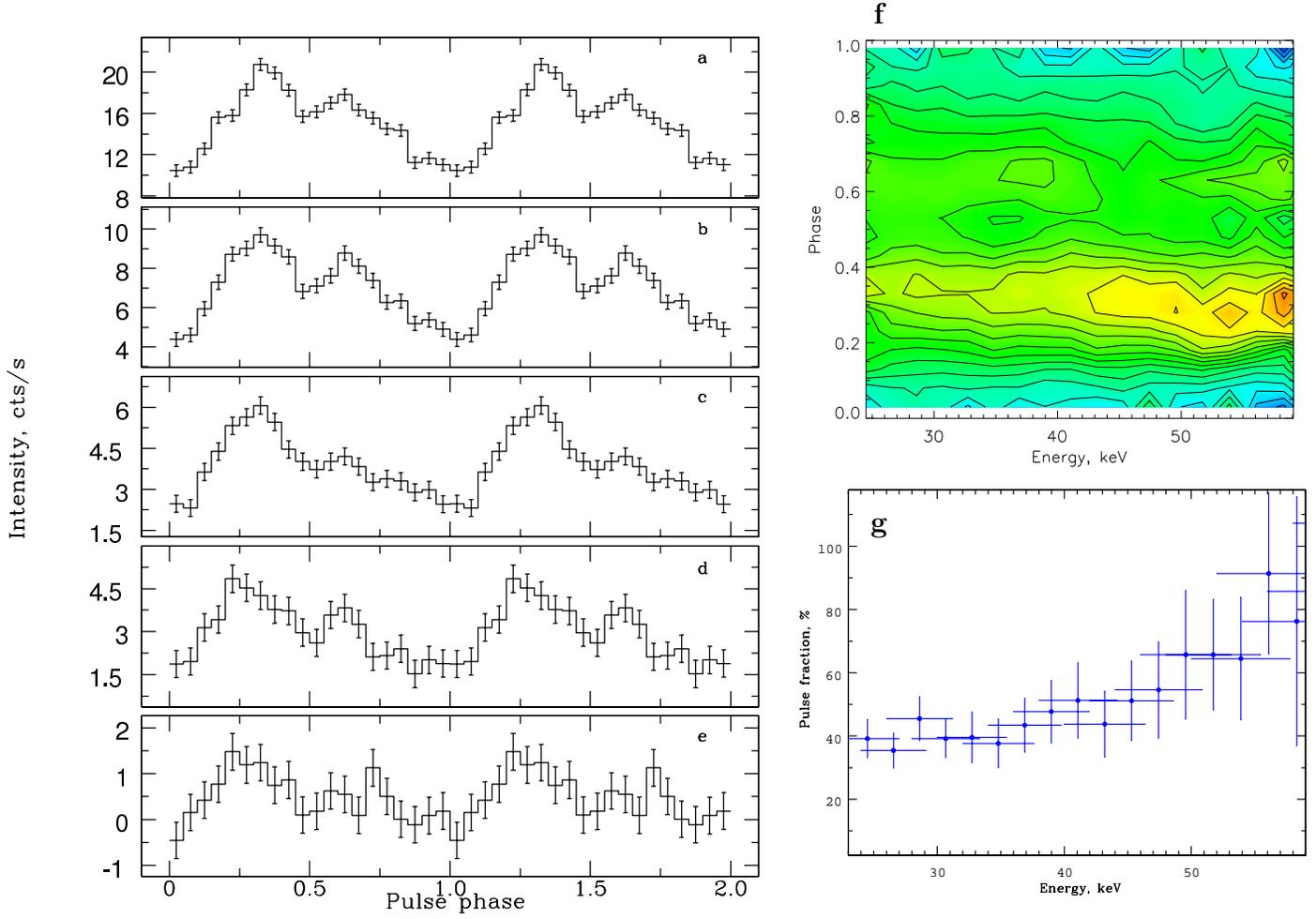}}}
\caption{ Same as Fig. 4 for the source OAO 1657-415, orbit 364.}\label{oaocont}
\end{center}
\end{figure}
\end{minipage}
%********************************************

\clearpage

%********************************************
\begin{minipage}[t]{9cm}
\begin{figure}[H]
\begin{center}
\rput(8.0,-2.6){\scalebox{1}{\includegraphics[width=7cm,bb=45 310 515 725,clip]{figures1/herx1_high_339_339_rev_contour_col.ps2}}}
\rput(5.5,0.8){\mbox{\bf f}}
\rput(5.5,-6.5){\mbox{\bf g}}
\end{center}
\end{figure}

\end{minipage}

\begin{minipage}[h]{9cm}
\begin{figure}[H]
\begin{center}
\rput(8.3,-7.6){\scalebox{1}{\includegraphics[width=8cm,bb=20 275 560 672,clip]{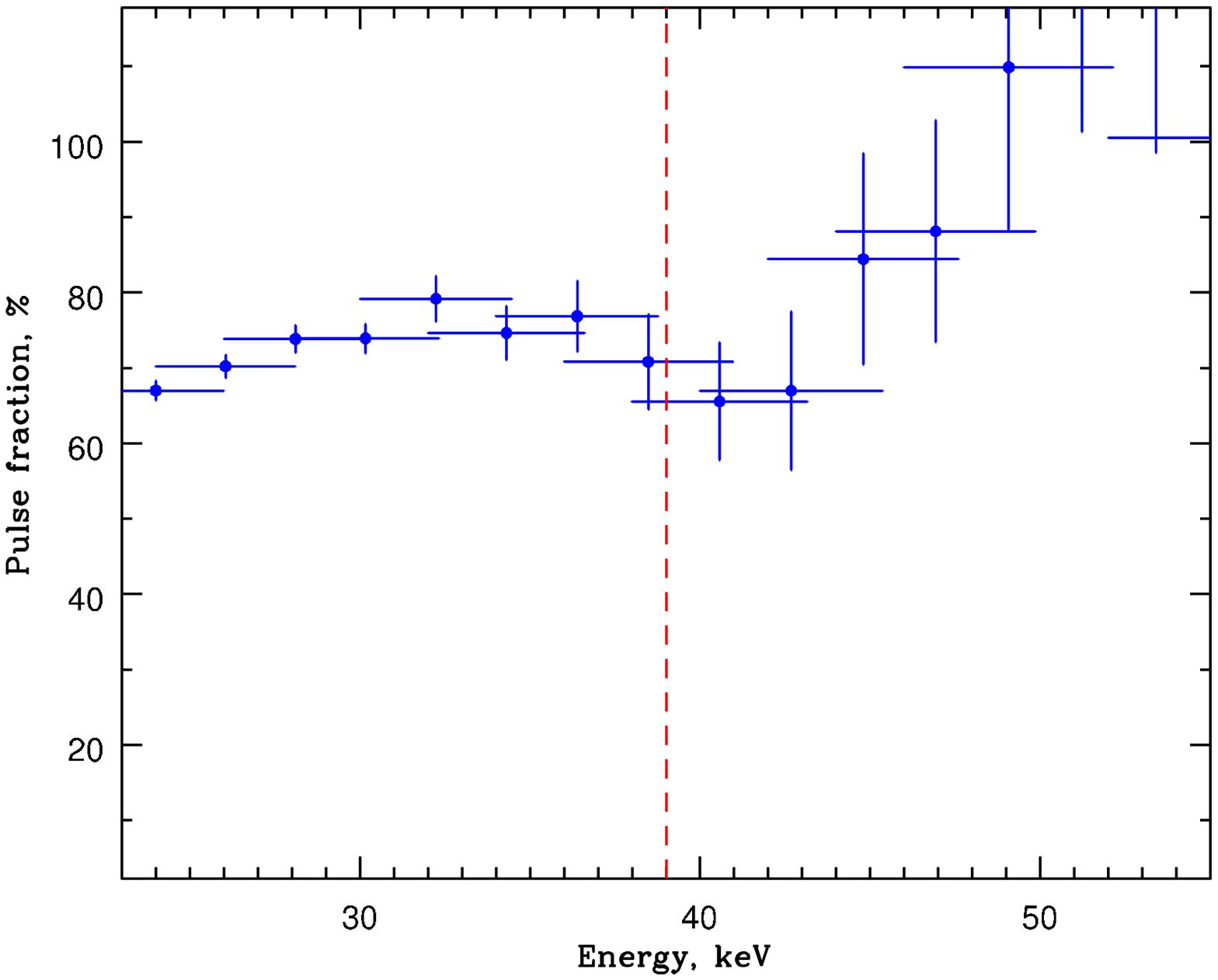}}}
\end{center}
\end{figure}

\end{minipage}

\begin{minipage}[t]{10cm}
\begin{figure}[H]
\begin{center}
\rput(-2,-3.3){\scalebox{1}{\includegraphics[width=11cm,bb=60 175 500 698,clip]{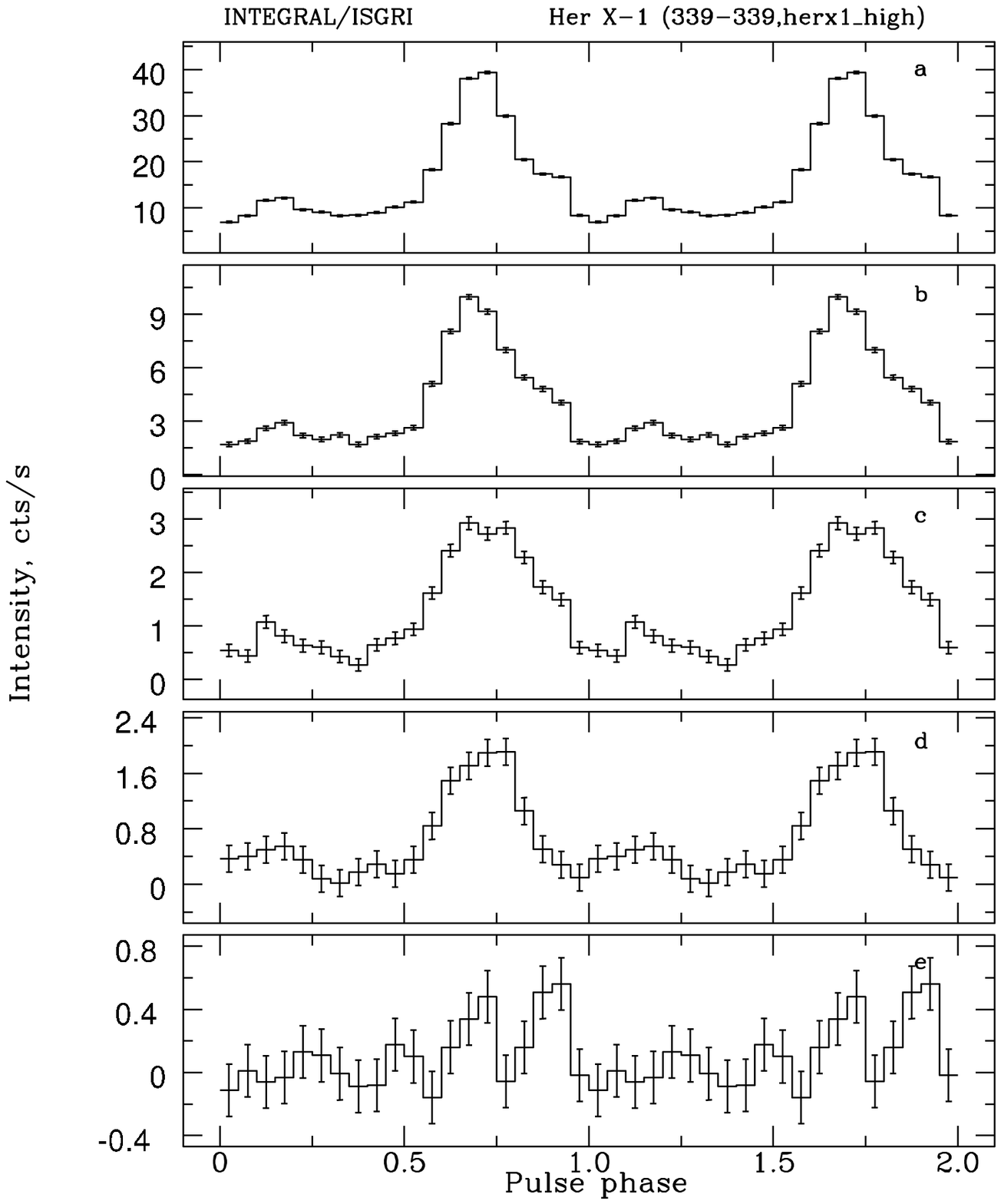}}}
\end{center}
\end{figure}

\end{minipage}

\vspace{85mm}

\begin{minipage}[b]{15cm}
\begin{figure}[H]
\begin{center}
\rput(-2,-3.3){\scalebox{1}{\includegraphics[width=1cm,bb=60 175 500 698,clip]{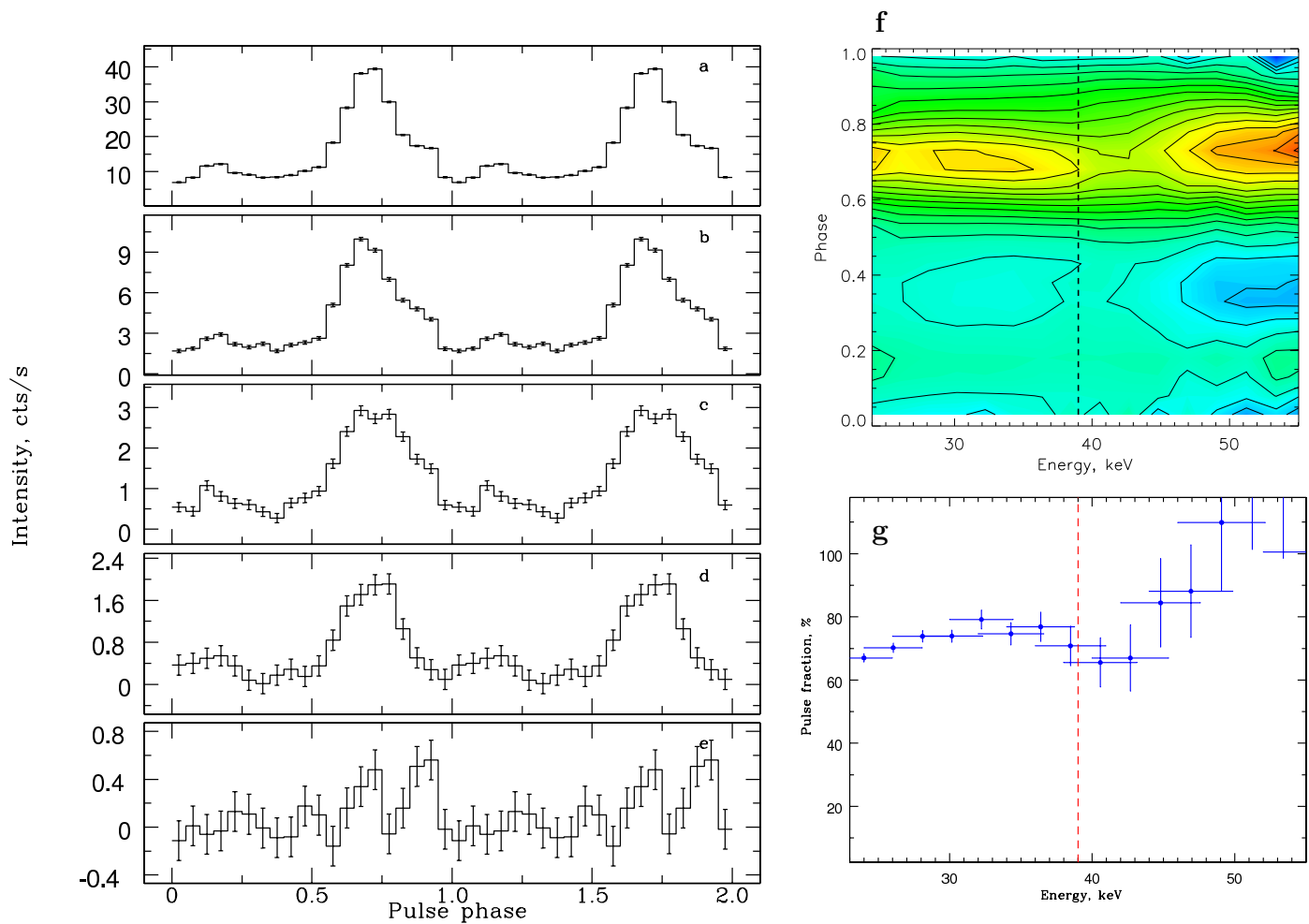}}}
\caption{ Same as Fig. 4 for the source Her X-1, orbit 339, a high state.}\label{herx1}
\end{center}
\end{figure}
\end{minipage}
%********************************************

\clearpage

%********************************************
\begin{minipage}[t]{9cm}
\begin{figure}[H]
\begin{center}
\rput(8.0,-2.6){\scalebox{1}{\includegraphics[width=7cm,bb=45 310 515 725,clip]{figures1/gx14.all_46_46_rev_contour_col.ps2}}}
\rput(5.5,0.8){\mbox{\bf f}}
\rput(5.5,-6.5){\mbox{\bf g}}
\end{center}
\end{figure}

\end{minipage}

\begin{minipage}[h]{9cm}
\begin{figure}[H]
\begin{center}
\rput(8.3,-7.6){\scalebox{1}{\includegraphics[width=8cm,bb=20 275 560 672,clip]{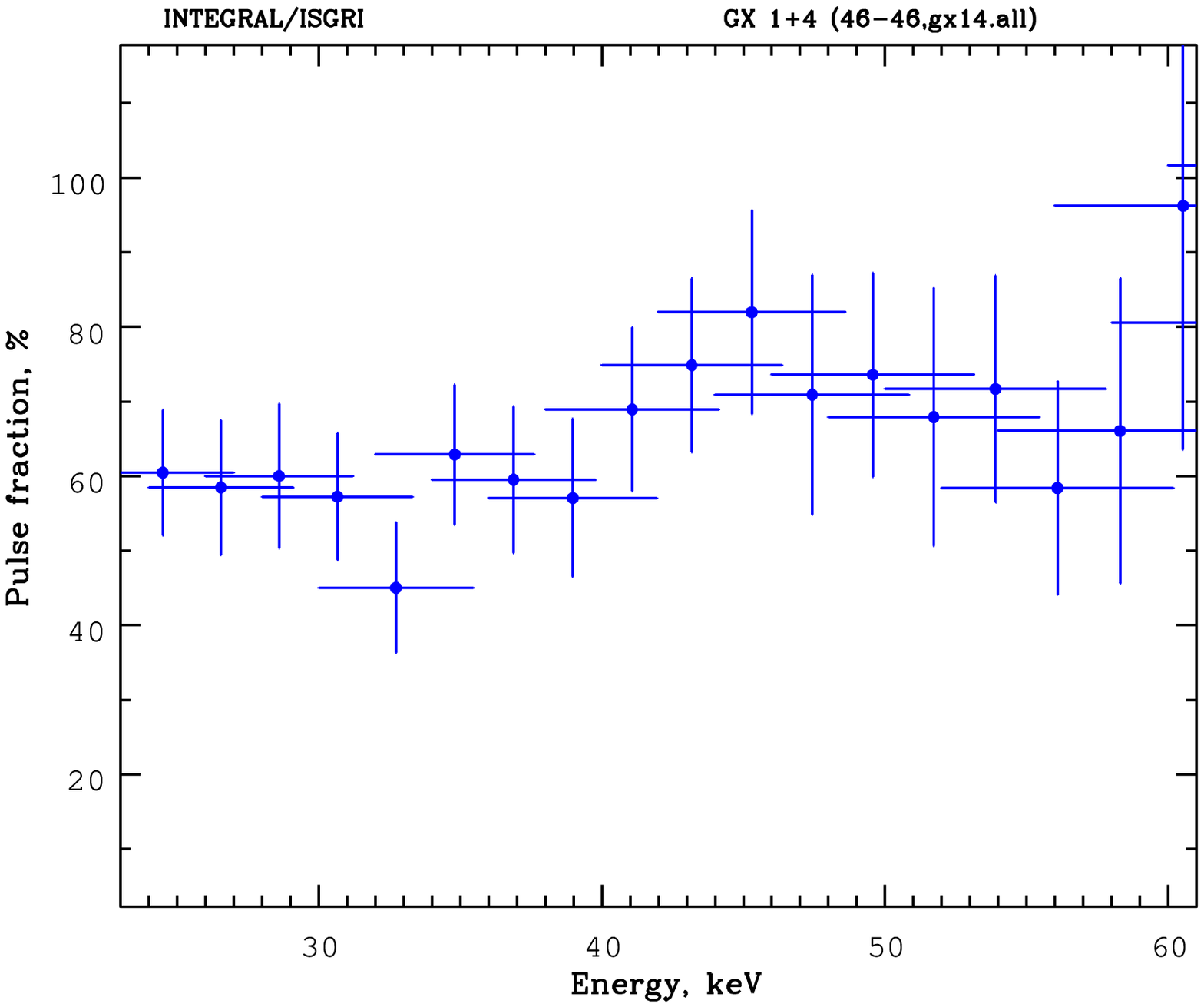}}}
\end{center}
\end{figure}

\end{minipage}

\begin{minipage}[t]{10cm}
\begin{figure}[H]
\begin{center}
\rput(-2,-3.3){\scalebox{1}{\includegraphics[width=11cm,bb=60 175 500 698,clip]{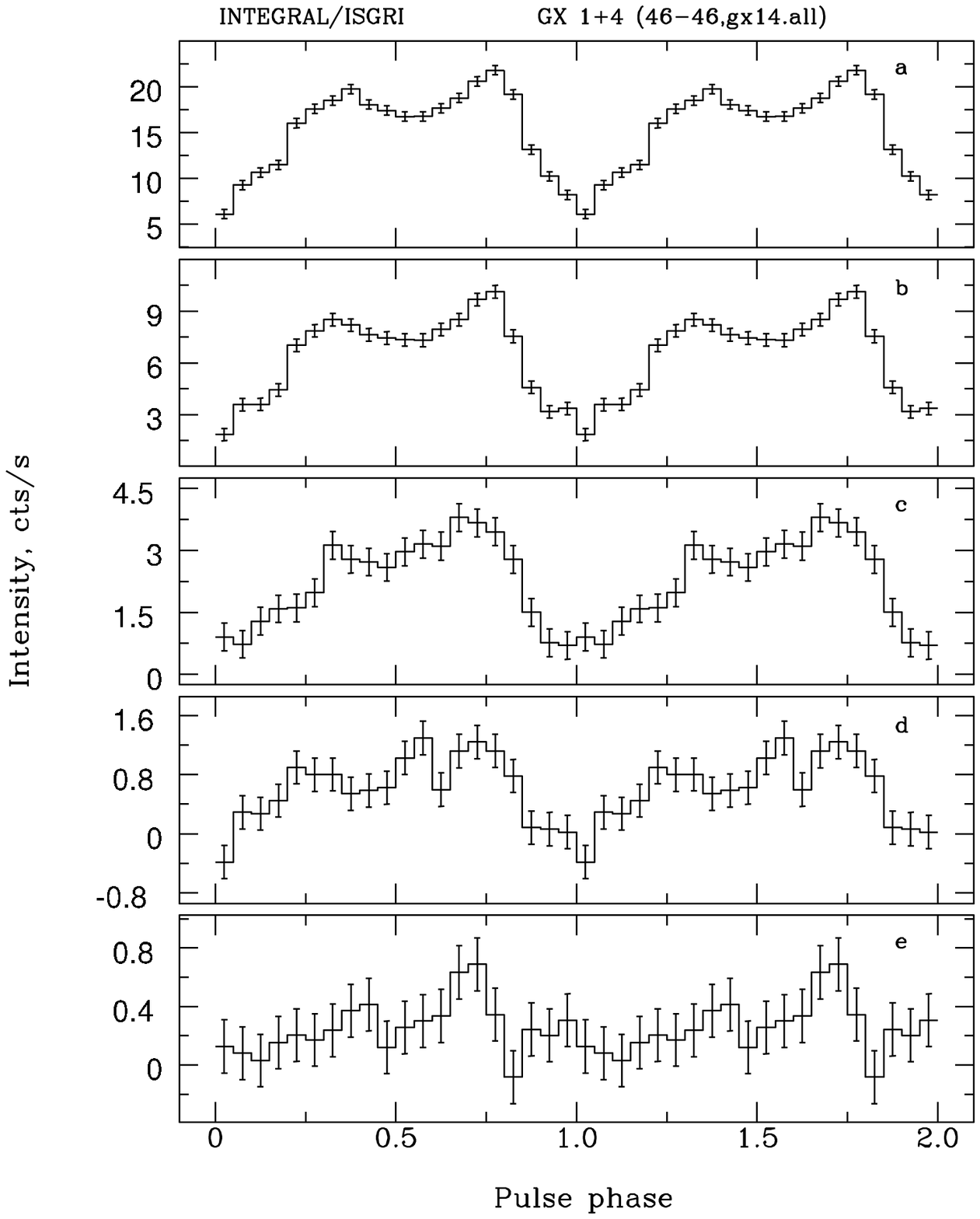}}}
\end{center}
\end{figure}

\end{minipage}

\vspace{85mm}

\begin{minipage}[b]{15cm}
\begin{figure}[H]
\begin{center}
\rput(-2,-3.3){\scalebox{1}{\includegraphics[width=1cm,bb=60 175 500 698,clip]{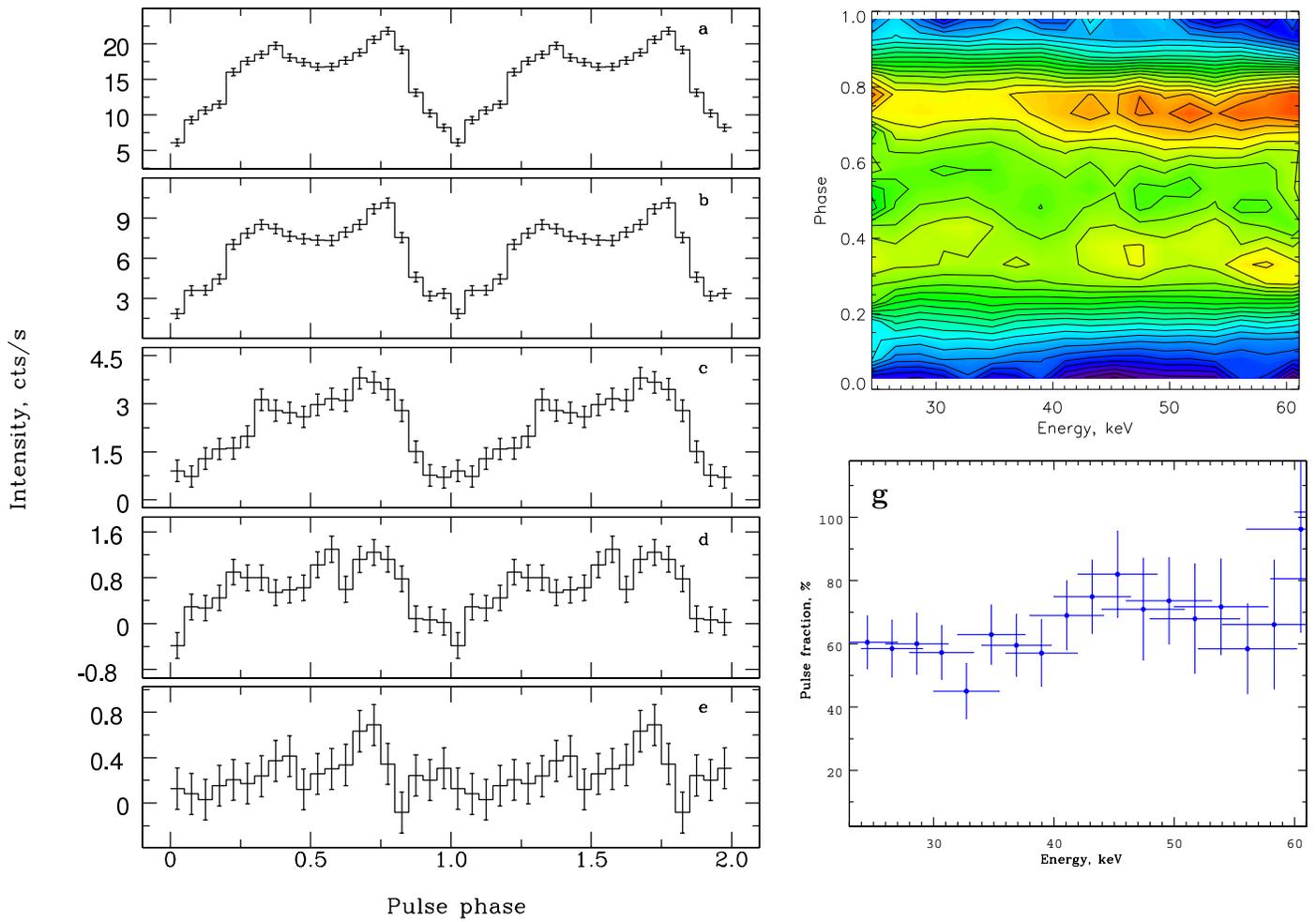}}}
\caption{ Same as Fig. 4 for the source GX 1+4, orbit 46.}\label{gx14two}
\end{center}
\end{figure}
\end{minipage}
%********************************************

\clearpage

%********************************************
\begin{minipage}[t]{9cm}
\begin{figure}[H]
\begin{center}
\rput(8.0,-2.6){\scalebox{1}{\includegraphics[width=7cm,bb=45 310 515 725,clip]{figures1/gx14.all_235_237_rev_contour_col.ps2}}}
\rput(5.5,0.8){\mbox{\bf f}}
\rput(5.5,-6.5){\mbox{\bf g}}
\end{center}
\end{figure}

\end{minipage}

\begin{minipage}[h]{9cm}
\begin{figure}[H]
\begin{center}
\rput(8.3,-7.6){\scalebox{1}{\includegraphics[width=8cm,bb=20 275 560 672,clip]{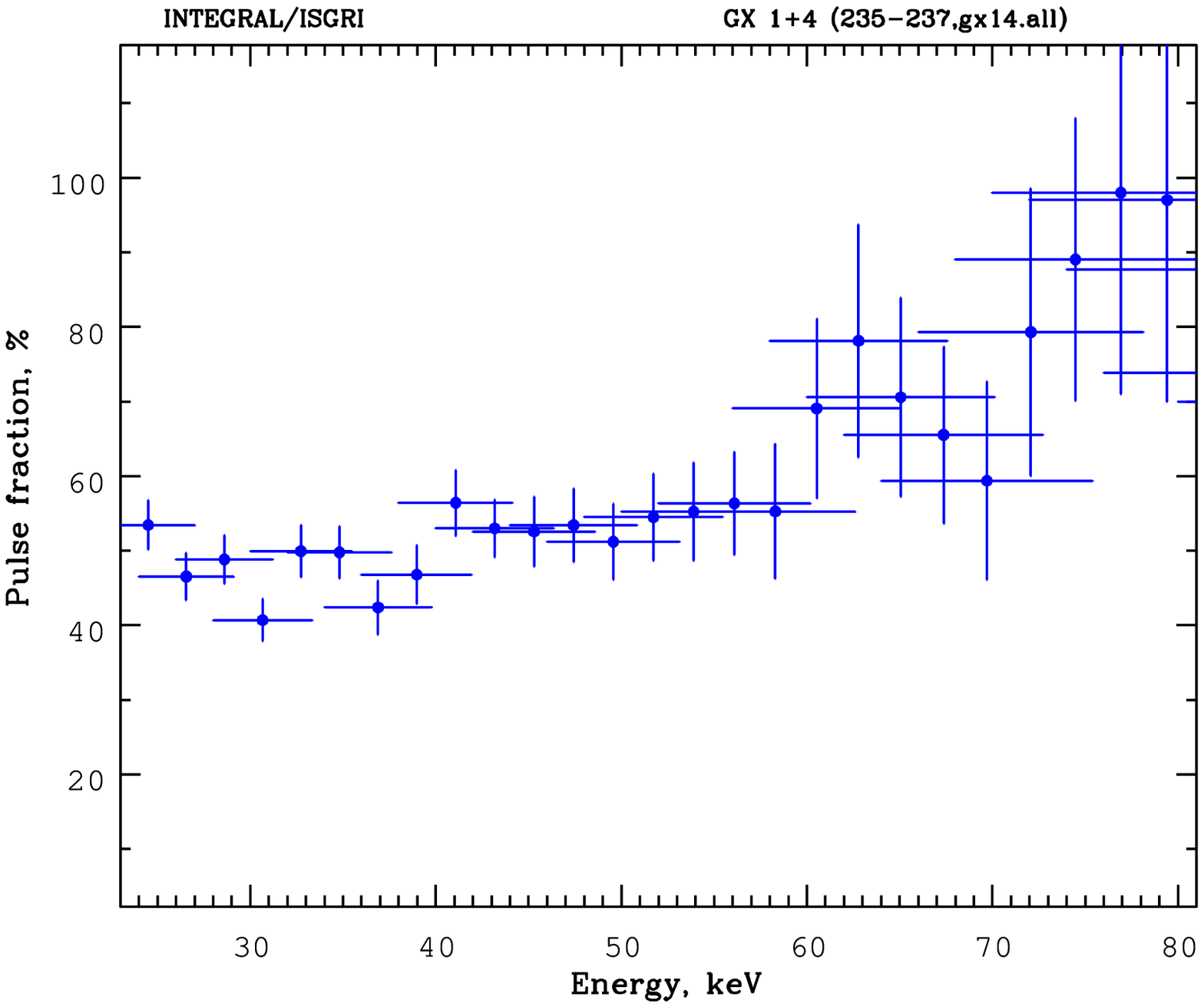}}}
\end{center}
\end{figure}

\end{minipage}

\begin{minipage}[t]{10cm}
\begin{figure}[H]
\begin{center}
\rput(-2,-3.3){\scalebox{1}{\includegraphics[width=11cm,bb=60 175 500 698,clip]{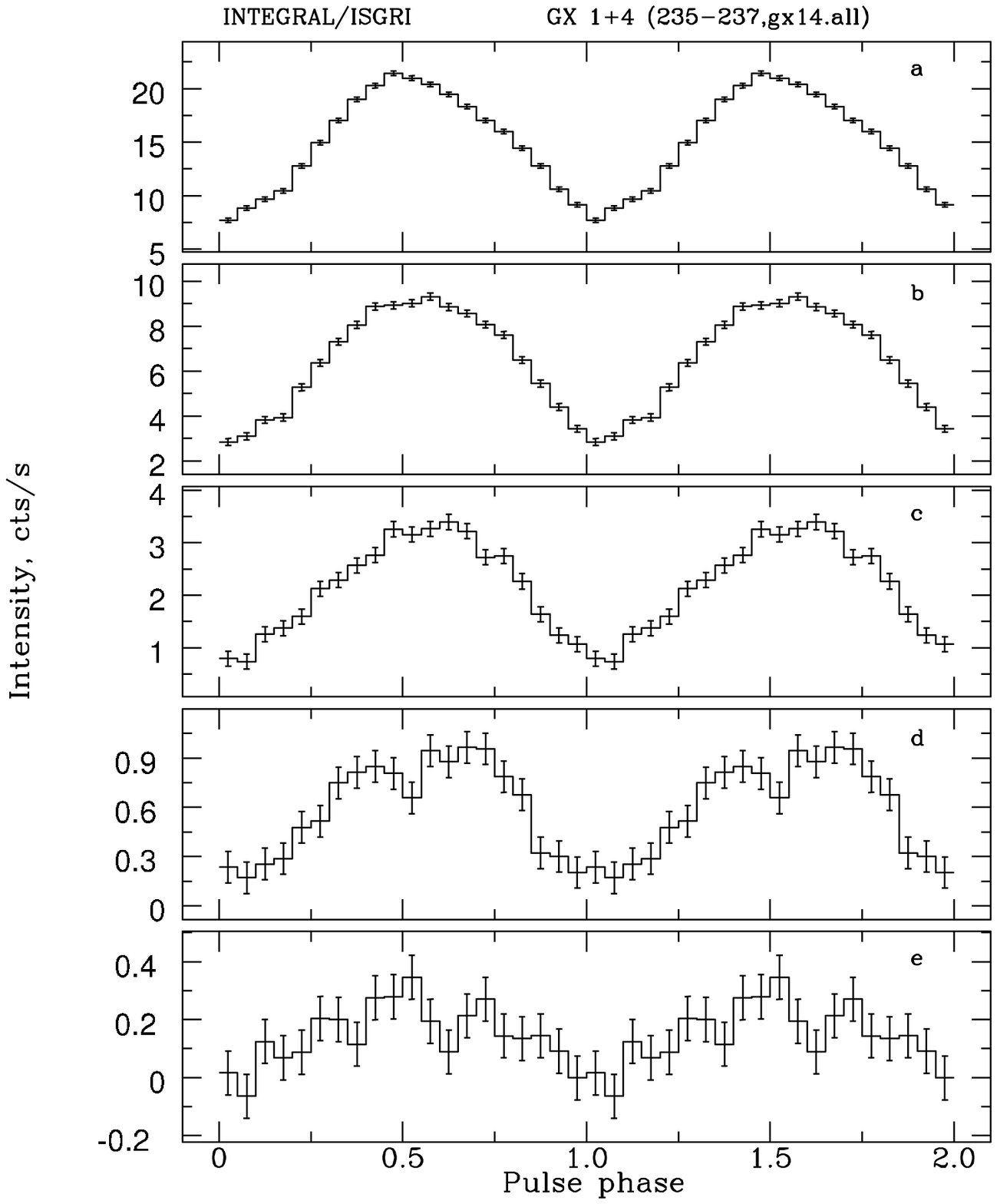}}}
\end{center}
\end{figure}

\end{minipage}

\vspace{85mm}

\begin{minipage}[b]{15cm}
\begin{figure}[H]
\begin{center}
\rput(-2,-3.3){\scalebox{1}{\includegraphics[width=1cm,bb=60 175 500 698,clip]{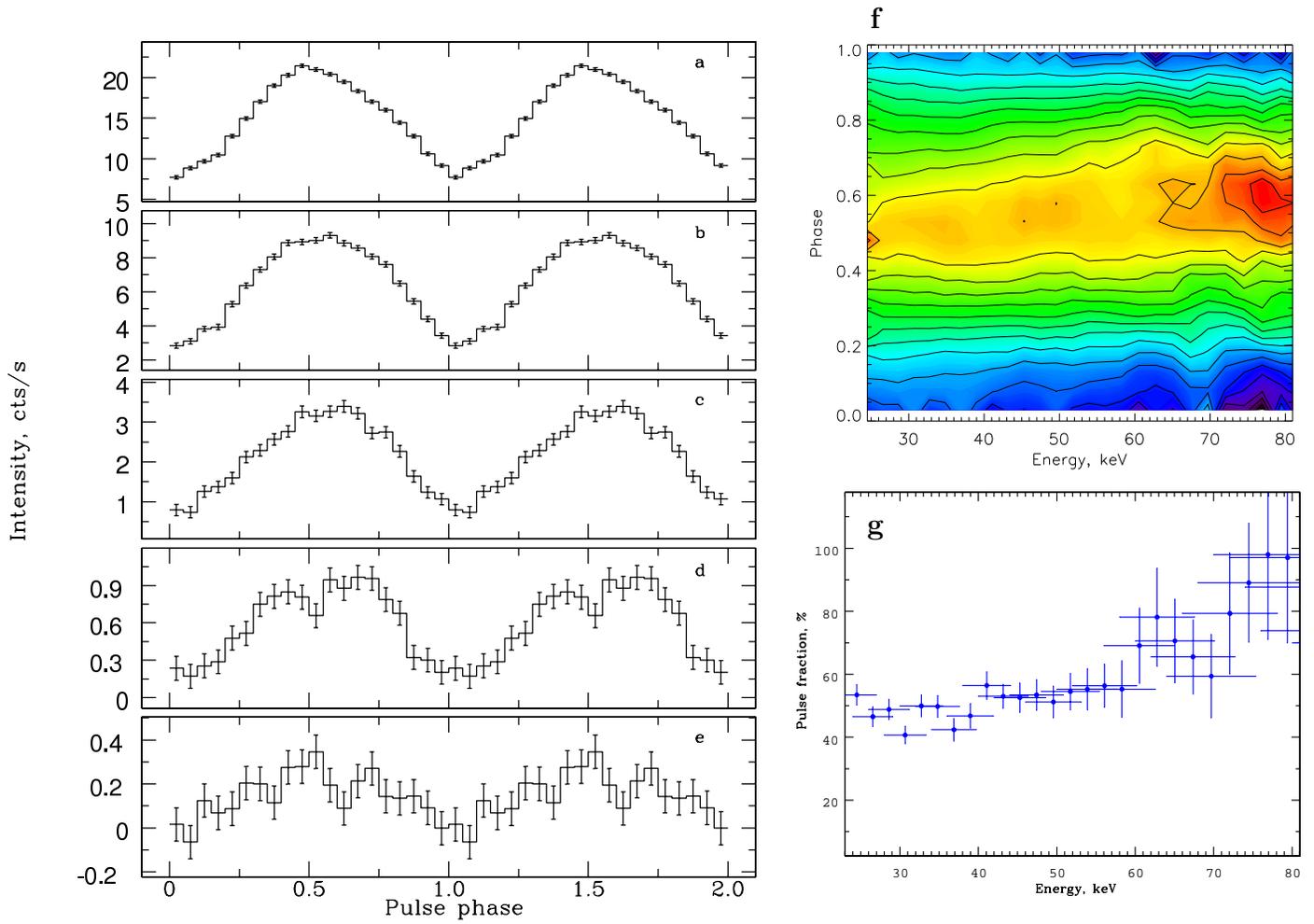}}}
\caption{ Same as Fig. 4 for the source GX 1+4, orbits 235-237.}\label{gx14one}
\end{center}
\end{figure}
\end{minipage}
%********************************************

\clearpage

%********************************************
\begin{minipage}[t]{9cm}
\begin{figure}[H]
\begin{center}
\rput(8.0,-2.6){\scalebox{1}{\includegraphics[width=7cm,bb=45 310 515 725,clip]{figures1/exo2030_high_190_190_rev_contour_col.ps2}
}}
\rput(5.5,0.8){\mbox{\bf f}}
\rput(5.5,-6.5){\mbox{\bf g}}
\end{center}
\end{figure}

\end{minipage}

\begin{minipage}[h]{9cm}
\begin{figure}[H]
\begin{center}
\rput(8.3,-7.6){\scalebox{1}{\includegraphics[width=8cm,bb=20 275 560 672,clip]{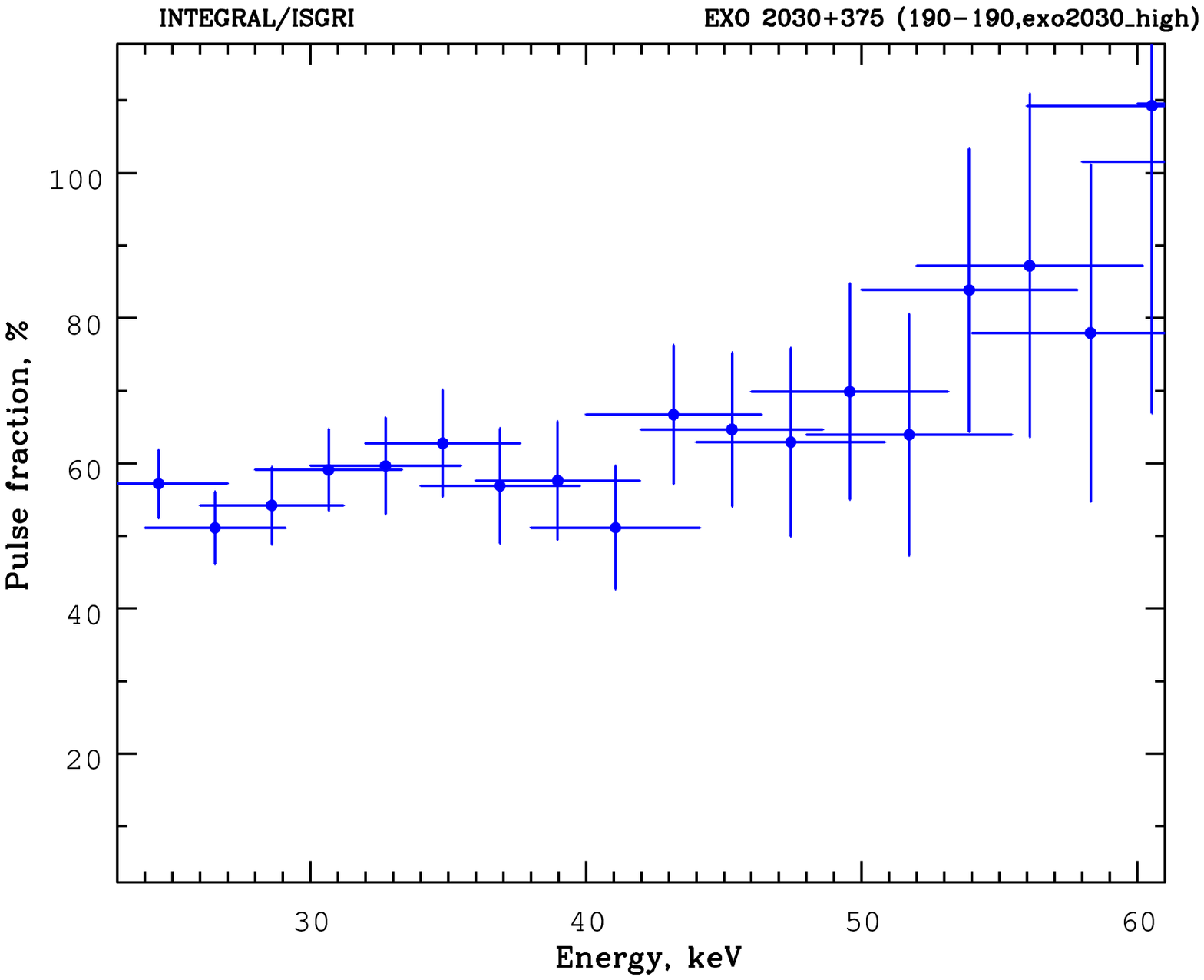}}}
\end{center}
\end{figure}

\end{minipage}

\begin{minipage}[t]{10cm}
\begin{figure}[H]
\begin{center}
\rput(-2,-3.3){\scalebox{1}{\includegraphics[width=11cm,bb=60 175 500 698,clip]{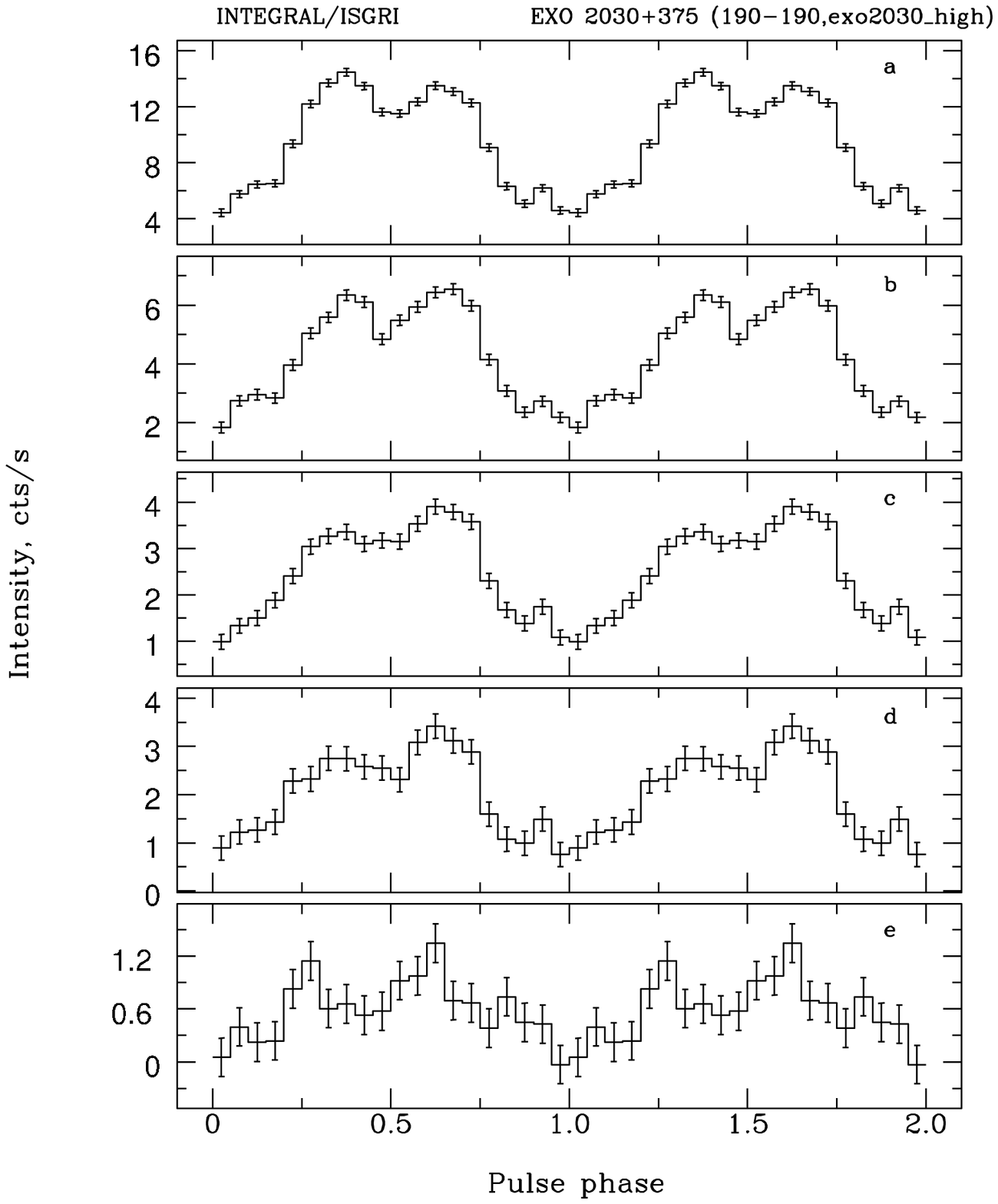}}}
\end{center}
\end{figure}

\end{minipage}

\vspace{85mm}

\begin{minipage}[b]{15cm}
\begin{figure}[H]
\begin{center}
\rput(-2,-3.3){\scalebox{1}{\includegraphics[width=1cm,bb=60 175 500 698,clip]{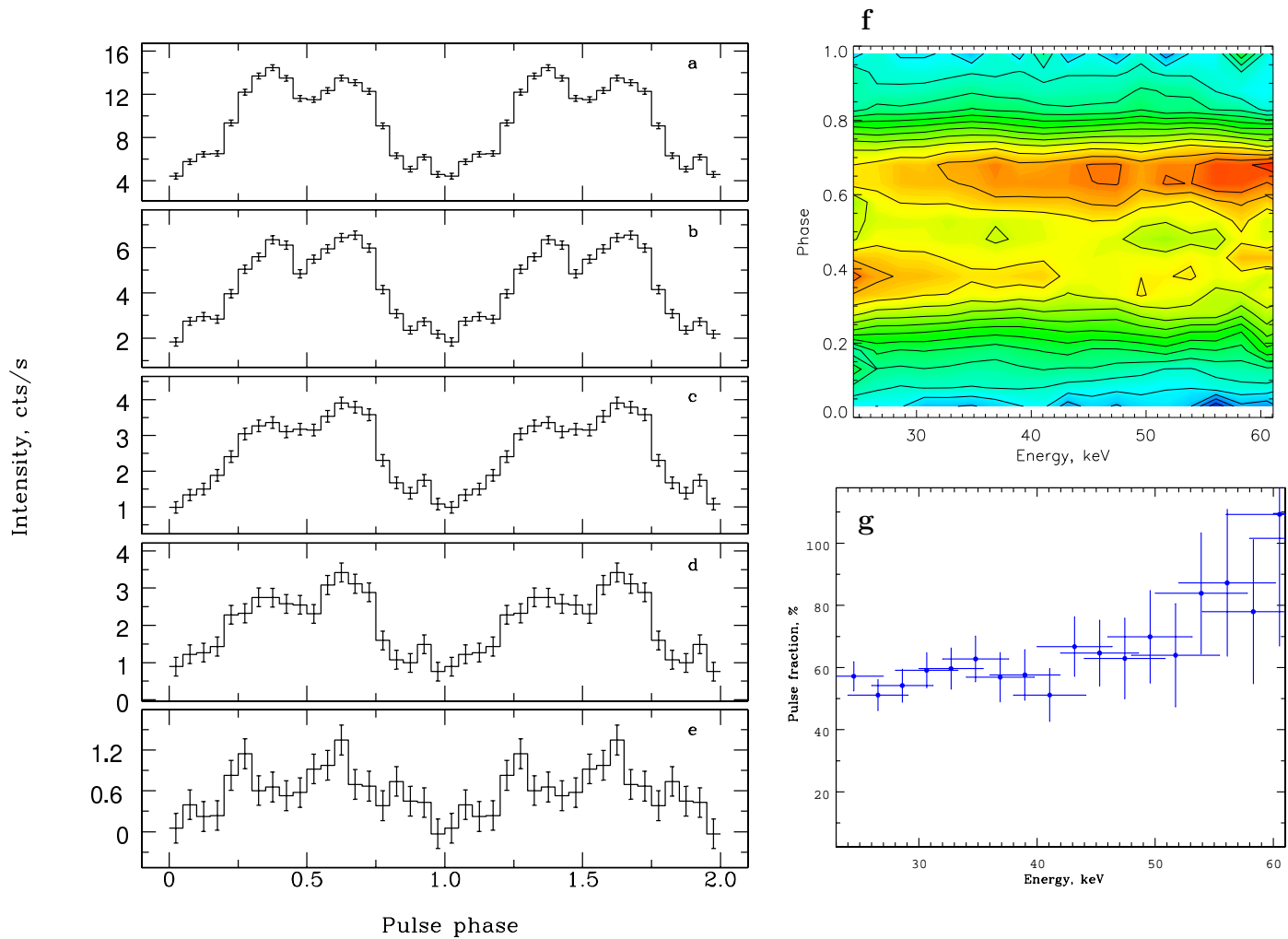}}}
\caption{ Same as Fig. 4 for the source EXO 2030+375, orbit 190, a low state.}\label{exolow}
\end{center}
\end{figure}
\end{minipage}
%********************************************

\clearpage

%********************************************
\begin{minipage}[t]{9cm}
\begin{figure}[H]
\begin{center}
\rput(8.0,-2.6){\scalebox{1}{\includegraphics[width=7cm,bb=45 310 515 725,clip]{figures1/exo2030_flare_460_465_rev_contour_col.ps2}
}}
\rput(5.5,0.8){\mbox{\bf f}}
\rput(5.5,-6.5){\mbox{\bf g}}
\end{center}
\end{figure}

\end{minipage}

\begin{minipage}[h]{9cm}
\begin{figure}[H]
\begin{center}
\rput(8.3,-7.6){\scalebox{1}{\includegraphics[width=8cm,bb=20 275 560 672,clip]{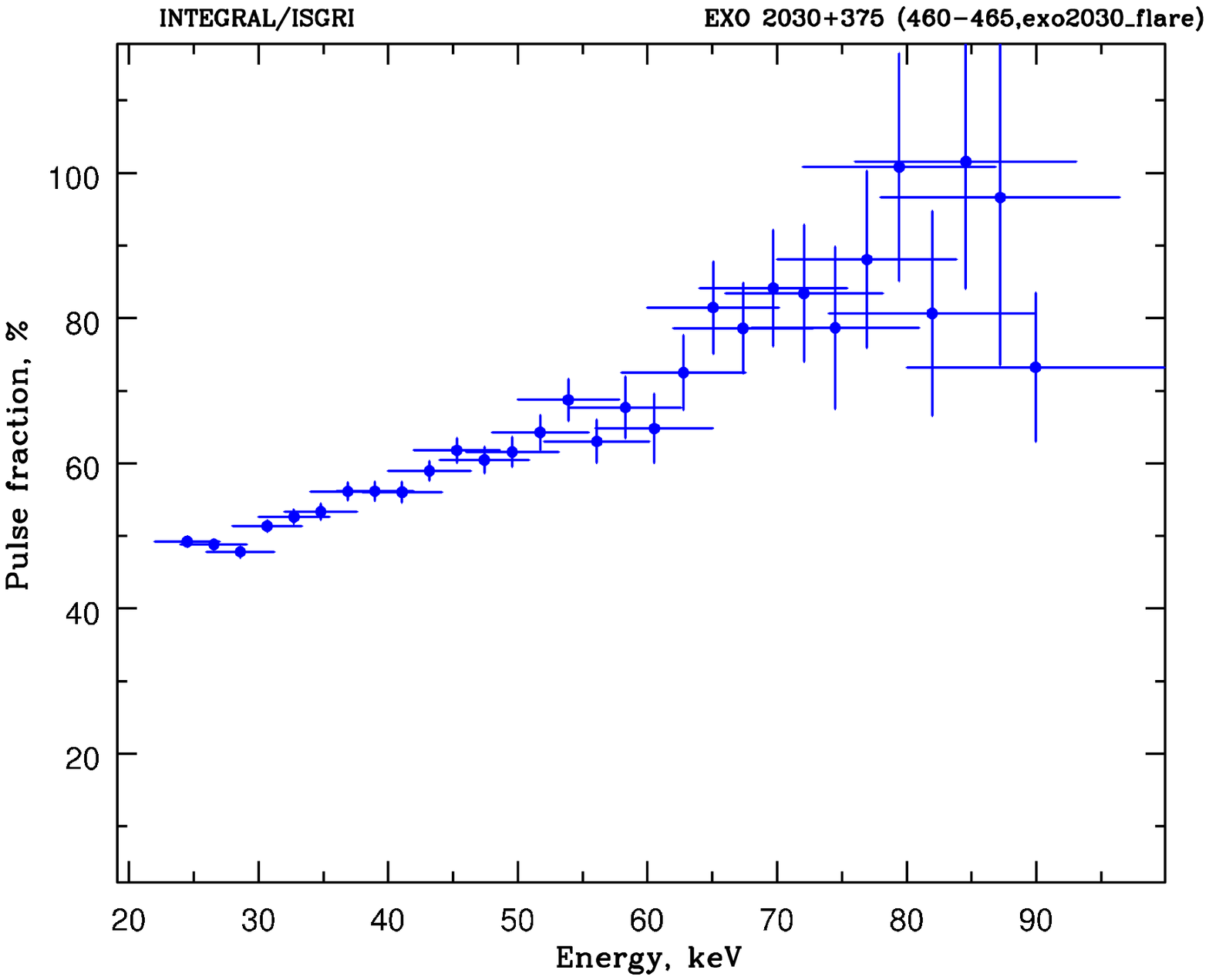}}}
\end{center}
\end{figure}

\end{minipage}

\begin{minipage}[t]{10cm}
\begin{figure}[H]
\begin{center}
\rput(-2,-3.3){\scalebox{1}{\includegraphics[width=11cm,bb=60 175 500 698,clip]{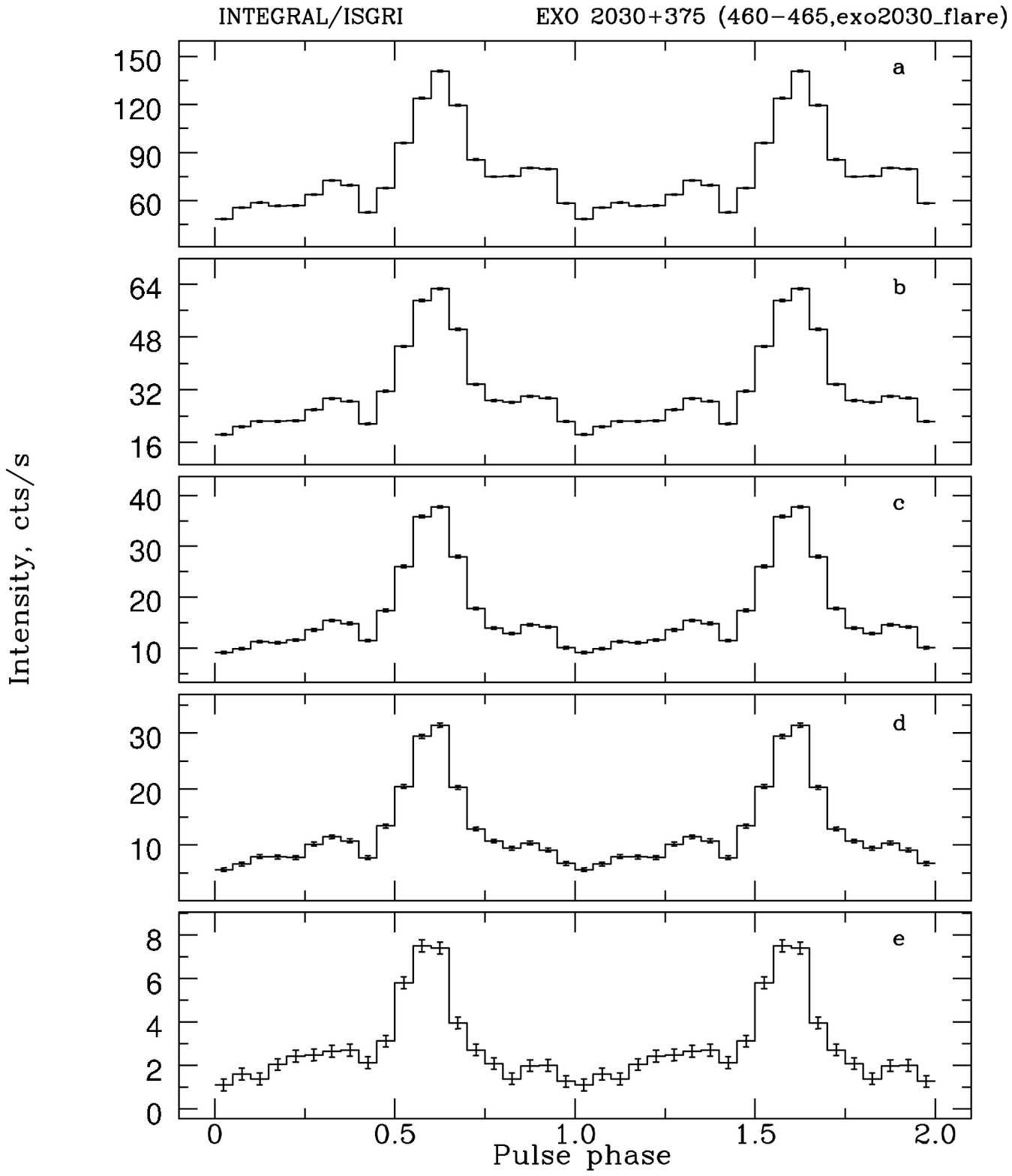}}}
\end{center}
\end{figure}

\end{minipage}

\vspace{85mm}

\begin{minipage}[b]{15cm}
\begin{figure}[H]
\begin{center}
\rput(-2,-3.3){\scalebox{1}{\includegraphics[width=1cm,bb=60 175 500 698,clip]{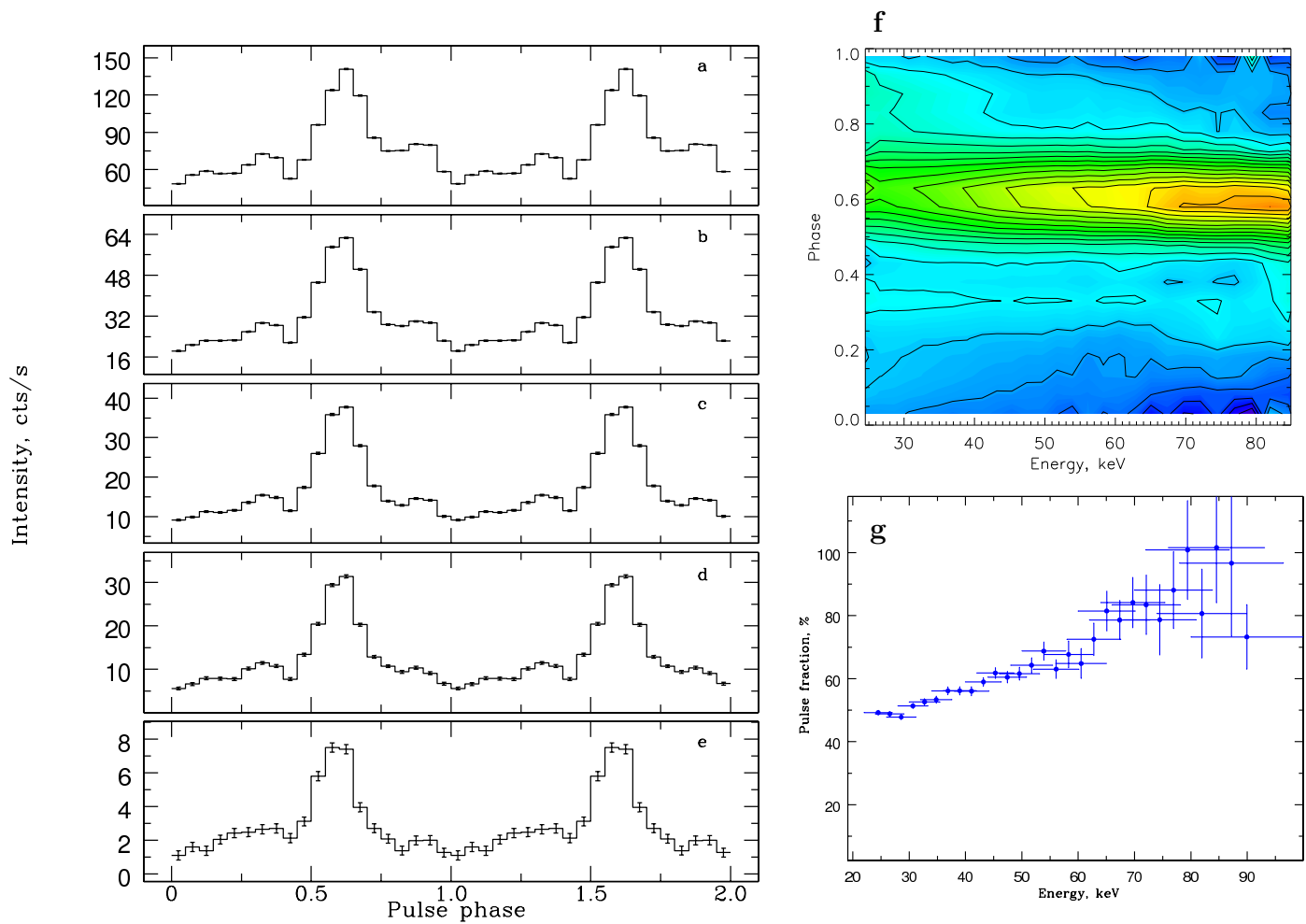}}}
\caption{ Same as Fig. 4 for the source EXO 2030+375, orbit 462, a high state.}\label{exohigh}
\end{center}
\end{figure}
\end{minipage}
%********************************************

\clearpage

%********************************************
\begin{figure*}[t]
\hspace{-10mm}\hbox{\includegraphics[width=8cm,clip]{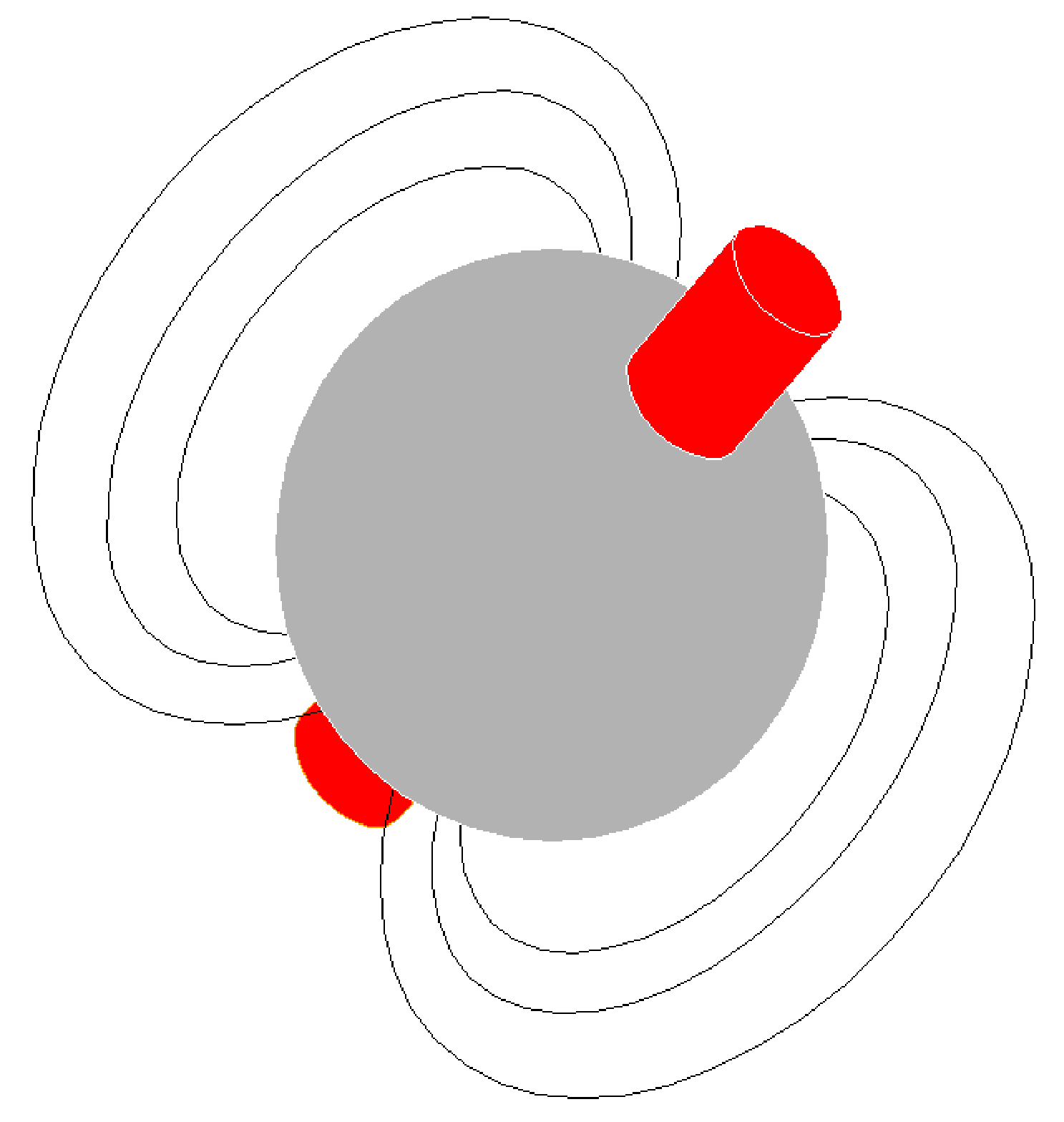}
\hspace{5mm}\includegraphics[width=8cm,clip]{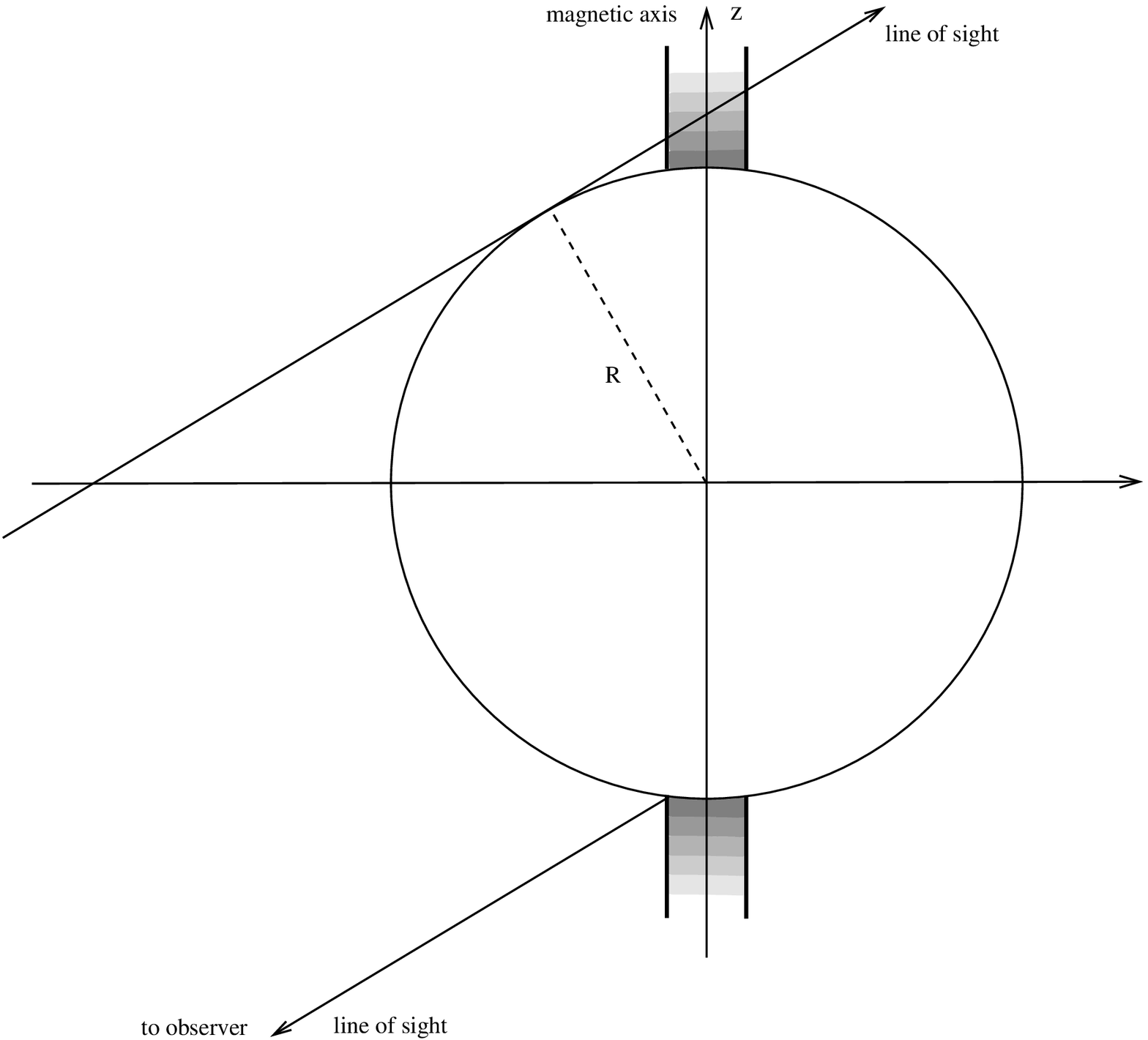}
}

\vfill

\caption{ Geometry of the radiation model for accretion columns of an X-ray pulsar (darker regions correspond to higher
temperatures) and relative positions of the observer, the rotation axis, and the magnetic axis (for more detail, see the text).}\label{toymodel}
\end{figure*}
%********************************************

\clearpage

%********************************************
\begin{figure*}[t]
\centerline{\includegraphics[width=16cm,bb=45 75 525 805,clip]{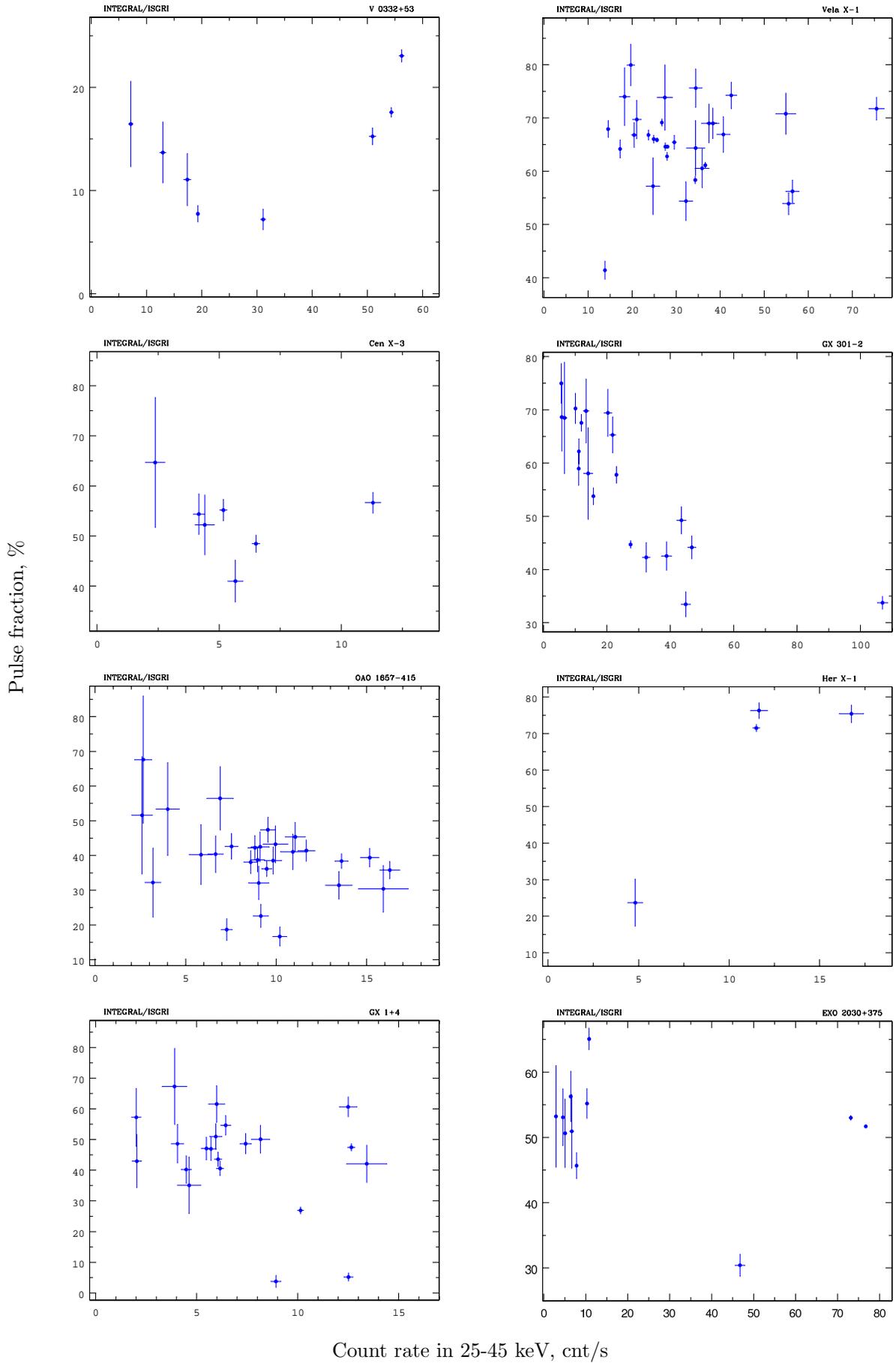}}

\vfill

\caption{ PF for X-ray pulsars in the 25-45 keV energy band versus flux recorded from them in this energy band.}\label{pf_flux}
\end{figure*}
%********************************************

\end{document}